\documentclass[twocolumn]{aastex631}

\usepackage{amsmath}	
\usepackage{amssymb}	
\usepackage{acronym} 
\usepackage{longtable}
\usepackage{multirow}
\usepackage{units}

\acrodef{ATNF}[ATNF]{Australia Telescope National Facility}
\acrodef{PHL}[PHL]{Parkes High-Latitude}
\acrodef{PMB}[PMB]{Parkes Multi-Beam}
\acrodef{SMB}[SMB]{Swinburne Intermediate-latitude}
\acrodef{MLP}[MLP]{multi-layer perceptron}
\acrodef{CNN}[CNN]{convolutional neural network}
\acrodef{DM}[$DM$]{dispersion measure}
\acrodef{SNR}[SNR]{Supernova Remnant}
\acrodef{GRBs}[GRBs]{Gamma-Ray Bursts}
\acrodef{FRBs}[FRBs]{Fast Radio Bursts}
\acrodef{ANNs}[ANNs]{artificial neural networks}
\acrodef{Adam}[Adam]{Adaptive Moment Estimation}
\acrodef{RMSE}[RMSE]{root-mean-square error}
\acrodef{MRE}[MRE]{mean relative error}
\acrodef{SKA}[SKA]{Square Kilometer Array}
\acrodef{ML}[ML]{machine learning}

\accepted{ApJ}

\shorttitle{Analyzing the Galactic pulsar distribution with ML}
\shortauthors{M. Ronchi et al.}

\begin{document}

\title{Analyzing the Galactic pulsar distribution with machine learning}


\correspondingauthor{Michele Ronchi, Vanessa Graber}

\author[0000-0003-2781-9107]{M. Ronchi}
\email{ronchi@ice.csic.es}
\affil{Institute of Space Sciences (ICE, CSIC), Campus UAB, Carrer de Can Magrans s/n, 08193, Barcelona, Spain}
\affil{Institut d'Estudis Espacials de Catalunya (IEEC), Carrer Gran Capit\`a 2--4, 08034 Barcelona, Spain} 

\author[0000-0002-6558-1681]{V. Graber}
\affiliation{Institute of Space Sciences (ICE, CSIC), Campus UAB, Carrer de Can Magrans s/n, 08193, Barcelona, Spain}
\affiliation{Institut d'Estudis Espacials de Catalunya (IEEC), Carrer Gran Capit\`a 2--4, 08034 Barcelona, Spain} 

\author[0000-0002-9575-6403]{A. Garcia-Garcia}
\affiliation{Institute of Space Sciences (ICE, CSIC), Campus UAB, Carrer de Can Magrans s/n, 08193, Barcelona, Spain}
\affiliation{Institut d'Estudis Espacials de Catalunya (IEEC), Carrer Gran Capit\`a 2--4, 08034 Barcelona, Spain} 
\affiliation{Departament de Física Aplicada, Universitat d'Alacant, 03690 Alicante, Spain} 

\author[0000-0003-2177-6388]{N. Rea}
\affiliation{Institute of Space Sciences (ICE, CSIC), Campus UAB, Carrer de Can Magrans s/n, 08193, Barcelona, Spain}
\affiliation{Institut d'Estudis Espacials de Catalunya (IEEC), Carrer Gran Capit\`a 2--4, 08034 Barcelona, Spain} 

\author[0000-0003-1018-8126]{J. A. Pons}
\affiliation{Departament de Física Aplicada, Universitat d'Alacant, 03690 Alicante, Spain}

\begin{abstract}
We explore the possibility of inferring the properties of the Galactic neutron star population through machine learning. In particular, in this paper we focus on their dynamical characteristics and show that an artificial neural network is able to estimate with high accuracy the parameters which control the current positions of a mock population of pulsars. For this purpose, we implement a simplified population-synthesis framework (where selection biases are neglected at this stage) and concentrate on the natal kick-velocity distribution and the distribution of birth distances from the Galactic plane. By varying these and evolving the pulsar trajectories in time, we generate a series of simulations that are used to train and validate a suitably structured convolutional neural network. We demonstrate that our network is able to recover the parameters governing the kick-velocity and Galactic height distribution with a mean relative error of about $10^{-2}$. We discuss the limitations of our idealized approach and study a toy problem to introduce selection effects in a phenomenological way by incorporating  the observed proper motions of 216 isolated pulsars. Our analysis highlights that increasing the sample of pulsars with accurate proper motion measurements by a factor of $\sim$10, one of the future breakthroughs of the Square Kilometer Array, we might succeed in constraining the birth spatial and kick-velocity distribution of the neutron stars in the Milky Way with high precision through machine learning.

\end{abstract}



\section{Introduction}
\label{sec:intro}

Neutron stars have been observed to travel through the Galaxy with typical velocities of around several hundreds of kilometers per second, reaching more than a thousand kilometers per second in some extreme cases \citep{Chatterjee2005,Hobbs2005,Hui2006,Pavan2014}. Accurate information on neutron star positions and velocities in the Milky Way usually comes from radio timing and interferometric observations \citep[see Chapters 8 and 9 in][and references therein]{Handbook2004,Liu2020} or high spatial resolution X-ray observations with, e.g., the {\em Chandra} X-ray observatory \citep{Motch2009}. These observations provide measurements of the pulsars' angular positions in the sky and their proper motions projected onto the celestial sphere. In some cases, the radio pulse \acf{DM} or the X-ray absorption density ($N_{\rm H}$) together with Galactic free electron-density and hydrogen-density models \citep{Balucinska-Church1992, Taylor1993, Cordes2002, Yao2017} can also yield a rough distance estimate. Moreover, in a few cases, a parallax measurement \citep{Deller2009, Matthews2016, Wang2017, Deller2019} or the presence of a \acf{SNR} \citep{Yao2017} might provide better distance measurements.

Such high proper velocities of the neutron star population as a whole exceed those of their progenitors (typically massive OB stars) \citep[see][and references therein]{Hansen1997, Lai2001}, and cannot be explained by the neutron stars' motion in the Galactic gravitational potential alone. The mechanisms providing such high velocities are still unclear but are likely related to the underlying supernova explosion. One possibility is that the central core of an exploding star receives a kick due to an asymmetric ejection of material from the star's outer layers; a direct result of momentum conservation \citep{Shklovskii1970, Dewey1987, Mandel2020}. Additionally, the anisotropic emission of neutrinos has been suggested to impart kicks on compact remnants \citep{Bisnovatyi-Kogan1993, Fryer2006, Tamborra2014, Nagakura2019}.

However, constraining the neutron stars' natal kick-velocity distribution from current observational data is not straightforward. Most pulsars, especially those with very high velocities, have moved far away from their birth places, and their proper motions have been modified by the Galactic gravitational potential. Thus, the current velocity of a pulsar may differ substantially from its velocity at birth. Knowing the exact pulsar age and its current 3D spatial velocity, we are in principle able to recover the initial conditions by integrating the pulsar's orbit back in time. However, in general we lack information about the pulsar's line-of-sight velocity, and accurate knowledge about its age, since the characteristic age estimated from the pulsar period and its derivative can differ significantly from the true age \citep[see e.g.][]{Kaspi2001, Vigano2013}. Furthermore, estimates of pulsar distances have typically large associated errors due to uncertainties in the underlying density models used to convert pulsar \acs{DM} or $N_{\rm H}$ into distance estimates \citep{Lorimer2006, He2013, Deller2019}.

Reconstruction of the three-dimensional initial position and velocity distribution of pulsars, and comparison with the observed Galactic neutron star population is therefore a complicated task that requires careful simulations as well as detailed estimates of the observational biases of multi-band surveys. Several studies have performed statistical and population synthesis analyses to recover the distributions of important neutron star parameters from the observed population \citep[see e.g.][]{Arzoumanian2002, Brisken2003, Hobbs2005, Faucher2006, Gullon2014, Verbunt2017, Cieslar2020}. While these models are broadly able to explain the observational data, high degrees of degeneracy between the different input parameters make it difficult to exactly pin down the distributions that control the pulsars' birth properties, such as their natal kick velocities. Nonetheless, disentangling the birth properties of the isolated neutron star population in our Galaxy is crucial as it has important implications for several lines of research, including formation mechanisms of these compact stars, the evolution of massive stars, as well as extreme events such as \acf{GRBs}, \acf{FRBs} and peculiar types of supernovae.

Constraining the birth properties of isolated neutron stars from observational data is the main motivator for our work. Instead of following earlier approaches that have employed standard statistical techniques, we focus on characterizing initial pulsar properties using machine learning techniques which have seen increasing interest in the astronomy and astrophysics communities in recent years. This paper, which will be the first in a series, is dedicated to the technical aspects of these efforts and aims to show that a machine-learning framework can be used to estimate parameters with high accuracy. For this feasibility study, we restrict ourselves to a simplified approach, where selection effects and observational biases are neglected and reduced physical models are sufficient. In particular, we focus on the dynamical properties of the pulsar population and explore the possibility of inferring the parameters that control a given Galactic pulsar kick-velocity and scale-height distribution at birth (the two quantities that largely control the spatial distribution of pulsars in the Milky Way) through neural networks. For this purpose, we implement a basic population synthesis code in \texttt{Python} and simulate the dynamical evolution of a synthetic population of isolated neutron stars for a variety of different birth position and natal kick distributions. These evolved mock populations are then fed into a suitably structured machine-learning pipeline with the aim of recovering the underlying parameters controlling the distributions. We show that this procedure is successful at estimating birth characteristics. Additionally, we link our framework to the observed sample of pulsars with measured proper motion in a phenomenological way and discuss implications for future pulsar survey efforts, i.e., with the Square Kilometer Array (SKA).

Our paper is structured as follows: In \S \ref{sec:popsynth} we describe the methods used to simulate and evolve a mock neutron star population in time. \S \ref{sec:MLsetup} contains a description of the machine-learning framework, including the generation of our data-sets (\S \ref{sec:dataset_creation}), the employed network architectures (\S \ref{sec:net_architecture}) as well as details of the training process (\S \ref{sec:train_process}). In \S \ref{sec:experiments} we present our experiments, which are discussed in detail and connected with observational data in \S \ref{sec:discussion}. Finally, we provide a summary and outlook in \S \ref{sec:summary}.


\section{Population Synthesis}
\label{sec:popsynth}

A widely used approach to investigate the properties of the observed neutron star population is through population synthesis \citep[see e.g.][]{Narayan1990, Faucher2006, Gonthier2007, Kiel2008, Kiel2009, Oslowski2011, Levin2013, Gullon2014, Bates2014, Cieslar2020}. These frameworks aim to simulate the evolution of a population of neutron stars from birth until today. The resulting mock population is then compared with the real observed population in order to constrain and validate the physical model assumptions that entered the simulation. In particular, the population synthesis approach relies on assumptions about the distributions of the birth properties of the mock neutron stars, and typically takes advantage of Monte--Carlo methods to construct the initial parameters of each simulated star. Starting from these initial conditions the mock population is then evolved over time according to some evolutionary prescriptions, and eventually contrasted with real data. For the development of our population synthesis framework we largely follow \citet{Faucher2006}, \citet{Gullon2014} and \citet{Cieslar2020}. The necessary ingredients are briefly summarized in the following.


\subsection{Age}
\label{subsec:age}

The age $t_{\rm age}$ of each neutron star is randomly drawn from a uniform probability distribution between $1$ and $\unit[10^7]{yr}$. By choosing a uniform distribution, we assume that the birth rate of neutron stars is constant in the chosen time range. For all simulations of the synthetic neutron star population we choose an average neutron star birth rate of 1 neutron star per century, compatible with the core-collapse supernova rate in the Galaxy \citep{Rozwadowska2021}. This yields a total of $10^{5}$ simulated neutron stars for each synthetic population, whose evolution we can compute within reasonable timescales. 


\subsection{Birth Position}
\label{subsec:position}

To define the initial positions we use both a Cartesian reference frame $(x,~y,~z)$ and a cylindrical reference frame $(r,~\phi,~z)$, whose origins are located at the center of the Galaxy. Here $r$ represents the distance in kiloparsec from the Galactic center, $\phi$ is the azimuthal angle in radians and $z$ is the distance from the Galactic plane. The two coordinate systems are related by the transformation: 
\begin{align}
    \begin{cases}
    x=r\cos \phi, \\
    y = r\sin \phi, \\
    z = z.
    \end{cases}
\end{align}
We assume that the Sun is located at the coordinates $x= \unit[0]{kpc}$, $y=R_{\odot}$, $z=z_{\odot}$, where $R_{\odot}= \unit[8.3]{kpc}$ and $z_{\odot}= \unit[0.02]{kpc}$ \citep[see][and references therein]{Pichardo2012}. We calculate the initial position at birth of each neutron star in both cylindrical and Cartesian galactocentric reference frames. To do so, we execute the following steps:
\begin{itemize}
\item[\it{i)}] First, we draw a random distance $r$ from the Galactic center for each neutron star ranging between $10^{-4}$ and \unit[20]{kpc} according to a pulsar radial density distribution $P(r)$. In particular, we follow the Milky Way's pulsar surface density $\rho(r)$ defined by Eq.~(15) in \citet{Yusifov2004} to determine the probability density function for the radial distance: 
\begin{align}
  P(r) = 2 \pi r \rho(r),
\end{align}
with
\begin{align}
  \rho(r) = A \left( \frac{r+R_1}{R_{\odot}'+R_1}\right)^a \exp \left[ -b\left(\frac{r-R_{\odot}'}{R_{\odot}'+R_1} \right) \right],
    \label{eq:radialPDF}
\end{align}
where $A = \unit[37.6]{kpc^{-2}}$, $a = 1.64$, $b = 4.0$, $R_1 = \unit[0.55]{kpc}$ and $R_{\odot}' = \unit[8.5]{kpc}$ is the Sun's distance from the Galactic center. Although different from the $R_{\odot}$ value assumed above, we keep $R_{\odot}' = \unit[8.5]{kpc}$ in this parameterization in order to be consistent with the results of \citet{Yusifov2004}. We note that this is the distribution for evolved pulsars rather than that of their progenitors and \citet{Yusifov2004} find small discrepancies between this distribution and that of OB stars. However, \citet{Faucher2006} show that the evolved pulsar population is well described by birth positions drawn from Eq.~\eqref{eq:radialPDF} and argue that differences fall within the current uncertainties of pulsar distance measurements. Given the lack of a more realistic description, we therefore adopt the above prescription.

\vspace{0.3cm}

\item[{\it ii)}] Neutron stars are born mainly within the Galactic spiral arms, as these regions are rich in massive OB stars \citep{Chen2019}. We implement a model for the galactic spiral structure that includes four arms with a logarithmic shape function which gives the azimuthal coordinate $\phi$ as a function of the distance from the Galactic center:
\begin{equation}	\label{eq:arm_structure1}
	\phi(r) = k \ln \left( \frac{r}{r_0} \right) + \phi_0. 
\end{equation}
Our values of the model parameters, i.e., the winding constant $k$, the inner radius $r_0$ and the inner angle $\phi_0$ are reported in Table \ref{tab:spiral_arms_params_YMW17} and evaluated from Table 1 in \citet{Yao2017} in order to match the same functional form as defined in Eq.~\eqref{eq:arm_structure1}. For our analysis, we follow \citet{Faucher2006} and do not include the Local arm, whose density is much smaller than that of the four major arms \citep{Cordes2002, Yao2017}. For each star we then randomly select one of the four spiral arms with equal probability, and evaluate the angular coordinate $\phi$ for its given $r$ according to Eq.~\eqref{eq:arm_structure1}.

\begin{deluxetable}{ccccc}

\tablecaption{Parameters of the Milky Way spiral arm structure: the winding constant $k$, the inner radius $r_0$ and the inner angle $\phi_0$  (adapted from Table 1 in \citet{Yao2017}; see \S\ref{subsec:position} for more details).
\label{tab:spiral_arms_params_YMW17}}
\tabletypesize{\small}
\tablecolumns{5}
\tablenum{1}
\tablewidth{0pt}
\tablehead{
\colhead{\multirow{2}{*}{Arm number}} &
\colhead{\multirow{2}{*}{Name}} &
\colhead{$k$} &
\colhead{$r_0$} &
\colhead{$\phi_0$} \\
& 
& 
\colhead{[rad]} & 
\colhead{[kpc]} & 
\colhead{[rad]} 
}
\startdata
1 & Norma & 4.95 & 3.35 & 0.77 \\
2 & Carina-Sagittarius & 5.46 & 3.56 & 3.82 \\
3 & Perseus & 5.77 & 3.71 & 2.09 \\
4 & Crux-Scutum & 5.37 & 3.67 & 5.76 
\enddata

\end{deluxetable}

The spiral pattern of the Galaxy is not static and as a first approximation can be considered as a rigid structure which rotates with an approximated period $T = \unit[250]{Myr}$ \citep{Vallee2017, Skowron2019}. Knowing the age of an object and assuming a rotational angular velocity of $\Omega = 2\pi/T$ for the spiral structure, we can derive the angular position at birth of each neutron star. Note that the Galaxy rotates in the clockwise direction, i.e., toward decreasing $\phi$ angles.

After obtaining the corresponding angular coordinate for each neutron star birth position, we add noise to both coordinates $r$ and $\phi$ to smear out the distribution and avoid artificial features near the Galactic center. For this purpose we add a correction $\phi_{\rm corr} = \phi_{\rm rand} \exp{\left( -0.35 r/ {\rm kpc} \right)}$ to the $\phi$ coordinate, where $\phi_{\rm rand}$ is randomly drawn from a uniform distribution in the interval $\left[ 0, 2\pi \right)$, and to the $r$ coordinate a correction $r_{\rm corr}$ randomly drawn from a normal distribution centered at 0 with standard deviation $\sigma = 0.07r$. Although this prescription was introduced by \citet{Faucher2006} (see their Section 3.2.1) in a somewhat arbitrary manner, the resulting stellar distribution broadly agrees with that observed for very young high-mass stars as shown in \citet{Reid2019}.

Then the birth position in polar coordinates of each neutron star is given by $(r+r_{\rm corr},~\phi(r)+\Omega t_{\rm age}+\phi_{\rm corr})$ with units [kpc, rad].

\vspace{0.3cm}

\item[{\it iii)}] To determine the height $z$ in kiloparsec from the Galactic plane of each neutron star, we adopt an exponential disk model as given by \citet{Wainscoat1992}. It is shaped by the characteristic scale-height parameter $h_{\rm c}$:
\begin{align}
  P(z) = \frac{1}{h_{\rm c}} \exp\left(-\frac{ \lvert z \rvert}{ h_{\rm c} } \right). 
\end{align}
For our machine-learning experiments, we will vary the scale height in the range $[0.02,2]$ $\unit[]{kpc}$ to simulate neutron star populations with different spread in galactic height. This range encompasses the value $h_{\rm c} = \unit[0.18]{kpc}$, which was adopted by \citet{Gullon2014} to match radio pulsar observations, and which is also compatible with the population of young massive stars in the Galactic disk \citep{Li2019}. We will consider $h_{\rm c} = \unit[0.18]{kpc}$ below, whenever a fiducial scale height is required for our synthetic pulsar population. The coordinate $z$ of each neutron star is then randomly drawn according to this height probability distribution in a range of $10^{-4}$ to $\unit[5]{kpc}$. We choose a maximal distance of $\unit[5]{kpc}$ from the galactic plane to model a fixed galactic volume for all of our simulation runs, while also ensuring sufficient resolution for the objects in the galactic disc for those models with small scale heights. Subsequently, for each star we randomly choose if $z$ is positive or negative determining in this way a position above or below the Galactic plane.
\end{itemize}


\subsection{Initial Velocity}
\label{subsec:velocity}

We assume that the initial velocity of the neutron stars in the Galaxy is given by two contributions: the progenitor velocity in the Galactic gravitational potential and a kick speed imparted onto the neutron stars as a result of the supernova explosion.
We consider a progenitor circular orbital speed given by the following relation:
\begin{align}
	v_{\rm orb} = \sqrt{ r \frac{\partial \Phi_{\rm MW} \left( r,z \right)}{\partial r} },
\end{align}
where $\Phi_{\rm MW}$ is the Milky Way gravitational potential discussed below.
We assume that each neutron star has an initial kick velocity $\boldsymbol{v}_{\rm k}$, whose 3D magnitude $v_{\rm k}$ is randomly drawn from a Maxwell distribution, shaped by the dispersion parameter $\sigma_{\rm k}$:
\begin{equation}	\label{eq:pdf_maxwell_kick}
	P(v_{\rm k}) = \sqrt{ \frac{2}{\pi} }  \frac{v_{\rm k}^2}{\sigma_{\rm k}^3} \exp\left(-\frac{v_{\rm k}^2}{ \sigma_{\rm k}^2 } \right).
\end{equation}
For our machine-learning purposes, we will vary $\sigma_{\rm k}$ in the range $[1, 700]$ $\unit[]{km \, s^{-1}}$ and randomly draw 3D velocity magnitudes from the resulting distribution in the range $[0, 2500]$ $\unit[]{km \, s^{-1}}$. This spread allows us to easily accommodate the fastest observed neutron stars whose velocities have been estimated to surpass 1000 $\unit[]{km \, s^{-1}}$ \citep[see for example][]{Chatterjee2005,Hui2006,Pavan2014}. Based on pulsar timing measurements, \cite{Hobbs2005} have suggested that $\sigma_{\rm k} = \unit[265]{km \, s^{-1}}$ provides a viable explanation for the proper motions of observed neutron stars. We will use this as a fiducial value below. For a given kick velocity magnitude we then associate a random direction to this velocity in order to evaluate the three components ($v_{{\rm k}, r}$, $v_{{\rm k}, \phi}$, $v_{{\rm k}, z}$) in galactocentric cylindrical coordinates. Therefore, the three components of the total initial velocity of each neutron star in the galactocentric reference frame are computed as ($v_{{\rm k}, r},~v_{\rm orb} + v_{{\rm k}, \phi},~v_{{\rm k}, z})$.  

\begin{deluxetable}{cc}

\tablecaption{Parameters of the Milky Way gravitational potential taken from Table 1 in \citet{Marchetti2019}, see also \citet{Bovy2015}.
\label{tab:MW_pot_params_M19}}
\tabletypesize{\small}
\tablecolumns{2}
\tablenum{2}
\tablewidth{0pt}
\tablehead{
\colhead{Component} &
\colhead{Parameters} }
\startdata
nucleus (n) & $M_{\rm n} = 1.71 \times 10^{9} M_{\odot}$ \\
 & $r_{\rm n} = 0.07$ kpc \\
bulge (b) & $M_{\rm b} = 5.0 \times 10^{9} M_{\odot}$ \\
 & $r_{\rm b} = 1.0$ kpc \\
disk (d) & $M_{\rm d} = 6.8 \times 10^{10} M_{\odot}$ \\
 & $a_{\rm d} = 3.00$ kpc \\
 & $b_{\rm d} = 0.28$ kpc \\
halo (h) & $M_{\rm h} = 5.4 \times 10^{11} M_{\odot}$ \\
 & $r_{\rm h} = 15.62$ kpc 
\enddata

\end{deluxetable}


\subsection{Galactic Potential}
\label{subsec:potential}

As typical for spiral galaxies like the Milky Way, we assume an axisymmetric Galactic potential $\Phi_{\rm MW}$ \citep{Carlberg1987, Bovy2015} that does not incorporate the impact of the spiral arms themselves. We specifically consider a four-component Galactic potential model consisting of a nucleus ($\Phi_{\rm n}$), a bulge ($\Phi_{\rm b}$), a disk ($\Phi_{\rm d}$) and a halo ($\Phi_{\rm h}$) as discussed in \citet{Marchetti2019}:
\begin{equation}	
	\Phi_{\rm MW} = \Phi_{\rm n} + \Phi_{\rm b} + \Phi_{\rm d} + \Phi_{\rm h}.
\end{equation}
The nucleus and the bulge are described by a Hernquist potential \citep{Hernquist1990}:

\begin{equation}	
	\Phi_{\rm n} = -\frac{ G M_{\rm n}}{ r_{\rm n} + r },
\end{equation}
\begin{equation}	
	\Phi_{\rm b} = -\frac{ G M_{\rm b}}{ r_{\rm b} + r }. 
\end{equation}
The disk has a Miyamoto-Nagai disk potential \citep{Miyamoto1975}:
\begin{equation}	
	\Phi_{\rm d} = -\frac{ G M_{\rm d}}{ \sqrt{ K^2 + r^2} },
\end{equation}
where $K = a_{\rm d} + \sqrt{z^2+b_{\rm d}^2}$ is the shape parameter with $a_{\rm d}$ as the scale length and $b_{\rm d}$ the scale height of the disk. 
\noindent
The halo has a Navarro-Frenk-White potential \citep{Navarro1996}:
\begin{equation}	
	\Phi_{\rm h} = -\frac{ G M_{\rm h}}{ r } \ln{ \left( 1 + \frac{r}{r_{\rm h}}\right) }.
\end{equation}
The parameters of this model are reported in Table \ref{tab:MW_pot_params_M19} and were derived by \citet{Bovy2015} through a fit of the mass profile of the Milky Way. We assume that these contributions to the galactic potential are stationary in time, i.e., they do not evolve over the time span we consider for the dynamical evolution.


\subsection{Dynamical Evolution}
\label{subsec:evolution}

Given the initial conditions defined above, i.e., the initial position, initial velocity and the Galactic gravitational potential, we can solve the equations of motion to determine the neutron stars' dynamical evolution. The system of dynamical equations that requires solving to determine the orbits of the neutron stars in the galactic potential is given by the Newtonian equations of motion: $\ddot{\boldsymbol{r}} = -\boldsymbol{\nabla} \Phi_{\rm MW}$.
In cylindrical galactocentric coordinates the three components of this vector equation take the form: 
\begin{equation}
	\begin{cases}	
	\ddot{r} - r \dot{\phi}^2 = -\frac{\partial \Phi_{\rm MW} }{ \partial r }, \\
	2\dot{r}\dot{\phi} + r \ddot{\phi} = - \frac{1}{r} \frac{\partial \Phi_{\rm MW} }{ \partial \phi }, \\
	\ddot{z} = - \frac{\partial \Phi_{\rm MW} }{ \partial z }. 
	\end{cases}
\end{equation}
For each neutron star we numerically integrate the above equations in time from $t = \unit[0]{yr}$ to $t = t_{\rm age}$ using a discrete time step. 
We use the Python package \texttt{scipy.integrate.odeint} and to speed up the computational time also employ the module \texttt{jit} (``just in time") from the \texttt{Numba} package (\url{https://numba.pydata.org/}, \citealt{Lam2015})\footnote{\texttt{Numba} translates Python functions into optimized machine code at run-time, which allows us to achieve a speed-up by about a factor of 6.}. To asses the performance of our integration method we check that both the total energy (i.e., the potential plus kinetic energy) and the $z$-component of the total angular momentum of all the stars in our simulation are conserved. For simplicity we assume that all pulsars have the same mass and we find that both quantities are conserved with a relative error of $\lesssim 10^{-7}$. The output of the dynamical evolution consists of the position and velocity of each neutron star computed in both galactocentric (GC) and equatorial ICRS (International Celestial Reference System) frames. To transform between different coordinate systems we employed the method \texttt{coordinates} from the Python library \texttt{Astropy} \citep{astropy13, astropy18}, where we adopted a galactocentric distance of $R_{\odot}= \unit[8.3]{kpc}$ and Galactic height of $z_{\odot}= \unit[0.02]{kpc}$ for the Sun.


\begin{figure*}
\centering
\includegraphics[width = 0.245\textwidth]{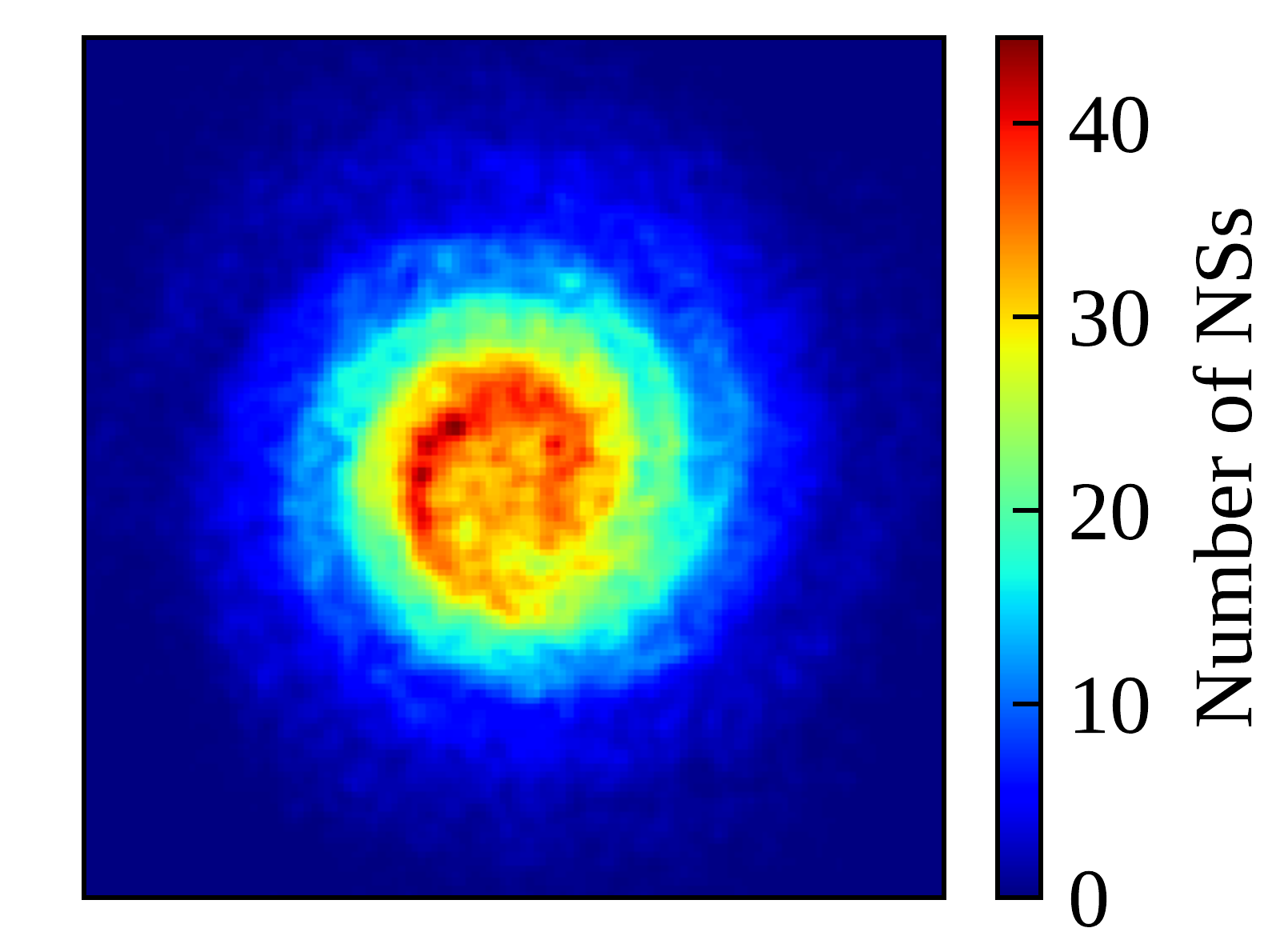}
\includegraphics[width = 0.245\textwidth]{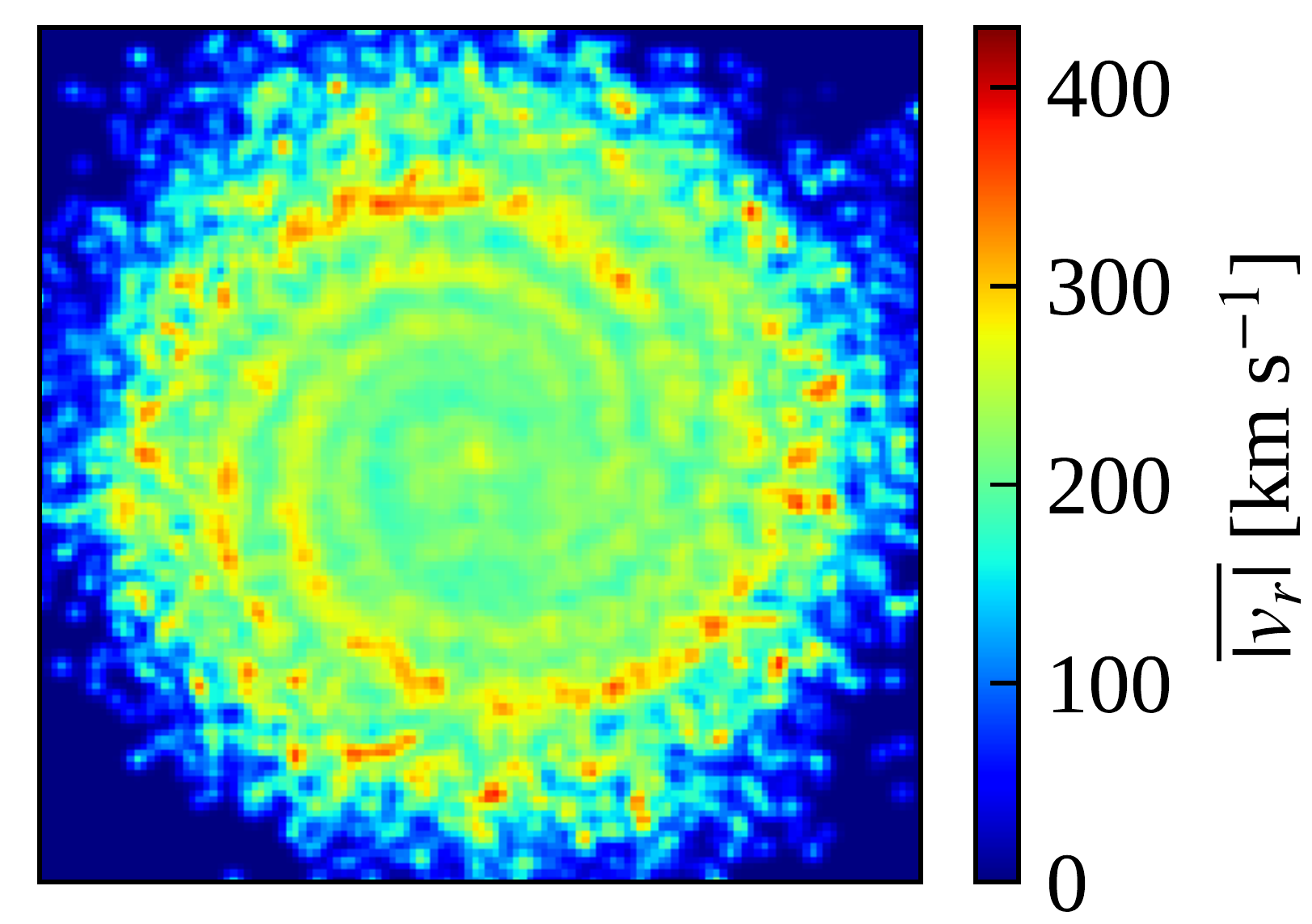}
\includegraphics[width = 0.245\textwidth]{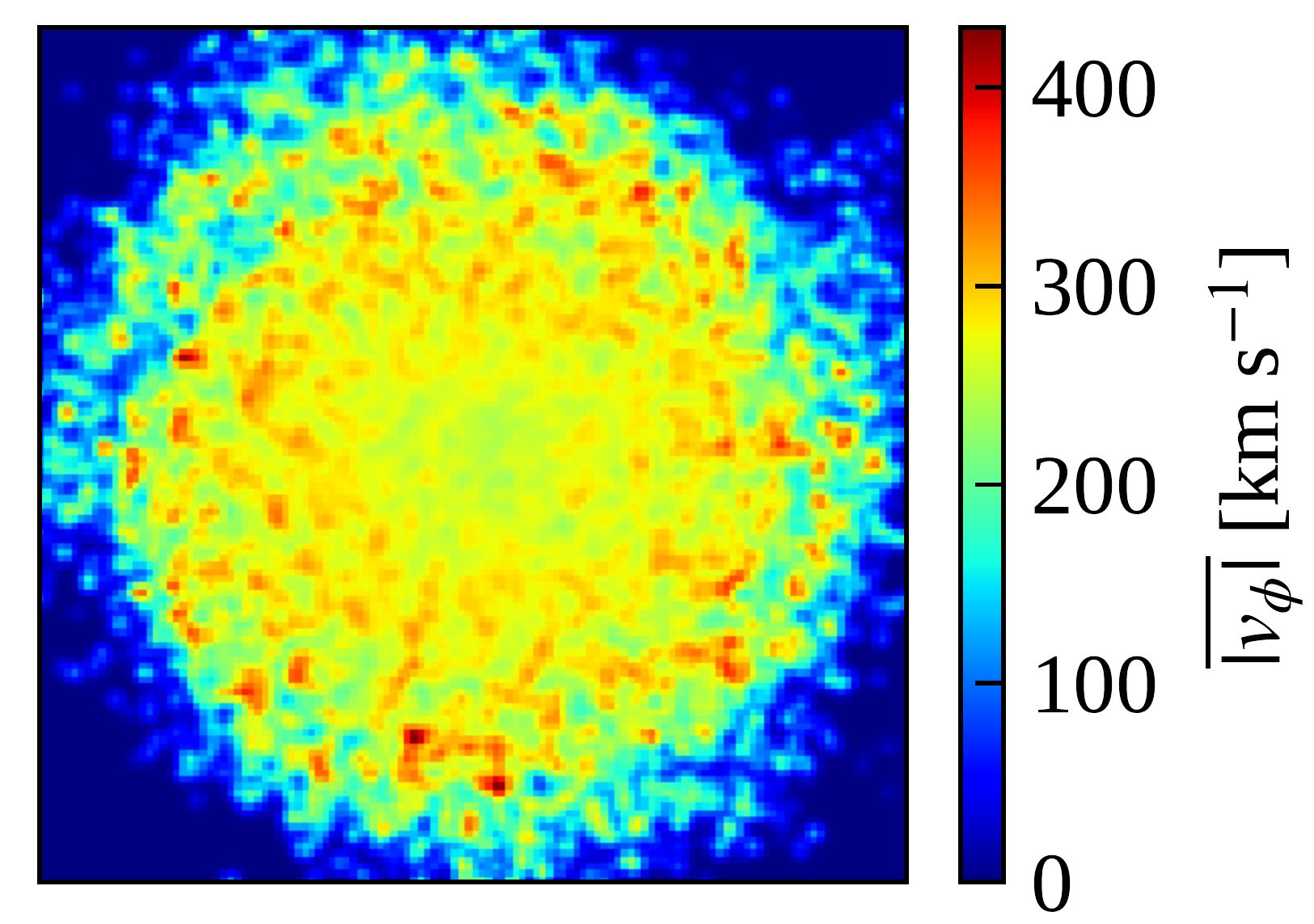}
\includegraphics[width = 0.245\textwidth]{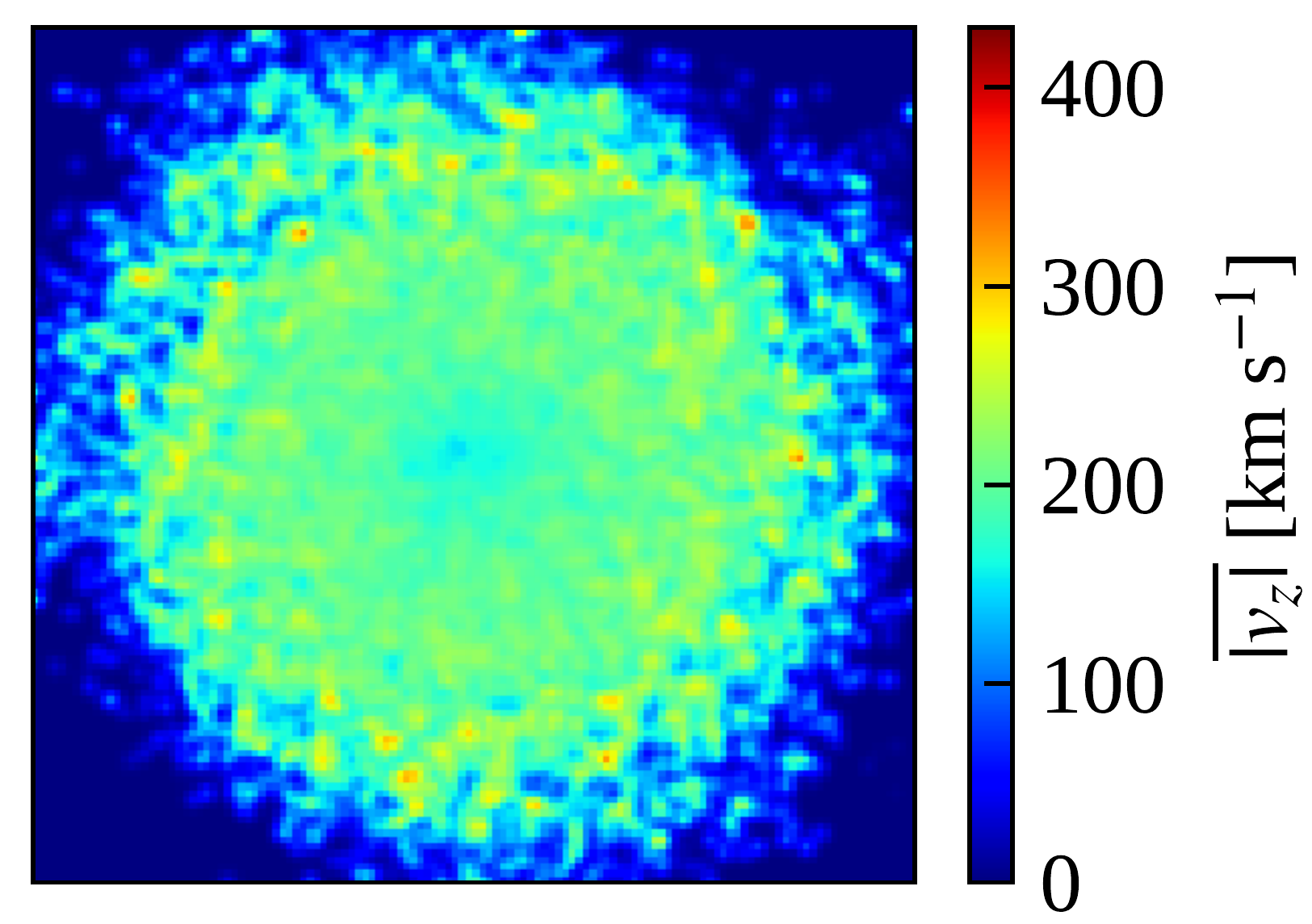}
\caption{\label{fig:gc_map}{Examples of $128 \times 128$ resolution maps in the galactocentric $xy$-plane extending from $-20$ to $\unit[20]{kpc}$ in both $x$ and $y$ direction and showing (in order from left to right) the density of simulated neutron stars, the distribution of average values of the $v_{r}$, $v_{\phi}$ and $v_{z}$ velocity components for a population of neutron stars simulated with $h_{\rm c} = \unit[0.18]{kpc}$ and $\sigma_{\rm k} = \unit[265]{km \, s^{-1}}$. For visualization purposes, we represent the data using a colormap to highlight the resulting structures; red regions are characterized by a higher density of stars or higher average magnitude of the velocity components, respectively, while blue areas correspond to lower densities and lower velocity magnitudes. We note that the spiral-arm pattern is still recognizable in the position-density map although high kick velocities tend to blur and disperse the stellar density distribution. In the $v_r$-velocity map, the inter-arm regions are visible as high-velocity areas, because during the dynamic evolution the space between the spiral arms is progressively filled with high velocity stars that have escaped from their birth places. The other two velocity components exhibit smoother behavior because the spiral-arm structure is smeared out.}}
\end{figure*}  
\begin{figure*}
\centering
\includegraphics[width = 0.325\textwidth]{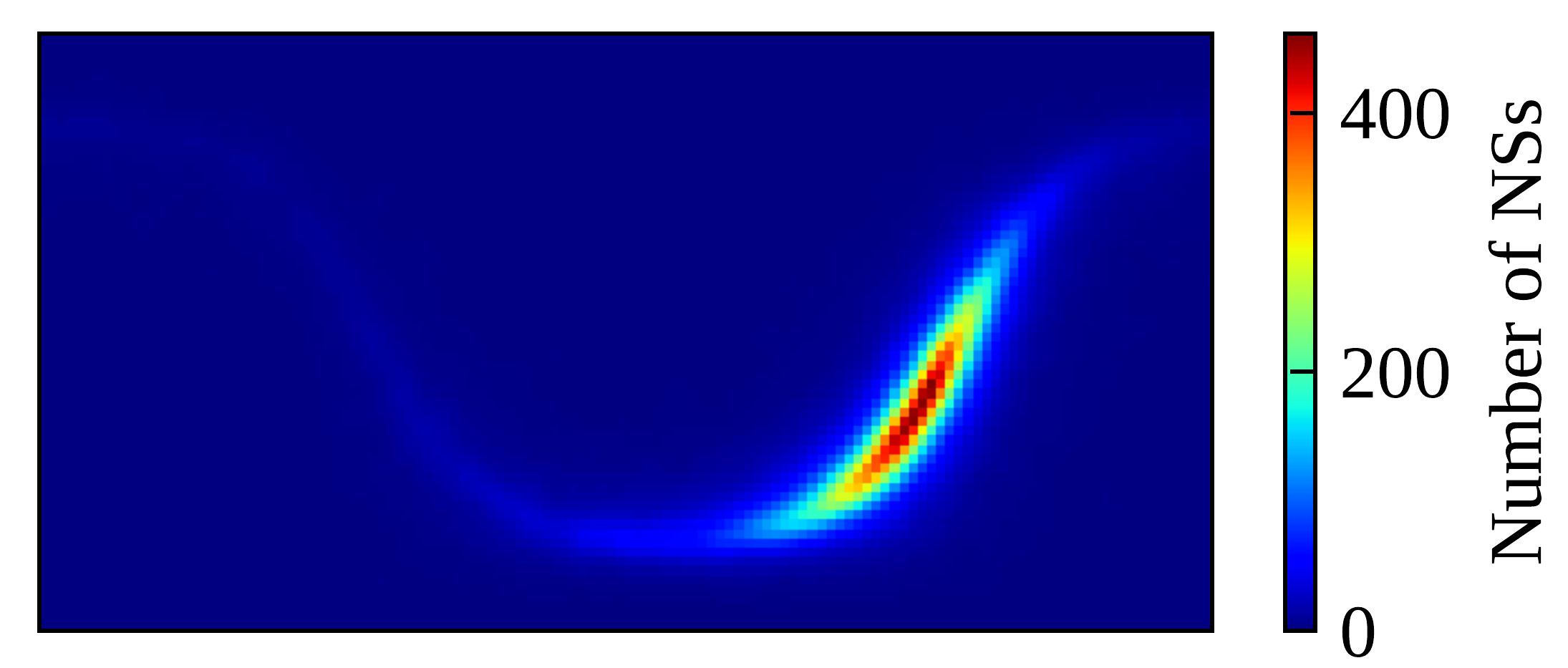}
\includegraphics[width = 0.325\textwidth]{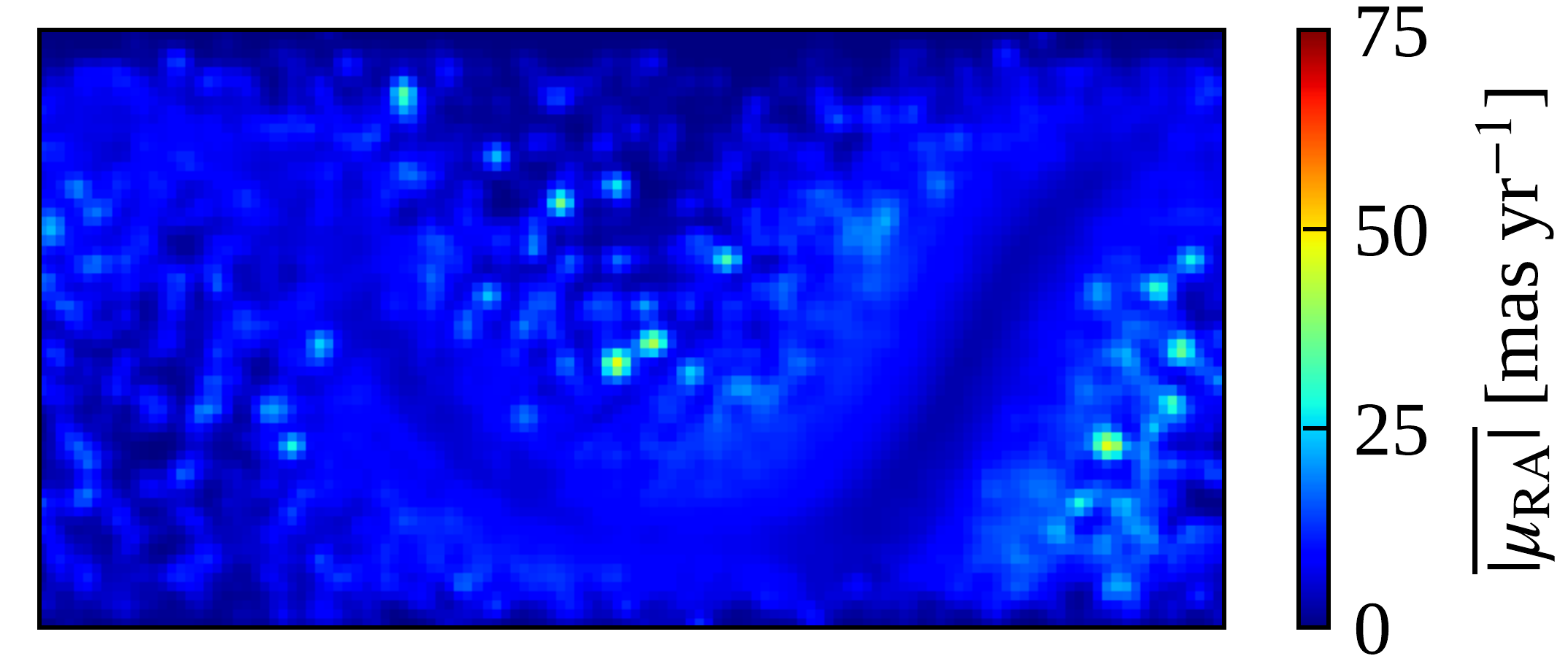}
\includegraphics[width = 0.325\textwidth]{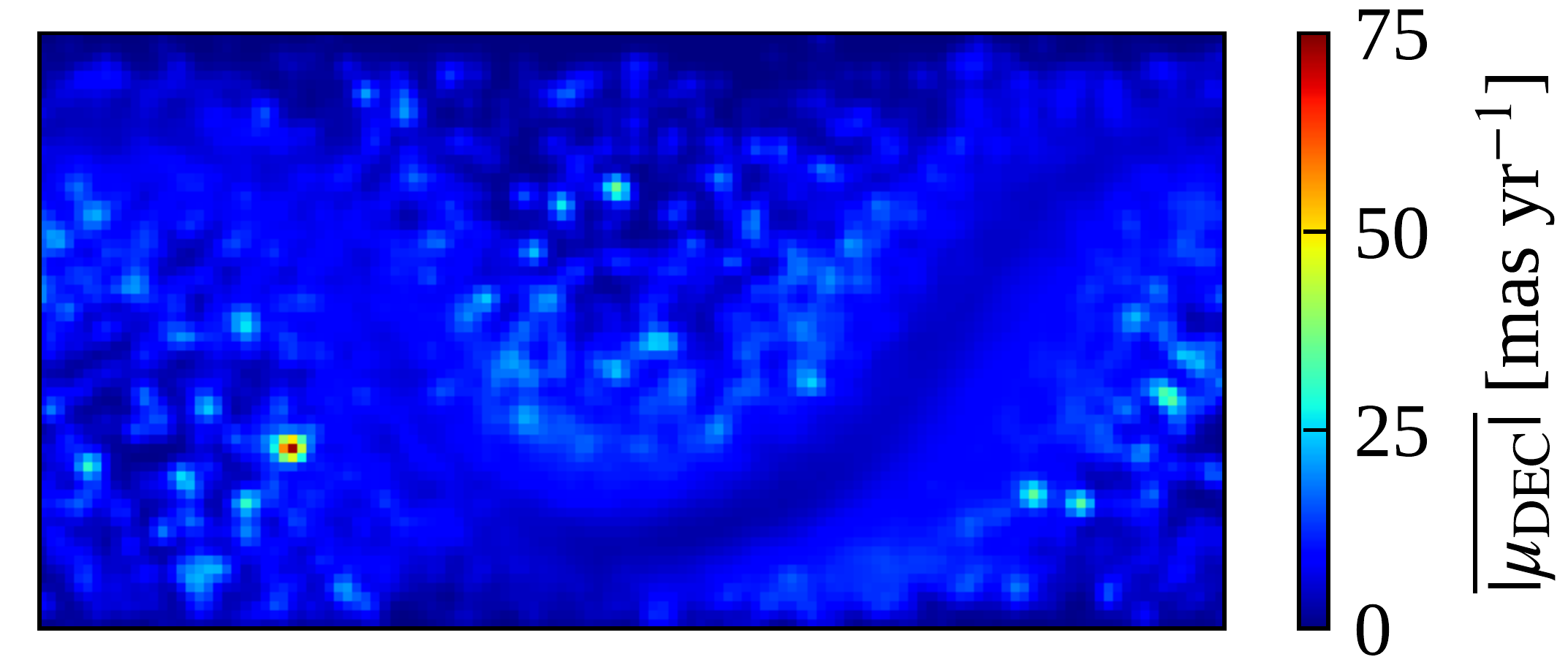}
\caption{\label{fig:icrs_map}{Examples of $128 \times 64$ resolution maps in the equatorial ICRS frame extending from $0^{\circ}$ to $360^{\circ}$ in RA and from $-90^{\circ}$ to $90^{\circ}$ in DEC and showing (in order from left to right) the density of simulated neutron stars, the distribution of average values for the $\mu_{\rm RA}$, $\mu_{\rm DEC}$ proper-motion components for a population of neutron stars simulated with $h_{\rm c} = \unit[0.18]{kpc}$ and $\sigma_{\rm k} = \unit[265]{km \, s^{-1}}$. For visualization purposes, we represent the data using a colormap to highlight the resulting structures}; red regions are characterized by a higher density of stars or a higher average magnitude of the velocity components, respectively, while blue areas correspond to lower densities and lower velocity magnitudes. Note that the Galactic silhouette is visible as a \textit{stream} in the position map with an enhanced stellar density close to the Galactic Center. Due to low-number statistics, the regions outside the Galactic stream in the proper-motion maps are dominated by statistical fluctuations, i.e., the corresponding high-velocity regions are attributed to a small number of high proper-motion stars that have escaped the disk. As a result, the disk itself is dominated by stars with lower proper motion. }
\end{figure*}  

\section{Machine-learning Set-up}
\label{sec:MLsetup}

In the past decade, the accumulation of extensive and heavy data-sets has been almost ubiquitous in astronomy and astrophysics. In order to take full advantage of these data and perform data-driven science that complements ongoing theoretical modeling efforts, new techniques and analysis pipelines that can handle these large amounts of data (and do so in an automated way) are required. Machine learning (\acs{ML}) has played an important role in developing such new algorithms \citep{Ball2010, Allen2019, Baron2019, Fluke2020}. For compact-object related science, \acs{ML} algorithms have for example been developed to classify new pulsars candidates \citep{Bethapudi2018, Balakrishnan2020, Lin2020} as well as transient radio events such as FRBs \citep{Agarwal2020}. Other approaches have aimed at forecasting and analyzing gravitational-wave signals in real time \citep{Cabero2020, Wei2020, Skliris2020, Gerosa2020}, interpreting gravitational wave-events in light of population synthesis \citep{Wong2019}, or reconstructing the neutron star equation of state from observed quantities \citep{Morawski2020}.

For our analysis, we will focus on \acf{ANNs}. \acs{ANNs} are algorithms inspired by the structure of biological brains that can be thought of as nets of interconnected neurons that exchange information from one to another. When the network receives an input, it is able to process it to produce an output, like a biological brain responds to external stimulation. In \acs{ANNs}, neurons are usually organized in a stack of layers. Each neuron in a layer receives input signals (typically real numbers) from the neurons in the previous layer and produces an output signal by applying a non-linear activation function to a linear combination of the input signals according to certain weights and a bias. The output is then passed to the neurons in the following layer, and so on until the final layer is reached and the output is generated. In our particular case, we will focus on supervised learning, where training the neural network consists of making it produce a specific target output when a particular input is passed through it. This is achieved by (i) labeling the input samples in the training data-set with a label indicating the property that the network has to learn (the so-called ground truth) and (ii) iteratively adjusting the values of weights and biases, also called network \text{parameters}, in order to minimize a specific loss function which measures the distance between the network output prediction and the target ground truth. 

Among their numerous applications, \acs{ANNs} have been employed in regression problems where the network is trained to infer the values of continuous variables for the given input data. This is the kind of problem we are after since we want our network to infer certain parameter values given the evolved neutron star population. In the remainder of this section, we discuss (i) the simulation data we create for our \acs{ML} experiments, then (ii) focus on the specific network architecture employed, and finally (iii) describe the details of our training process.


\subsection{Data-set Creation and Processing}
\label{sec:dataset_creation}

The goal of our \acs{ML} approach is to predict the parameters that control the dynamical properties of an evolved neutron star population. In particular, we focus on predicting the kick-velocity parameter $\sigma_{\rm k}$ and the scale-height parameter $h_{\rm c}$, which predominantly affect the distribution of pulsars in the Milky Way. To extract these from an evolved population, and follow a supervised learning approach, we first need to train a neural network on a series of simulated populations (created by exploring the ranges for $\sigma_{\rm k}$ and $h_{\rm c}$). Following the prescription described in \S \ref{sec:popsynth}, we perform the following simulation runs:
\begin{itemize}
    \item[\textbf{S1}] We generate 10 data-sets with an increasing number of samples (specifically 4, 8, 16, 32, 64, 128, 256, 512, 1024 and 20000 simulated populations) by uniformly varying the parameter $\sigma_{\rm k}$ of the kick velocity distribution in the range [1, 700] $\unit[]{km\,s^{-1}}$.\footnote{To generate our simulation data, we partially employ the package \texttt{Hydra} (\url{https://hydra.cc/}, \citealt{Yadan2019}), which allows us to easily sweep entire parameter ranges.} We also generate a test data-set with 1000 samples, each one simulated with $\sigma_{\rm k}$ randomly drawn from a uniform distribution in the same range of values. For these simulations, we keep the characteristic scale of the $z$-distribution fixed to its fiducial value $h_{\rm c} = \unit[0.18]{kpc}$.
    \item[\textbf{S2}] We fix the kick-velocity parameter to its fiducial value $\sigma_{\rm k} = \unit[265]{km\,s^{-1}}$ and generate a data-set of 20000 samples of simulated populations by uniformly varying the scale-height parameter $h_{\rm c}$ in the range [0.02, 2] $\unit[]{kpc}$. We also generate a test data-set with 1000 samples, each one simulated with $h_{\rm c}$ randomly drawn from a uniform distribution in the same range of values.
    \item[\textbf{S3}] We generate 6 data-sets, where we uniformly vary the kick-velocity parameter $\sigma_{\rm k}$ parameter in the range [1, 700] $\unit[]{km\,s^{-1}}$ \textit{as well as} the characteristic scale of the $z$-distribution $h_{\rm c}$ in the range [0.02, 2] $\unit[]{kpc}$. We choose the data-set sizes $16 = 4 \times 4$, $64 = 8 \times 8$, $256 = 16 \times 16$, $1024 = 32 \times 32$, $4096 = 64 \times 64$ and $16384 = 128 \times 128$ given by all the combinations of $\sigma_{\rm k}$ and $h_{\rm c}$ values. As an example: the 16 populations in the first set are obtained by combining each of the 4 values of the $\sigma_{\rm k}$ parameter with all 4 values of the $h_{\rm c}$ parameter. We also generate a test data-set with 1000 samples, each one simulated with both $\sigma_{\rm k}$ and $h_{\rm c}$ randomly drawn from uniform distributions in their respective parameter ranges specified above.
\end{itemize}
As addressed in detail in \S \ref{sec:experiments}, the smaller simulation data-sets will be used to explore the network behavior. The largest data-sets containing 20000 and 16384 samples, respectively, and the test data-sets with 1000 samples will be used to perform the final training experiments and assess the actual network accuracy in generalization scenarios.

After the runs have been performed, we transform the output of the simulation into a representation that can be interpreted by a \acs{ML} pipeline. Since \acs{ANNs} require the use of structured data, we represent the position and velocity output of the simulations in the form of 2D binned density and velocity maps both in the galactocentric and ICRS reference frames. The density maps give information about the density of neutron stars in the Galaxy by providing the number count of stars in each spatial bin. On the other hand, velocity maps contain information about the kinematic properties of the neutron stars by providing the average magnitude of the stellar velocity components in each spatial bin. In the galactocentric maps the Galaxy is represented face on and projected onto the $xy$-plane of the Cartesian galactocentric frame, extending from $\unit[-20]{kpc}$ to $\unit[20]{kpc}$ in $x$ and $y$ direction. The ICRS maps instead extend from $0^{\circ}$ to $360^{\circ}$ in right ascension (RA) and from $-90^{\circ}$ to $90^{\circ}$ in declination (DEC). To each map we apply a smoothing Gaussian filter (with radius $4.0 \cdot \sigma + 0.5$ and $\sigma = 1$) in order to add some blurring and avoid sharp features. By doing so, we reduce noisy high-frequency features and thus make the training more stable, presumably resulting in better generalization capabilities. Therefore, for each simulated population we have:
\begin{itemize}
    \item 1 density map in the galactocentric frame.
    \item 3 velocity maps, one for each component of the velocity in cylindrical galactocentric coordinates $v_{r}$, $v_{\phi}$ and $v_{z}$ in [$\unit[]{km \, s^{-1}}$].
    \item 1 density map in ICRS coordinates.
    \item 2 proper motion maps, one for each component of the angular proper motion projected on the celestial sphere $\mu_{\rm RA}$ and $\mu_{\rm DEC}$ in [$\unit[]{mas \, yr^{-1}}$].
\end{itemize}
This set of maps for each population is labeled with the corresponding values of the parameters $\sigma_{\rm k}$ and $h_{\rm c}$ used to simulate that specific population.

We will test the \acs{ML} performance on three different map resolutions. Thus, we generate each of the above data-sets with resolutions $32 \times 32$, $128 \times 128$ and $512 \times 512$ square bins. Note that in the ICRS maps the DEC coordinate axis range is half that of the RA coordinate axis. Hence, these maps have half the bins along the DEC coordinate and their resolution is $32 \times 16$, $128 \times 64$ and $512 \times 256$ square bins. For brevity hereafter we will refer to the three resolutions as 32, 128 and 512 resolutions for both galactocentric and ICRS maps, respectively. An example of the maps with resolution 128 for a simulation with fiducial values $h_{\rm c} = \unit[0.18]{kpc}$ and $\sigma_{\rm k} = \unit[265]{km \, s^{-1}}$ is shown in Figs.~\ref{fig:gc_map} and \ref{fig:icrs_map}. 

Before loading the maps into our \acs{ML} pipeline, they are normalized so that each bin contains a continuous value between 0 and 1. The same applies to the related labels so that their values range continuously between 0 and 1. The aim of normalization is to speed up the training process and make convergence easier since all inputs will provide signals of similar magnitude to the loss-function minimization. This is useful especially for multi-parameter and multi-channel training, that is when we train a network to predict more than one parameter or use channels that have different absolute magnitudes. In these cases, training without normalization might lead to slower, worse or even no convergence at all. Apart from the blurring and normalization described above, we do not apply any additional pre-processing steps to our input data.

\begin{figure*}
\centering
\includegraphics[width = \textwidth]{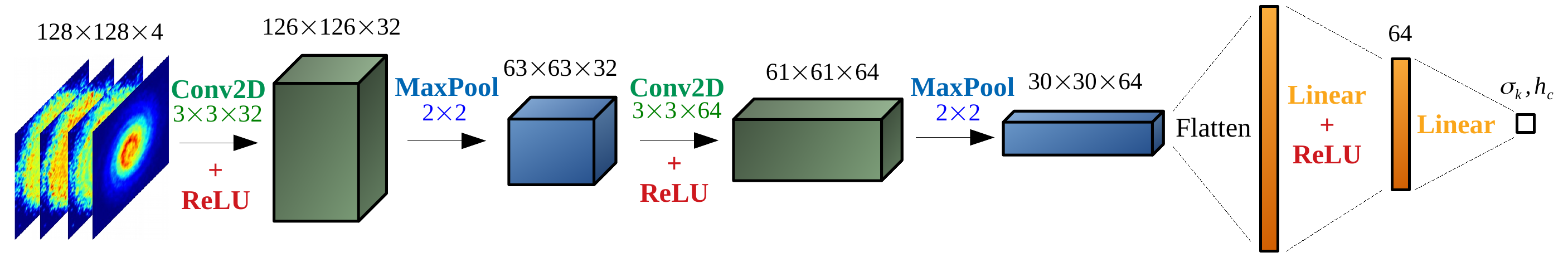}
\caption{\label{fig:CNN_architecture}{Schematic representation of our \acs{CNN} architecture for an input of galactocentric maps} with resolution $128 \times 128$ and 4 channels (1 density map plus 3 velocity maps). A module formed by two blocks that each contain a convolution layer and max pooling layer is followed by a fully connected linear network with one hidden layer. The final network output is either a single parameter or two parameters depending on the experiment specifics.}
\end{figure*}  


\subsection{Network Architecture}
\label{sec:net_architecture}

For the implementation of our \acs{ML} pipeline we use \texttt{PyTorch} \citep{Paszke2019}, an optimized tensor library for deep learning using GPUs and CPUs, that is written in \texttt{Python}. The simplest neural network that can be used for this task is a fully connected neural network also referred to as a \acf{MLP} with only two layers of neurons, which are referred to as the input and output layer, respectively. As this is the starting point to develop more complicated and advanced architecture models, we first test how this simple configuration behaves. The number of neurons for the input layer is equal to the total number of input features, i.e., $C \times W \times H$. Here, $C$ is the number of input channels (corresponding to the total number of maps used), while $W$ and $H$ are the number of bins in width and height, respectively. The number of neurons in the output layer is equal to the number of regression parameters that we would like to predict, i.e., 1 or 2 in our experiments. For the activation function we use the Rectified Linear Unit (ReLU) defined as ${\rm ReLU}(x) = \max{(0,x)}$. To obtain the output values, the ReLU activation function is applied to a linear combination of the input features with weights and a bias.

A more sophisticated model architecture is represented by a \acf{CNN}. \acs{CNN}s are a particular type of deep neural network that have proven to be very successful in regression and classification tasks when applied to structured and matrix-like 2D inputs \citep[see][for a review]{Rawat2017}. The basic structure of \acs{CNN}s consists of convolutional, pooling, and fully connected layers. Convolutional layers are multi-channel filters that slide along the 2D input maps and are able to extract feature maps. The role of the pooling layer is to down-sample the output of a convolutional layer. This inevitably causes a loss of information but in general helps to improve the training efficiency by increasing the size of the receptive field (i.e., the region of the input that produces the feature for each neuron) and reducing the number of trainable parameters. The fully connected layers collect all the output features from the convolution layers into a 1D input and return the final output prediction.

The detailed structure of the \acs{CNN} we built for our case study can be found in Table \ref{tab:CNN_architecture}. A schematic representation of its structure for a 4-channel input with galactocentric maps is also shown in Fig. \ref{fig:CNN_architecture} as an example. It consists of the following layers:  
\begin{itemize}
    \item A 2D convolution layer with kernel size $3 \times 3$, $C$ input channels, 32 output channels, stride 1 and no padding.
    \item A 2D Max pooling layer of size $2 \times 2$ with stride 2 and no padding.
    \item A 2D convolution filter with kernel size $3 \times 3$, 32 input channels, 64 output channels, stride 1 and no padding.
    \item A 2D Max pooling layer of size $2 \times 2$ with stride 2 and no padding.
    \item A fully connected linear layer with flattened input from the convolutional modules output and 64 output neurons.
    \item A fully connected linear layer with 64 input neurons and 1 or 2 output neurons (depending on the number of parameters we would like to predict).
\end{itemize}
For the convolutional and pooling layers the stride parameter regulates the amount of displacement in bins that the filter moves over the map at each step. Padding adds one or more bins at the border of the 2D maps, so that the filters can move and cover the whole map without leaving any bins out. We use a padding of 0 because the borders of the maps do not contain relevant information. 

The choice of this architecture was found by trial and error experiments where we started from a very simple structure and progressively increased the complexity, adding more and more layers to acquire the desired accuracy in predicting the input parameters.

\begin{deluxetable}{ccc}

\tablecaption{\acs{CNN} architecture. The total number of input and output features is reported.  $C$ is the number of used channels, while $W$ and $H$ represent the number of bins (i.e. the resolution) in width and height of the density and velocity maps, respectively. Input and output feature numbers have been rounded down to the lower integer.
\label{tab:CNN_architecture}}
\tabletypesize{\tiny}
\tablecolumns{3}
\tablenum{3}
\tablewidth{0pt}
\tablehead{
\colhead{Layer} &
\colhead{Input} &
\colhead{Output} }
\startdata
Conv2d + ReLU & $C \times W \times H$ & $32 \times (W-2) \times (H-2)$ \\
MaxPool2d & $32 \times (W-2) \times (H-2)$ & $32 \times \left( \frac{W}{2}-1 \right) \times \left( \frac{H}{2}-1 \right)$ \\
Conv2d + ReLU & $32 \times \left( \frac{W}{2}-1 \right) \times \left( \frac{H}{2}-1 \right)$  & $64 \times \left( \frac{W}{2}-3 \right) \times \left( \frac{H}{2}-3 \right)$ \\
MaxPool2d & $64 \times \left( \frac{W}{2}-3 \right) \times \left( \frac{H}{2}-3 \right)$ & $ 64 \times \left( \frac{W}{4}-\frac{3}{2} \right) \times \left( \frac{H}{4}-\frac{3}{2} \right)$ \\
Linear + ReLU & $64 \times \left( \frac{W}{4}-\frac{3}{2} \right) \times \left( \frac{H}{4}-\frac{3}{2} \right)$ & 64 \\
Linear & 64 & 1(2) 
\enddata

\end{deluxetable}


\subsection{Training Process}
\label{sec:train_process}

For the training of the network, we use the \acf{RMSE} both for the loss function and validation metric, i.e., to compute the distance between the network predictions and the ground truths of the $h_{\rm c}$ and $\sigma_{\rm k}$ parameters. In general validation occurs at the same time as training and consists of testing the network over a data-set different from the training set. This is needed to asses the ability of the network to generalize what it is learning to an unknown data-set. The minimization of the loss function occurs through gradient descent and backpropagation \citep{Kelley1960, Ruder2017}, i.e., computation of the loss-function gradients with respect to all network parameters (weights and biases). These gradients taken with a negative sign indicate the directions towards which the network parameters should be updated so that the loss is reduced, and hence the network predictions move closer to the true, expected labels. In this regard, a crucial aspect to ensure the best performance of a neural network is to properly initialize the weights and biases. For this purpose we use the Kaiming initialization method \citep{Kaiming2015} in order to avoid exploding or vanishing gradients during the training. 

The training process itself is regulated by several hyperparameters. The first one is the learning rate, which is a coefficient for the weight updates. In general, a larger learning rate results in updates of larger magnitude, which could in turn lead to faster convergence, but might also reduce the stability of the training process and thus increase the risk of overshooting the minima of the loss landscape. A second hyperparameter is the batch size, which defines the number of samples that are packed together and passed through the network before an optimization step is performed. In general, for bigger training data-sets, a larger batch size helps to increase the efficiency and stability of the training process, since the gradient-descent steps are averaged over many samples and noise is reduced. For the gradient-descent optimizer we use the \acf{Adam} \citep{Kingma2014}. As its name suggests, \acs{Adam} adds an adaptive momentum term to the gradient descent to automatically modify the learning rate and accelerate convergence. When using the \acs{Adam} optimizer the chosen initial value of the learning rate represents only an upper limit. 

We fix the maximum number of learning epochs to 1024. Every epoch the network performs a series of optimization steps by going through the whole training data-set once. Then epoch-averaged loss and validation metric values are computed. If the validation metric value has improved with respect to the previous epoch the current status of the optimized network is saved. We set an early stop of 128 epochs, so that if the validation metric does not improve over this epoch span, the training process automatically stops and the weights of the best epoch are stored. This prevents the network from overfitting the training samples, which would reduce its ability to generalize over unknown data.


\begin{figure*}
\centering
\includegraphics[width = 0.245\textwidth]{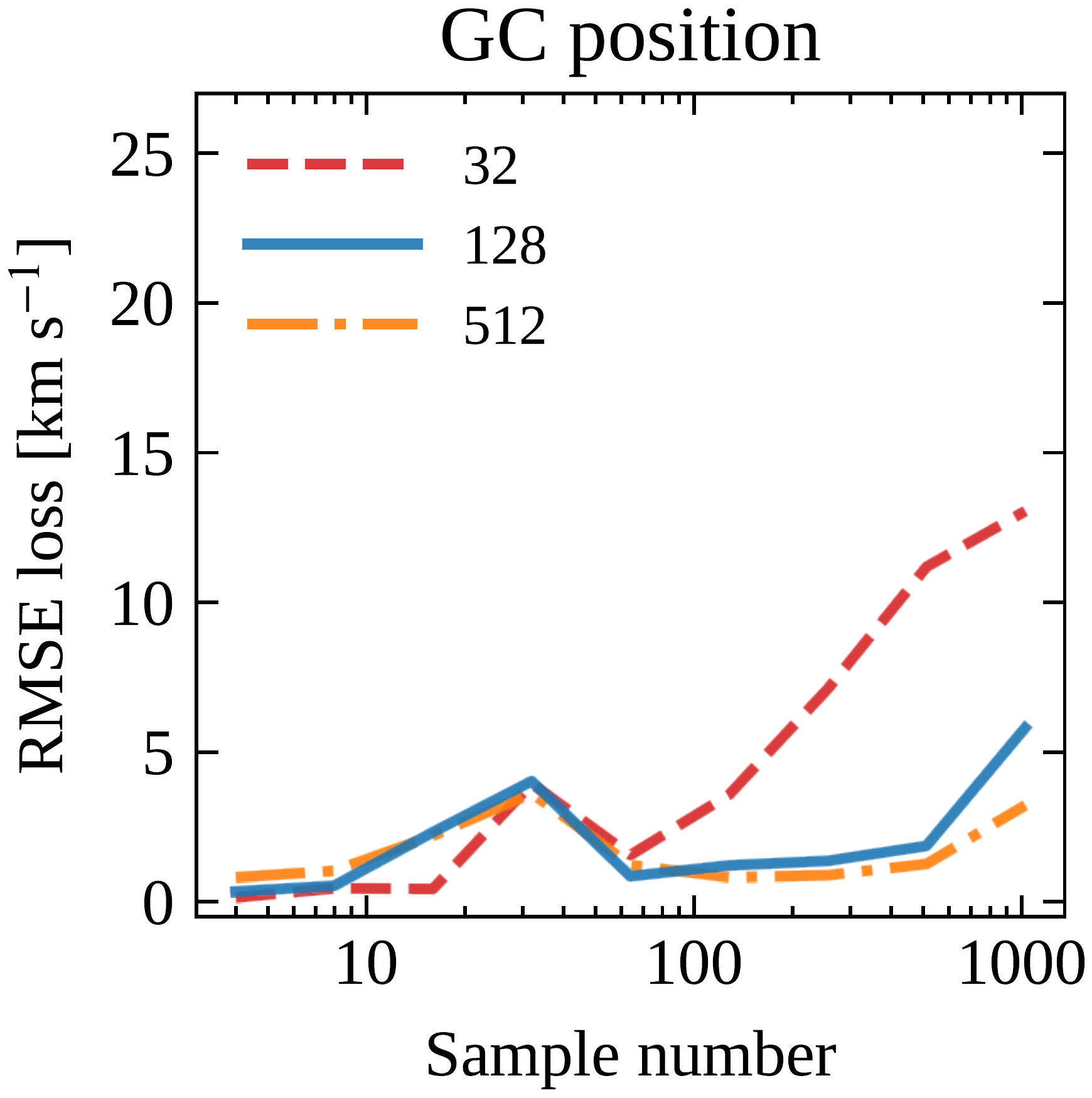}
\includegraphics[width = 0.245\textwidth]{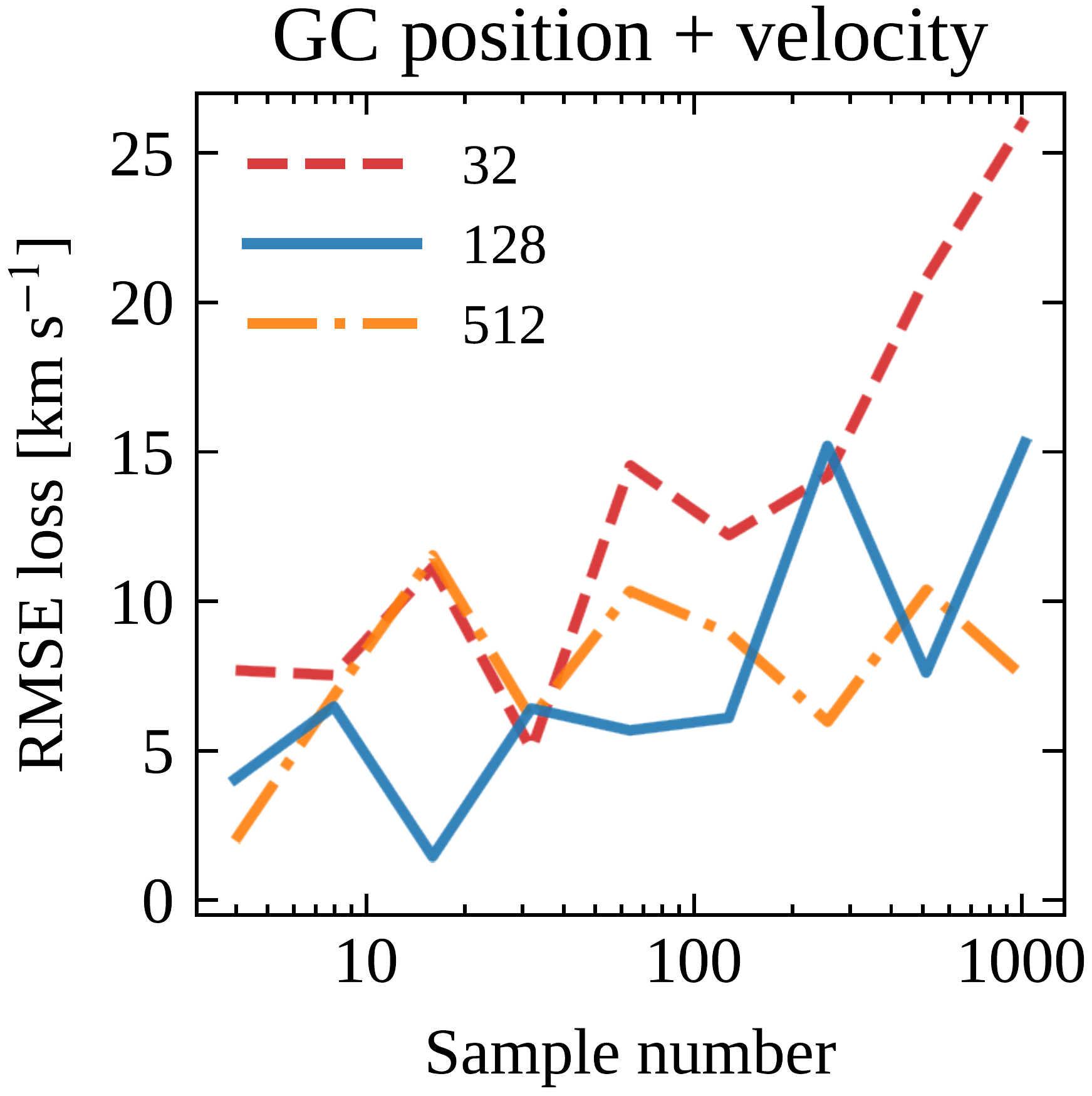}
\includegraphics[width = 0.245\textwidth]{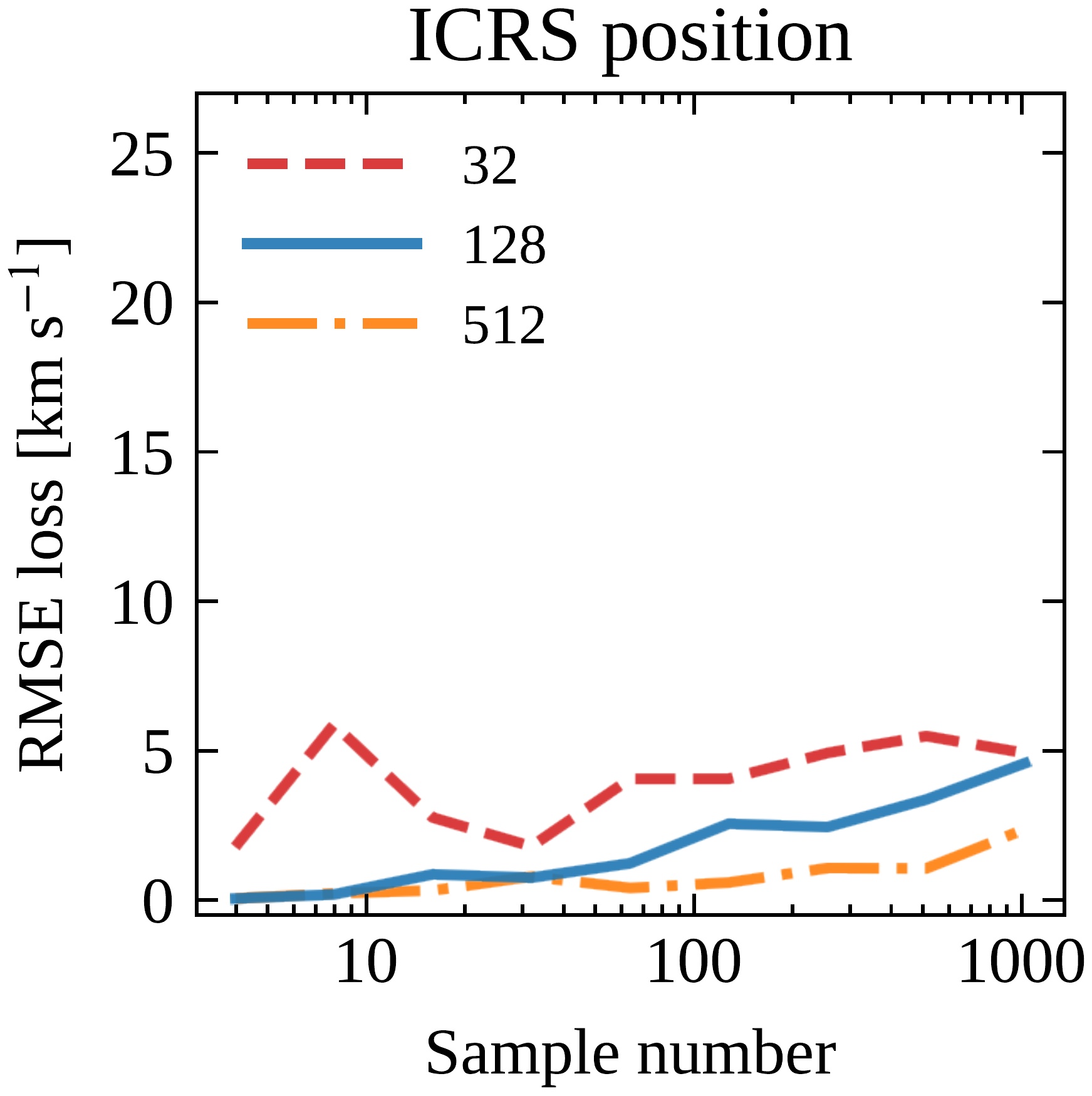}
\includegraphics[width = 0.245\textwidth]{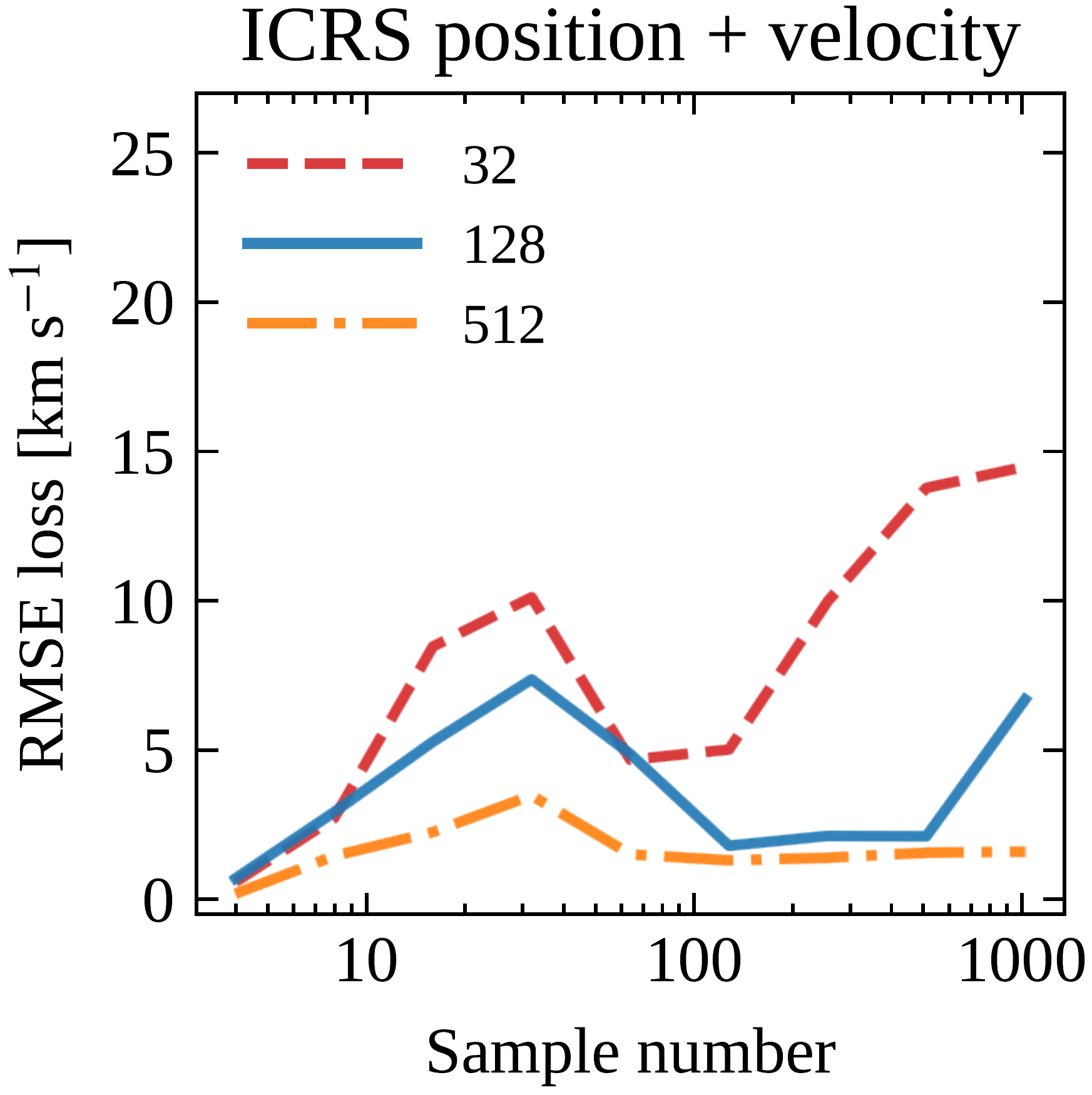}
\caption{\label{fig:lnn_rmse_experiment}{\acs{MLP} best training \acs{RMSE} values for the training process on the single parameter $\sigma_{\rm k}$ of the Maxwell kick-velocity distribution, as a function of the training data-set size and the resolution ({\it red}, {\it blue}, and {\it orange} curves for 32, 128 and 512 respectively) using the four different input configurations T1 (GC position), T2 (GC position + velocity), T3 (ICRS position) and T4 (ICRS position + velocity). }}
\end{figure*}  
\begin{figure*}
\centering
\includegraphics[width = 0.245\textwidth]{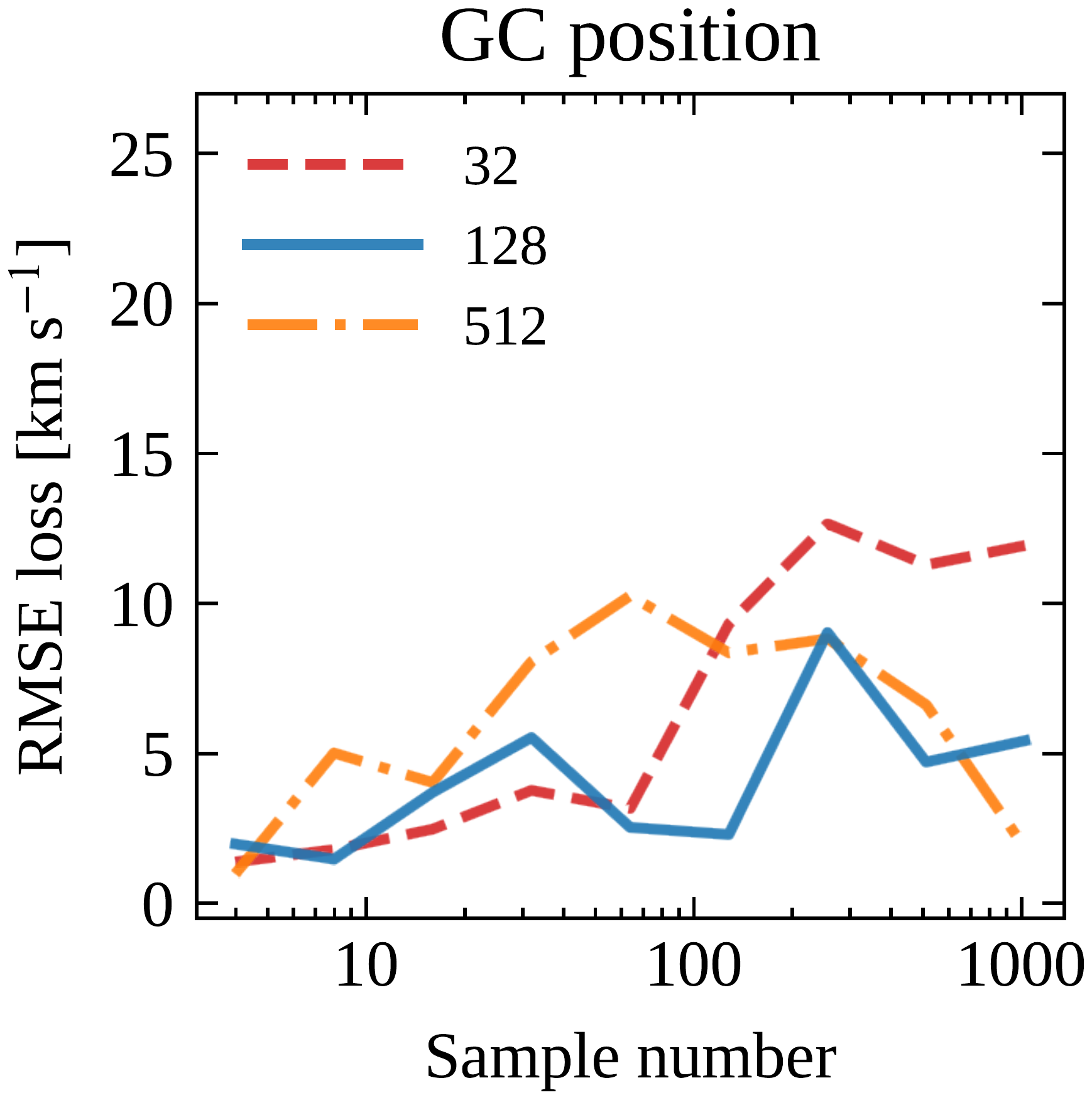}
\includegraphics[width = 0.245\textwidth]{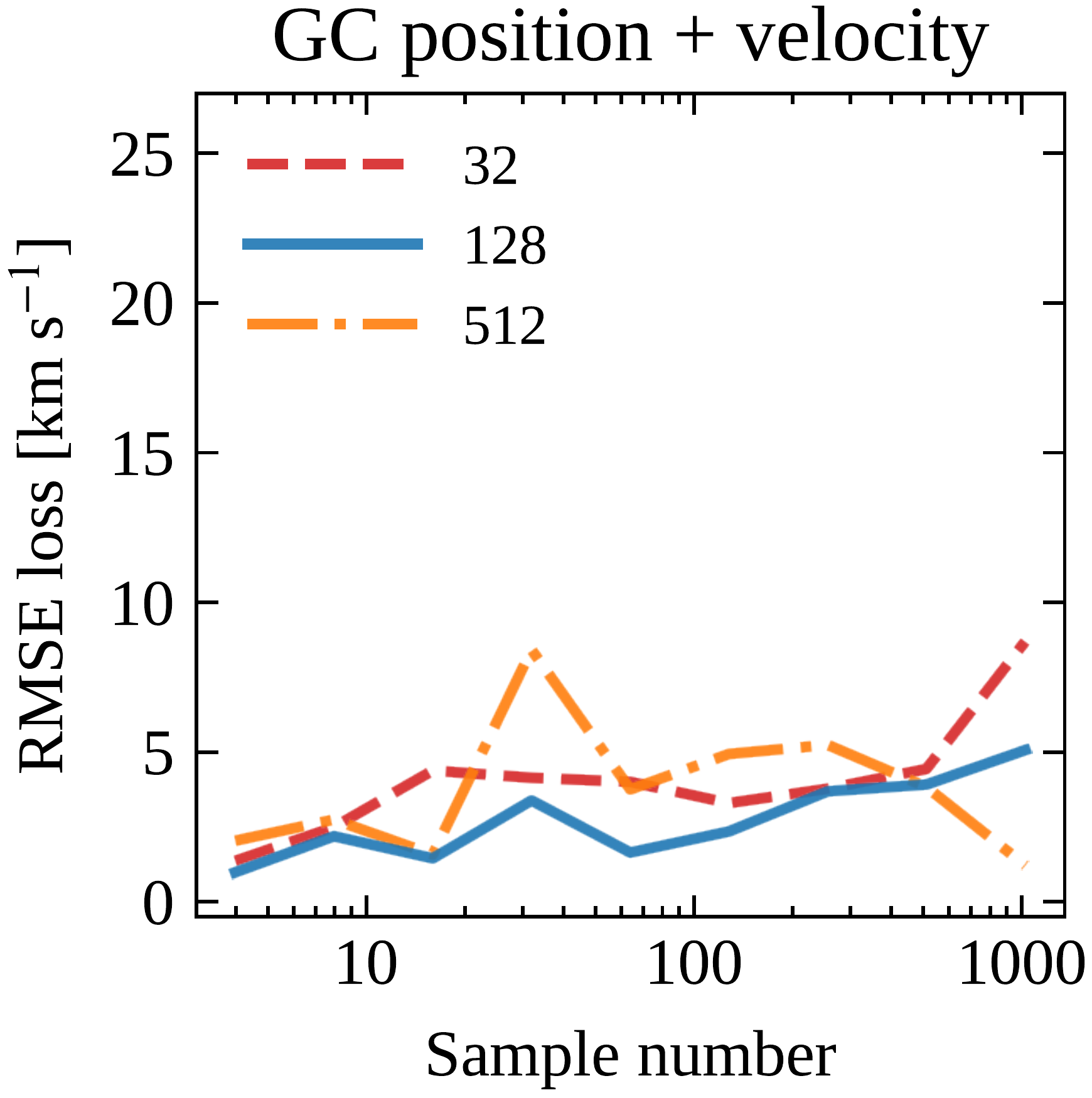}
\includegraphics[width = 0.245\textwidth]{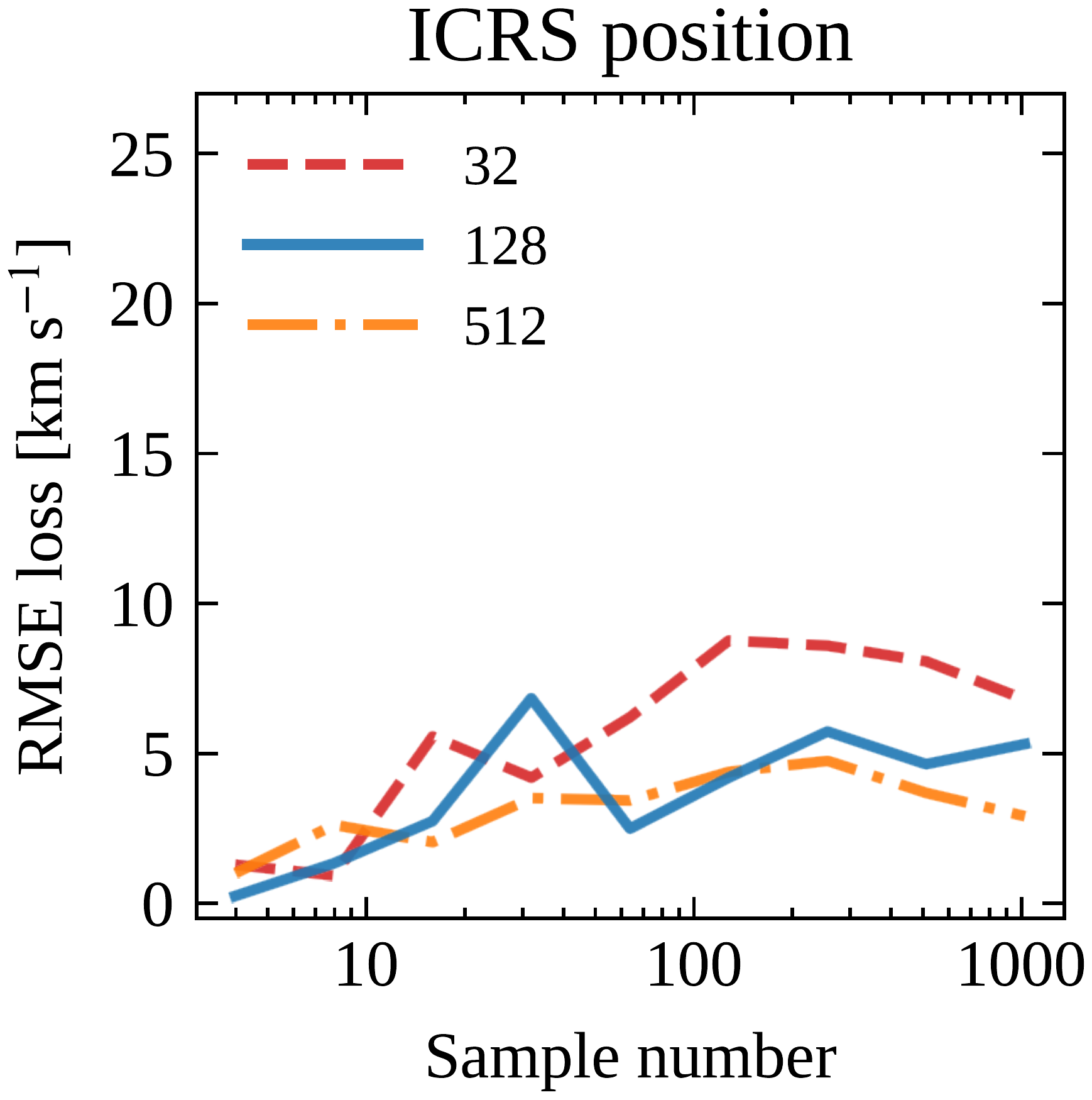}
\includegraphics[width = 0.245\textwidth]{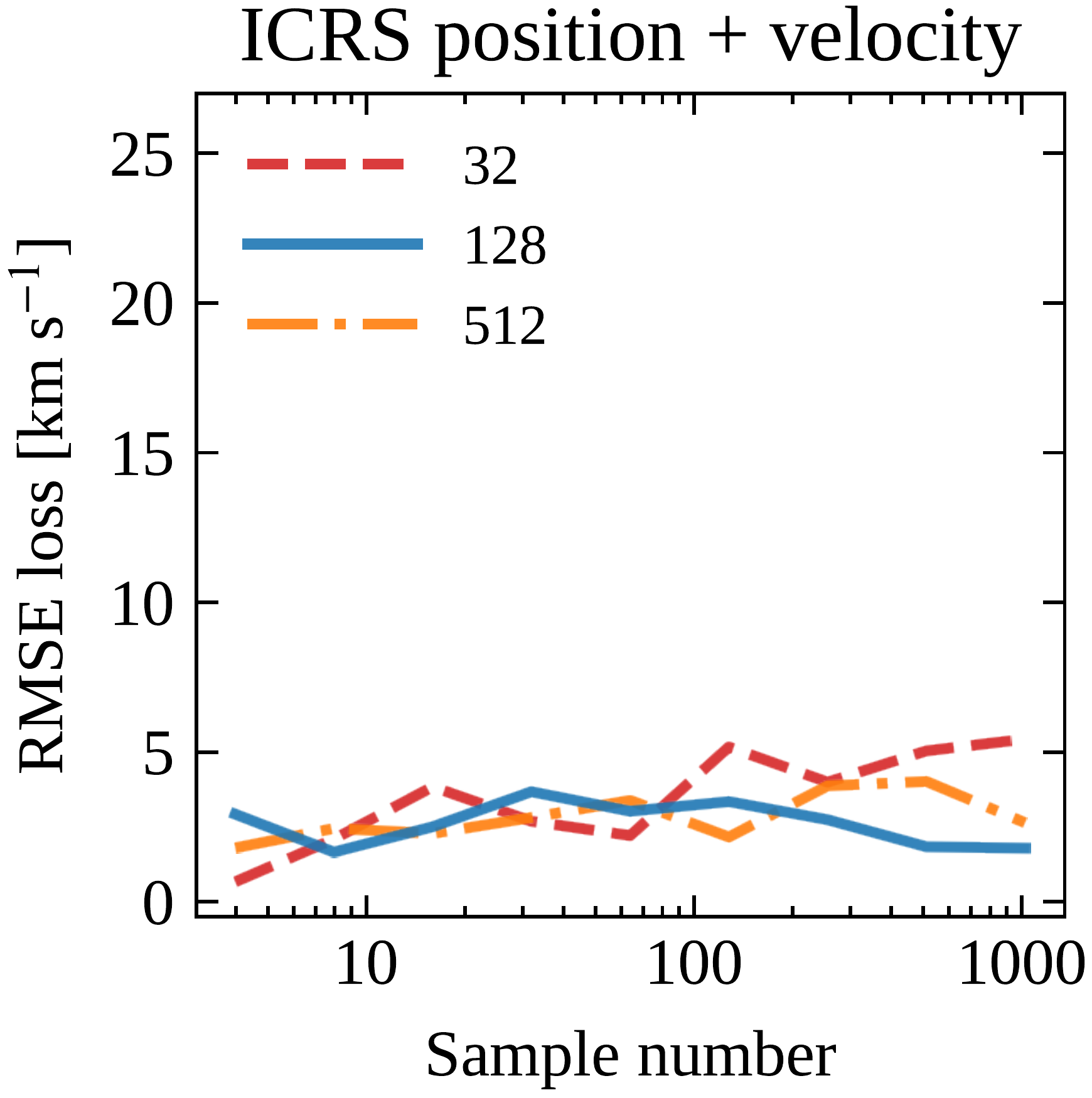}
\caption{\label{fig:cnn_rmse_experiment}{\acs{CNN} best training \acs{RMSE} values for the training process on the single parameter $\sigma_{\rm k}$ of the Maxwell kick-velocity distribution, as a function of the training data-set size and the resolution ({\it red}, {\it blue}, and {\it orange} curves for 32, 128 and 512 respectively) using the four different input configurations T1 (GC position), T2 (GC position + velocity), T3 (ICRS position) and T4 (ICRS position + velocity). }}
\end{figure*}  

\section{Experiments}
\label{sec:experiments}

As a first step we perform several tests to analyze which configuration of the input feature maps provides the best training experience and which proposed neural network architecture, either the \acs{MLP} or the \acs{CNN}, behaves better. Regarding the input configuration we would like to understand (i) what the best type of maps is (galactocentric vs equatorial ICRS); (ii) how many input channels are needed to obtain good results (do density maps provide enough information alone or does the addition of velocity maps improve the results?); (iii) which resolution of the input maps provides the best result. To do so, we first compare the behavior of the \acs{MLP} and the \acs{CNN} when trained to predict a single parameter. We explore different types of input signals by varying the resolution of the density and velocity maps as well as the number of input channels. Once we find the optimal configuration for the input maps and the best performing network, we proceed to test its generalization power. A similar strategy is then followed for the two-parameter prediction.

\subsection{Single-parameter Predictions}
\subsubsection{Data-representation and Architecture Comparison}
\label{sec:1par_tests}

We focus on predicting the parameter $\sigma_{\rm k}$ of the Maxwell kick velocity function and employ the density and velocity maps generated from simulation run S1. We keep aside the data-set with 20000 samples as it will be used to asses the generalization power of the best performing network (see \S \ref{sec:1par_generalization_result}). Thus, we have training sets with increasing number of samples (from 4 to 1024) and increasing map resolution (32, 128, 512). For each of these training sets we compare the performance on four different kinds of input information:
\begin{itemize}
    \item[\textbf{T1}] Galactocentric position information: 1 density map in galactocentric coordinates (1 channel). 
    \item[\textbf{T2}] Galactocentric position and velocity information: 1 density map plus 3 velocity maps in galactocentric coordinates (4 channels).
    \item[\textbf{T3}] ICRS position information: 1 density map in equatorial ICRS coordinates (1 channel).
    \item[\textbf{T4}] ICRS position and velocity information: 1 density map plus 2 proper motion maps in equatorial ICRS coordinates (3 channels).
\end{itemize}
We fix the network architectures and the training set-up as described in \S \ref{sec:net_architecture} and \S \ref{sec:train_process}. For these initial experiments we do not incorporate validation but only focus on the training results for different types of data-sets, as we are not yet interested in evaluating the generalization power of the networks. To assess the convergence of the training runs we set a threshold for the $\sigma_{\rm k}$ \acs{RMSE} training loss to $\unit[10]{km \, s^{-1}}$. If the final \acs{RMSE} value is higher than this threshold the training is repeated up to a maximum of 8 times. If convergence is not reached after 8 trials we take the trained model with the lowest final \acs{RMSE}. For each training experiment we also monitor the computational time needed to perform a single optimization step on a single data batch (see Appendix \ref{app:timing_tests} for more details). 

Initially we perform a search for the starting value of the learning rate that provides the best results. In particular for the \acs{MLP} we find that to ensure a decaying \acs{RMSE} value during training, the initial learning rate needs to be decreased as the input-map resolution increases. Therefore, after several tests we set the initial learning rate to $10^{-4}$, $10^{-5}$ and $10^{-6}$, respectively, for the 32, 128 and 512 resolution maps. The \acs{CNN} instead is more flexible and stable and all three initial learning rates are suitable for every resolution. In this case we set it to $10^{-4}$ to obtain training convergence in the smallest number of epochs.

Next, we tune the batch size to make the learning process more stable and efficient as the number of training samples increases. In general, we keep the batch size to 1 for the data-set sizes 4, 8, 16, 32 and progressively increase it to 4, 8, 16, 32, 64 as the data-set size increases to 64, 128, 256, 512, 1024, respectively. Only when testing the performance of the \acs{MLP} on the T2 input configuration we need to further fine tune the batch size in order to reach acceptable values of the \acs{RMSE}. In all other cases the batch sizes mentioned above work well. 

\begin{figure*}
\centering
\includegraphics[width = 0.32\textwidth]{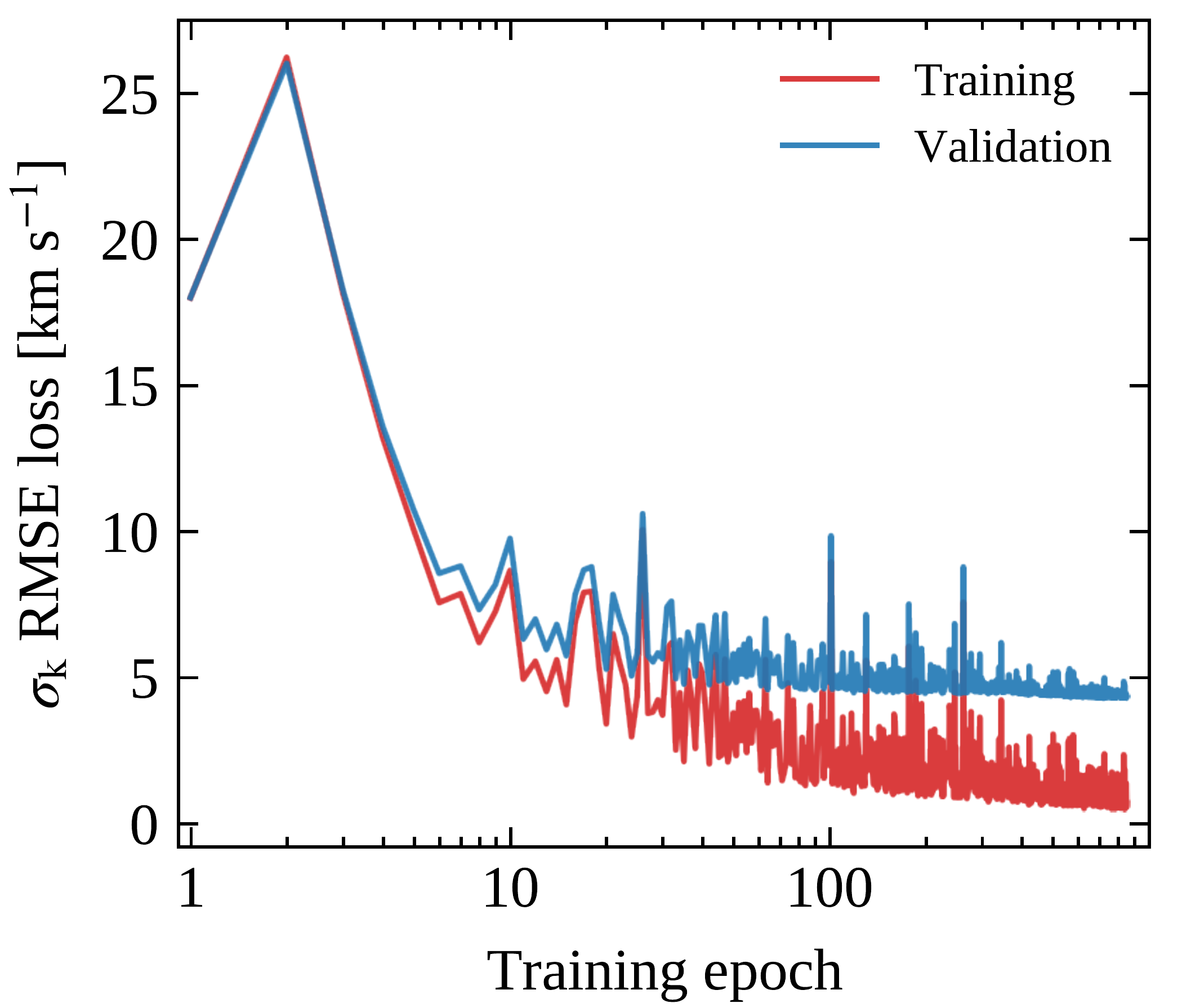}
\includegraphics[width = 0.32\textwidth]{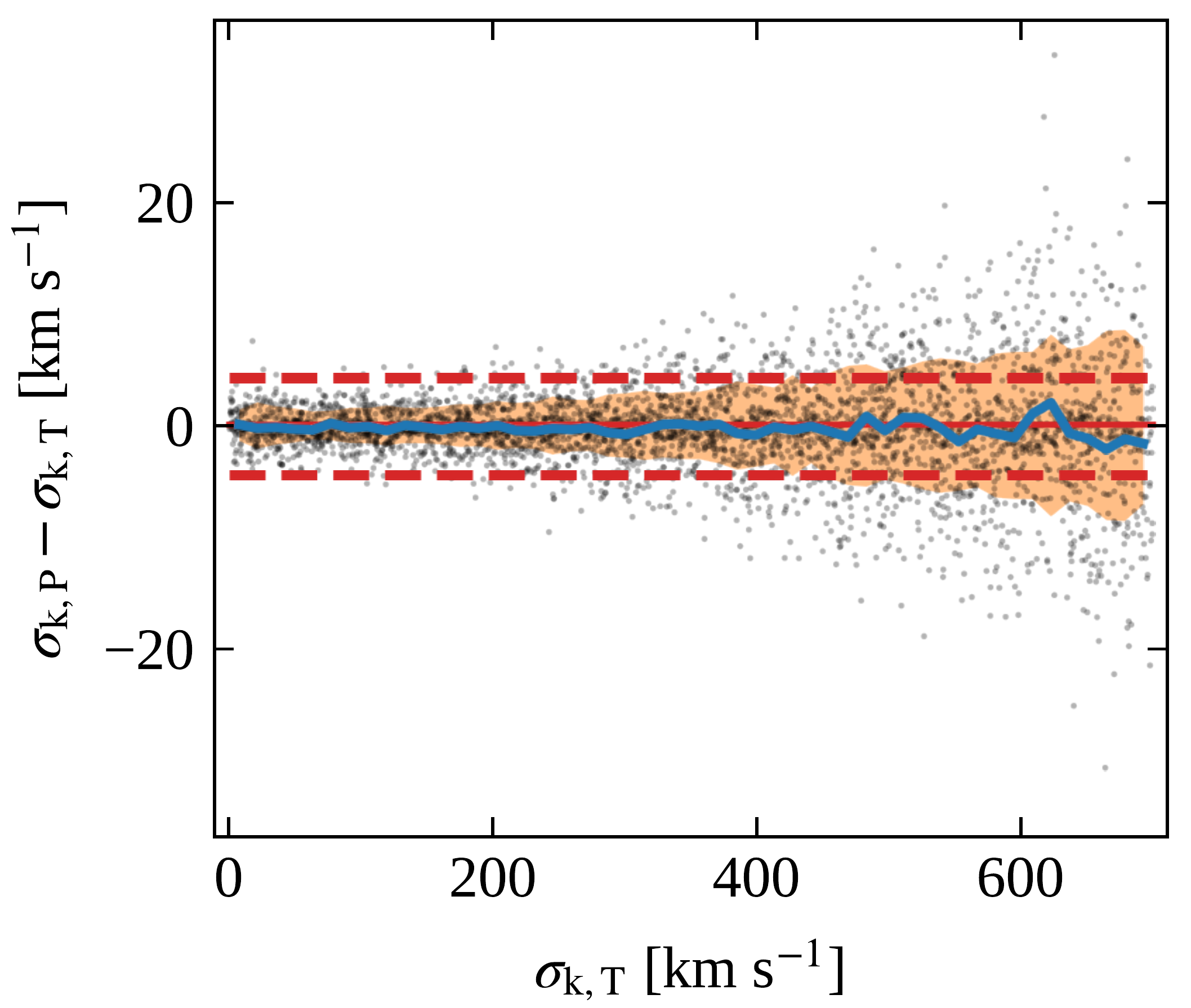}
\includegraphics[width = 0.32\textwidth]{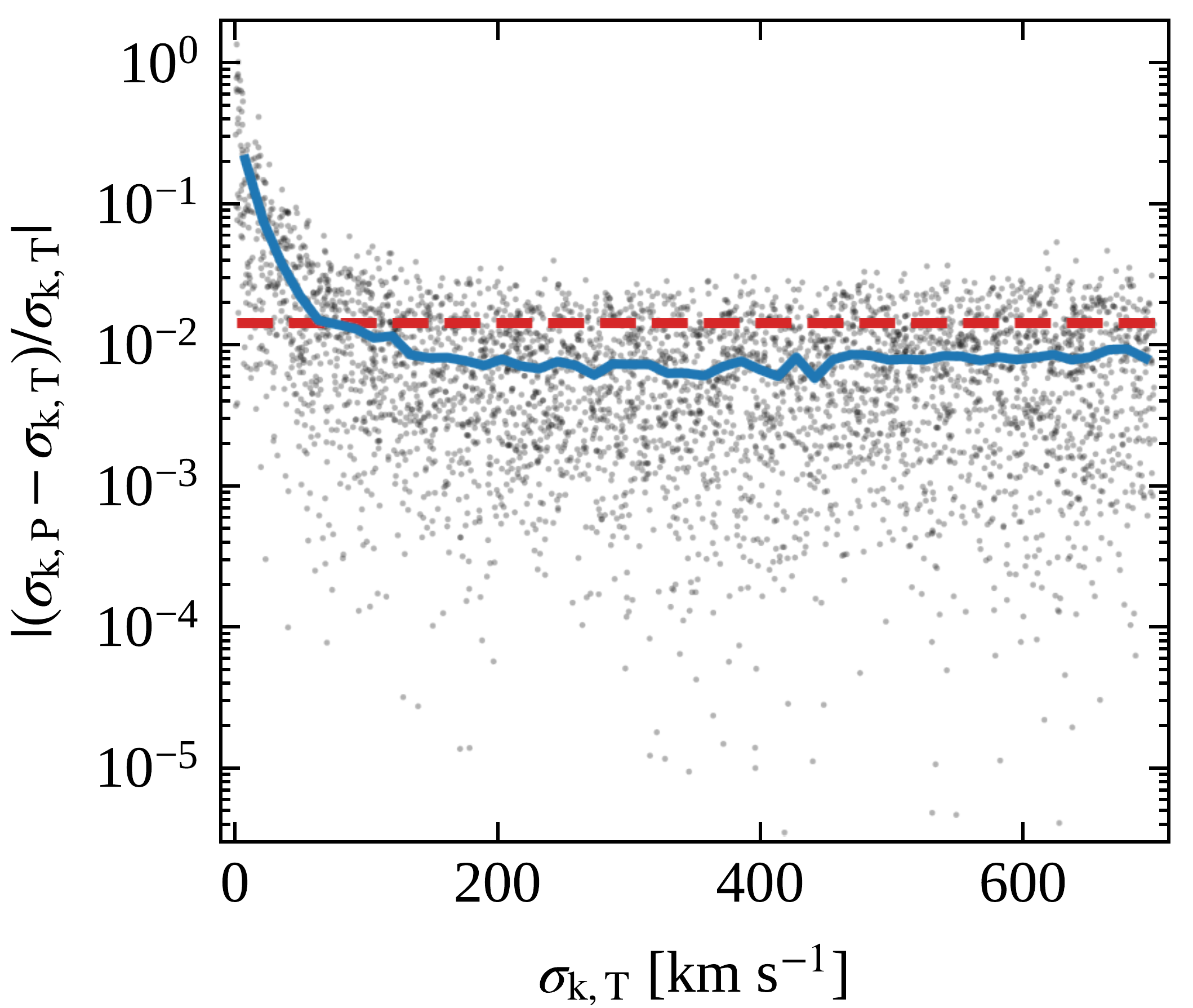}\\
\vspace{0.25cm}
\includegraphics[width = 0.32\textwidth]{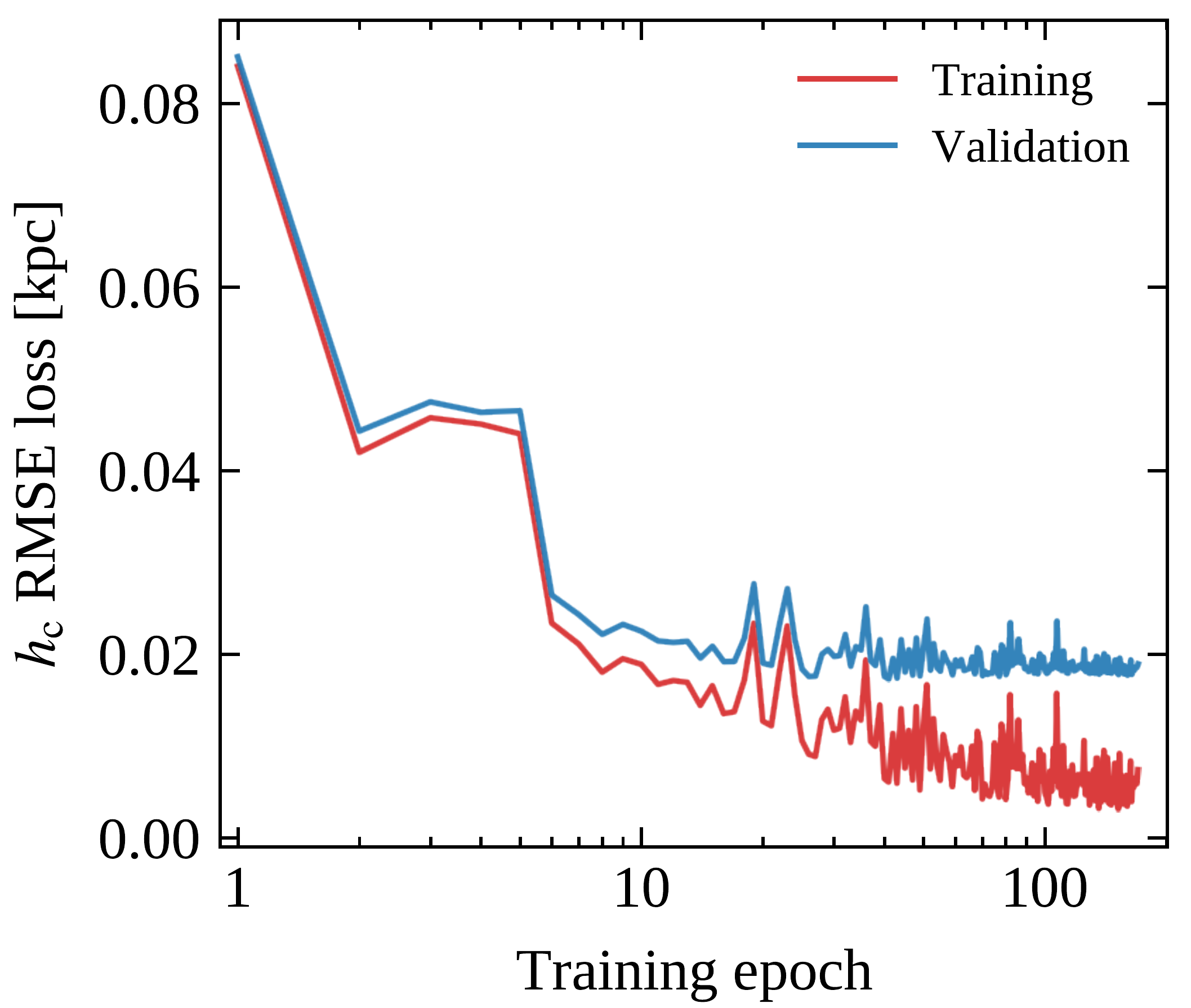}
\includegraphics[width = 0.32\textwidth]{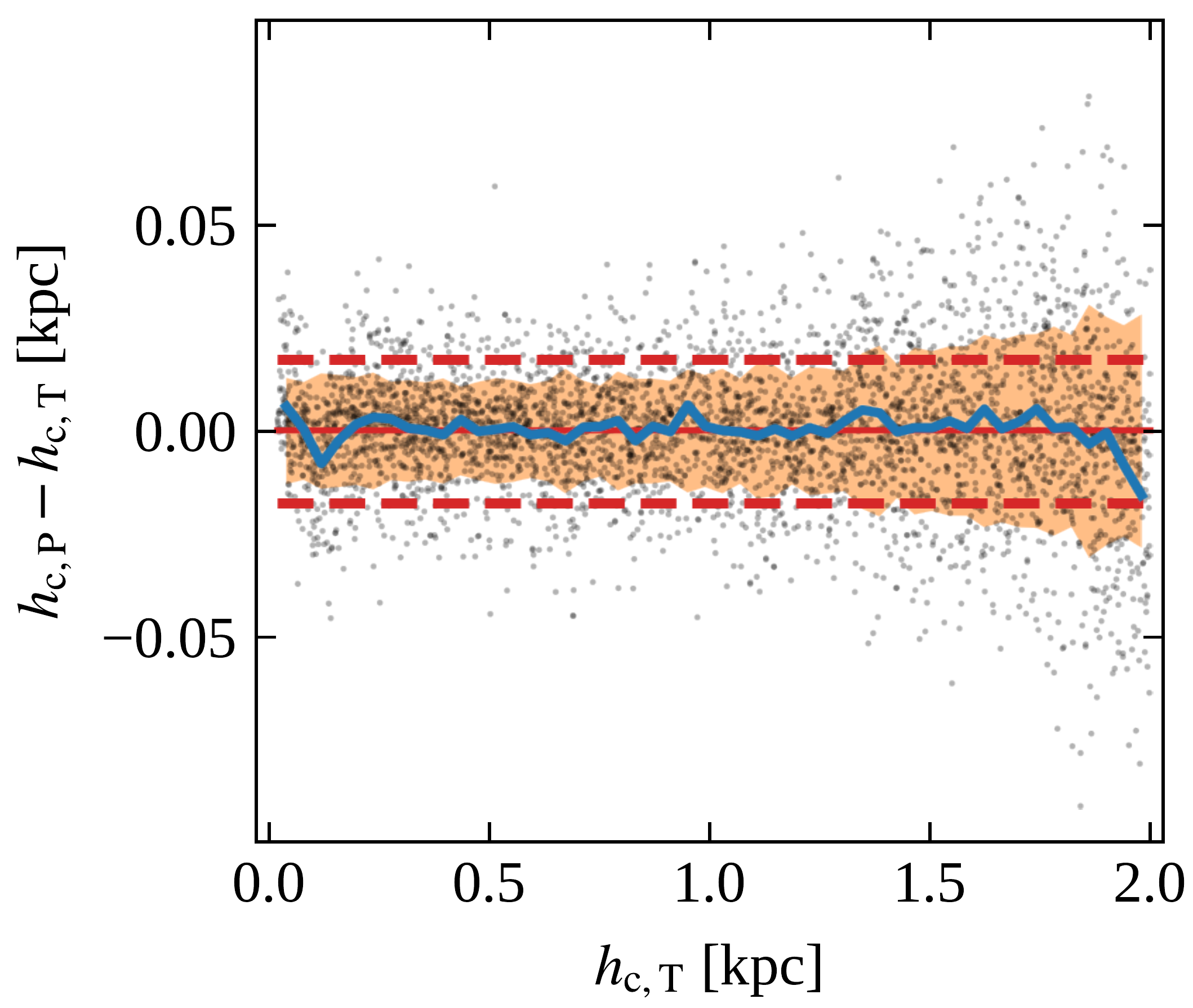}
\includegraphics[width = 0.32\textwidth]{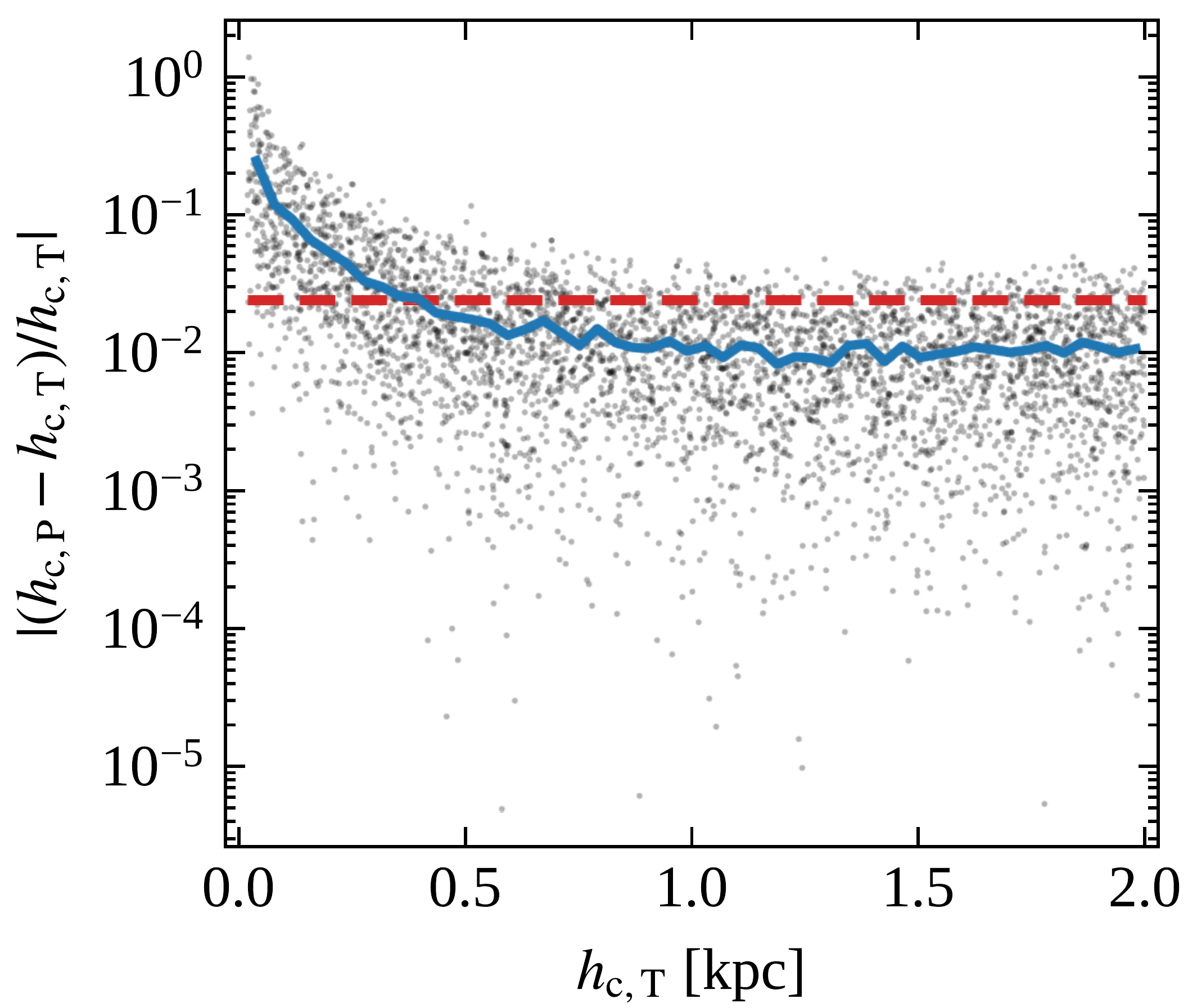}
\caption{Results of the \acs{CNN} single-parameter prediction for the kick-velocity parameter $\sigma_{\rm k}$ and scale height $h_{\rm c}$ for the corresponding validation sets. {\it Top (bottom) left panel}: evolution of the \acs{RMSE} training ({\it red}) and validation ({\it blue}) losses as a function of the training/validation epoch for the $\sigma_{\rm k}$ ($h_{\rm c}$) parameters. {\it Top (bottom) central panel}: residuals of the prediction as a function of the target $\sigma_{\rm k}$ ($h_{\rm c}$) value; the subscripts P and T refer to the predicted and target values, respectively. The red dashed lines delimit the $68$\% uncertainty region corresponding to a \acs{RMSE} $=\unit[4.4]{km \, s^{-1}}$ ($\unit[0.017]{kpc}$) computed over the whole range $[1, 700]$ $\unit[]{km \, s^{-1}}$ ($[0.02, 2]$ $\unit[]{kpc}$). The {\it orange} region delimits the $68$\% uncertainty region computed as a running \acs{RMSE} for which we have divided the full $\sigma_{\rm k}$ ($h_{\rm c}$) range into $50$ bins of equal size. The {\it blue} line shows the trend of the average residuals which are well centered around the value 0. {\it Top (bottom) right panel}: relative error of the prediction as a function of the target $\sigma_{\rm k}$ ($h_{\rm c}$) value. The red dashed line corresponds to a \acs{MRE} $=0.014$ ($0.024$) computed over the whole range $[1, 700]$ $\unit[]{km \, s^{-1}}$ ($[0.02, 2]$ $\unit[]{kpc}$). The {\it blue} line shows the trend of the running \acs{MRE} computed over $50$ bins of equal size into which we have divided the full $\sigma_{\rm k}$ ($h_{\rm c}$) range.}
\label{fig:1par_cnn_inference_sigmak_hc}
\end{figure*}  

The results of our experiments using the \acs{MLP} and the \acs{CNN} on the training data-sets from S1 with configuration T1, T2, T3 and T4 are shown in Figs.~\ref{fig:lnn_rmse_experiment} and \ref{fig:cnn_rmse_experiment}, respectively (see Appendix \ref{app:timing_tests} for the timing results), highlighting the best converged \acs{RMSE} values as a function of the training data-set size and the resolution.

First of all, we note that for both model architectures, ICRS maps allow us to obtain slightly better results in terms of the best \acs{RMSE} values. An explanation for this could be that only the ICRS maps contain 3D information of the Galaxy, i.e., they encode the stellar height distribution with respect to the galactic disk which correlates with the kick velocity magnitude. In fact the higher the kick velocity of newborn neutron stars the more spread out their distribution in galactic height $z$ at the end of the dynamical evolution. On the other hand galactocentric maps show the Galaxy represented face on and the information on the stars' height $z$ with respect to the Galactic plane is hidden. Therefore, it is likely easier for the networks to distinguish populations with different $\sigma_{\rm k}$ by processing the ICRS maps.

We also note that the addition of the velocity or proper motion information has opposite effects on the two model architectures. In particular, on one side it worsens the \acs{MLP} performance, as the best \acs{RMSE} almost doubles. On the other side, it helps the \acs{CNN} in reducing the overall \acs{RMSE}.
We interpret this as an indication that, as the complexity of the input data increases, a deeper (with more layers) and more sophisticated model architecture like the \acs{CNN} is more suitable to process the input data and extract meaningful features to perform the regression task. However, the improvement is not dramatic, which could suggest that the density maps already provide enough information to distinguish, with sufficient precision, populations simulated with different $\sigma_{\rm k}$ values.

Concerning the resolution, we observe that higher resolutions allow us to reach slightly lower \acs{RMSE} values with both the \acs{MLP} and the \acs{CNN} but at the expense of longer computation time (see Figs.~\ref{fig:lnn_time_experiment} and \ref{fig:cnn_time_experiment} in Appendix~\ref{app:timing_tests}). We therefore consider the small differences between the best \acs{RMSE} values obtained with 128 and 512 resolutions not sufficient to justify the choice of the higher resolution. Hence, the use of the 128 resolution appears to be a good compromise to ensure good accuracy and reasonably fast training.

In light of these results for the single-parameter estimation, we conclude that the optimal representation to be used for training is composed of the ICRS density plus proper motion maps with 128 resolution. Moreover, as the \acs{CNN} obtains the best results and appears more stable and flexible when compared to the simple \acs{MLP} (especially for the multi-channel input features), we employ our \acs{CNN} architecture in the following experiments, which should also be less prone to overfitting.


\subsubsection{Generalization Results} 
\label{sec:1par_generalization_result}

As the next step we separately train the \acs{CNN} to predict the $\sigma_{\rm k}$ and $h_{\rm c}$ parameters by using the two big data-sets with 20000 simulations each (see S1 and S2). As input features we use the 3-channel ICRS representation with one density map and two proper motion maps with 128 resolution that ensure the best results as suggested by our earlier experiments. We randomly split both data-sets into training and validation subsets with a relative percentage of $80/20\%$, respectively. Therefore, training is performed over 16000 simulations, randomly sampled from the entire data-sets, while validation is performed over the remaining 4000 simulations. We adopt an initial learning rate of $10^{-4}$, a batch size of 64, set the total number of learning epochs to 1024 and an early stop at 128 epochs to avoid overfitting. The evolution of the training and validation losses is shown in the left panels of Fig.~\ref{fig:1par_cnn_inference_sigmak_hc}. 

The predictions of the trained network on the validation set for $\sigma_{\rm k}$ and $h_{\rm c}$ are summarized in Fig.~\ref{fig:1par_cnn_inference_sigmak_hc} and Table \ref{tab:genralization_results}. In the first case, the network is able to predict the value of the kick-velocity parameter $\sigma_{\rm k}$ for the simulations in the validation data-set with a \acs{RMSE} uncertainty of $\unit[4.4]{km \, s^{-1}}$, computed over the whole range $[1, 700]$ $\unit[]{km \, s^{-1}}$. This is indicated by the red dashed lines in the residuals plot (see top central panel of Fig.~\ref{fig:1par_cnn_inference_sigmak_hc}), which delimit the $68$\% uncertainty region. In the second case, the network is able to predict the value of the scale-height parameter $h_{\rm c}$ for the simulations in the validation data-set with a \acs{RMSE} uncertainty of $\unit[0.017]{kpc}$, computed over the whole range $[0.02, 2] $ $\unit[]{kpc}$ (see bottom central panel of Fig.~\ref{fig:1par_cnn_inference_sigmak_hc}). Note that in both experiments, the residuals spread out as the target values increase. We visualize this by computing a running \acs{RMSE} with increasing target values. As is shown by the orange regions in the residuals plots, the running \acs{RMSE} increases from $1.4$ to $\unit[7.1]{km \, s^{-1}}$ for $\sigma_{\rm k}$ and from $0.013$ to $\unit[0.028]{kpc}$ for $h_{\rm c}$, respectively. However, the average residuals are consistent with $0$ over the entire target value ranges as marked by the blue line in the residual plots, showing no anomalous trends in the predicted values.

In the right panels of Fig.~\ref{fig:1par_cnn_inference_sigmak_hc} we also show the relative error to highlight the precision with which the network is able to predict the $\sigma_{\rm k}$ and $h_{\rm c}$ parameters for a given target value. We observe that the precision of the predictions improves with increasing $\sigma_{\rm k}$ and $h_{\rm c}$ and eventually stabilizes to a relative error of around $0.01$. This is highlighted by the blue lines, which show the trend of the \acf{MRE} as a function of the target parameters. The red dashed lines correspond instead to the \acs{MRE} computed on the whole parameter ranges and are equal to $0.014$ and $0.024$ for $\sigma_{\rm k}$ and $h_{\rm c}$, respectively. The fact that the precision of the predictions decreases at the lower end of the target ranges suggests that (for our chosen network set-up) the input maps become harder to distinguish as the neutron stars' initial kick velocities and their galactic birth heights decrease in magnitude.

We then evaluate the generalization capability of the two trained networks on the corresponding test sets with 1000 samples each. We find RMSEs of $\unit[4.8]{km \, s^{-1}}$ and $\unit[0.019]{kpc}$ and MREs of $0.018$ and $0.029$ for $\sigma_{\rm k}$ and $h_{\rm c}$, respectively. To further assess the confidence intervals of both estimators, we evaluate the RMSE and the MRE of the parameter values predicted by the two networks over 1000 bootstrapped sets of the related test sets. We find that the RMSE variation is around $3$\%, while the MRE variation is around $11$\% for both predicted parameters. These results indicate that the trained networks are able to generalize well over unseen data-sets.


\subsection{Two-parameter Predictions}
\subsubsection{Data-representation Comparison}

To see how the \acs{CNN} behaves when two parameters, i.e., the kick-velocity parameter $\sigma_{\rm k}$ and the characteristic scale height $h_{\rm c}$, are inferred simultaneously, we use the data-set of maps generated from simulation run S3 (see \S \ref{sec:dataset_creation}). In this case, we have training sets with increasing number of samples, 16 = 4x4, 64 = 8x8, 256 = 16x16, 1024 = 32x32, 4096 = 64x64. We leave aside the largest data-set with 16384 = 128x128 simulations for our final generalization experiment. Given the results of the single-parameter training experiments, we choose the 128 resolution maps and compare the \acs{CNN}'s performance on the four kinds of input information T1, T2, T3 and T4 as we did for the single-parameter case (see \S \ref{sec:1par_tests}). We fix the \acs{CNN} architecture and the training set-up as described in \S \ref{sec:net_architecture} and \S \ref{sec:train_process}, respectively. As for the single-parameter comparison, we only focus on the training behavior for this initial comparison. We keep track of the \acs{RMSE} training losses for both parameters separately, but training proceeds by minimizing the total loss computed on both parameters. We set the convergence threshold for $\sigma_{\rm k}$ to $\unit[10]{km \, s^{-1}}$ and for $h_{\rm c}$ to $\unit[0.5]{kpc}$; if convergence is not reached for both parameters after 8 training trials we quote the experiment with the best performance. As before, we also keep track of the computational time for a single optimization step (see Appendix \ref{app:timing_tests}). The initial learning rate is set to $10^{-4}$, while the batch size is changed according to the data-set size. In particular, we use a batch size of 1, 4, 16, 64, 128 for the data-set sizes 16, 64, 256, 1024 and 4096, respectively. 

The results of our experiments using the \acs{CNN} for the two-parameter prediction on the training data-sets from S3, with configuration T1, T2, T3 and T4 are summarized in Fig. \ref{fig:2par_cnn_tests} (see Appendix \ref{app:timing_tests} for the timing results). As in the single-parameter case we find that the best results are provided by the ICRS maps, which allow us to reach the lowest training \acs{RMSE} losses for both parameters. In particular, for the ICRS 3-channel input, the \acs{CNN} is able to reach a training \acs{RMSE} loss $\lesssim \unit[5]{km \, s^{-1}}$ for the $\sigma_{\rm k}$ parameter, comparable with the single-parameter case. For $h_{\rm c}$, the \acs{CNN} reaches a training \acs{RMSE} $\lesssim \unit[0.1]{kpc}$. However, we note a drop in accuracy for the $\sigma_{\rm k}$ parameter when only the ICRS density maps are used. As already mentioned in \S \ref{sec:1par_tests}, information on the stars' $z$-coordinate is encoded in the ICRS maps. As we simultaneously vary the initial kick velocities and the galactic heights of the pulsars' birth places, the degeneracy between the effects of these two parameters becomes relevant. This makes a distinction of the impact of one parameter over the other more difficult for the network, when only ICRS density maps are provided. Adding two extra channels that contain information about the stars' proper motion thus helps to improve the accuracy on the kick-velocity parameter estimation. The results of these initial explorations are promising and indicate that our simple \acs{CNN} architecture has good predictive power for both parameters, if provided with all three input channels in the ICRS representation.

\begin{figure}
\centering
\includegraphics[width = 0.23\textwidth]{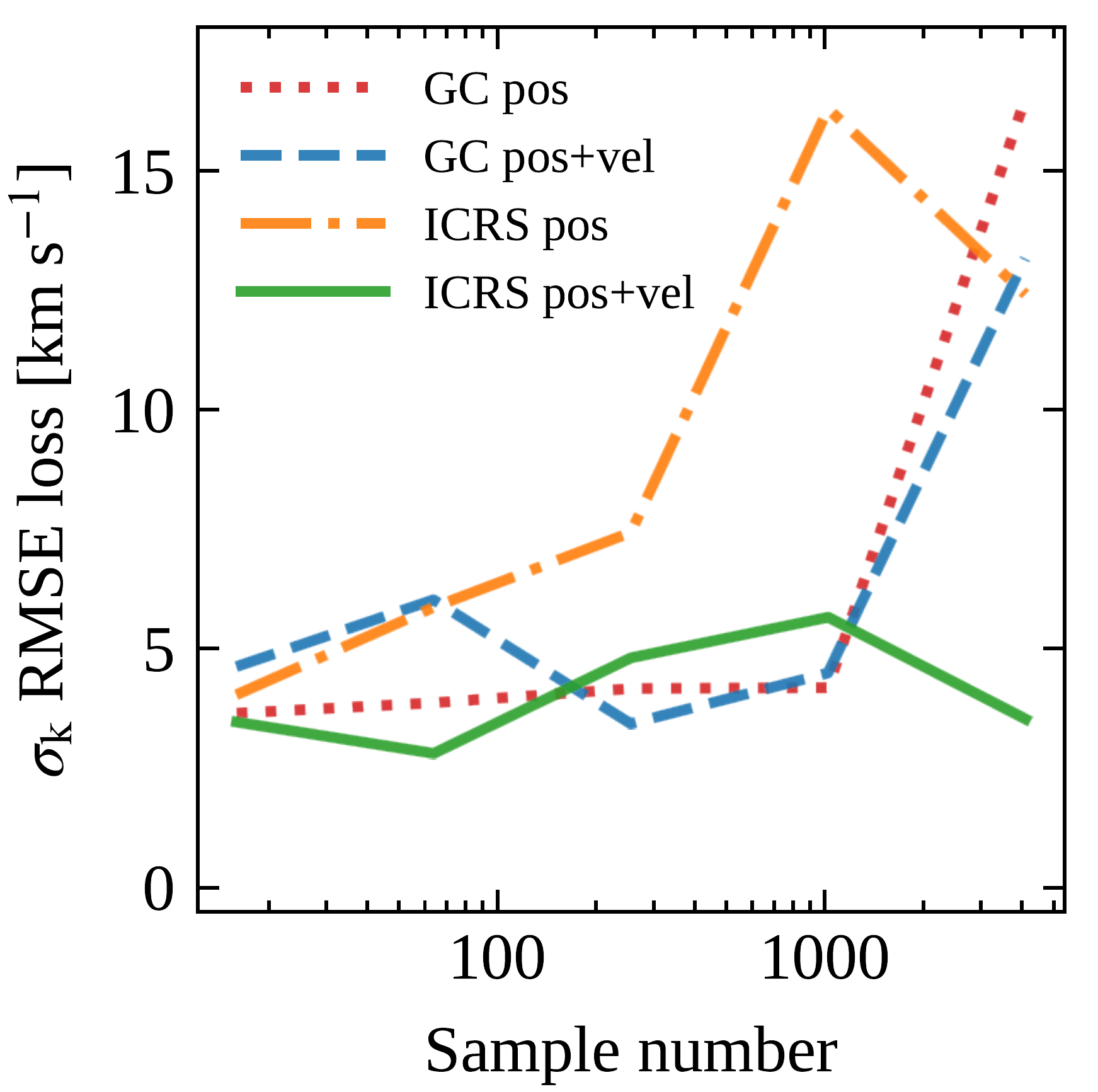}
\includegraphics[width = 0.23\textwidth]{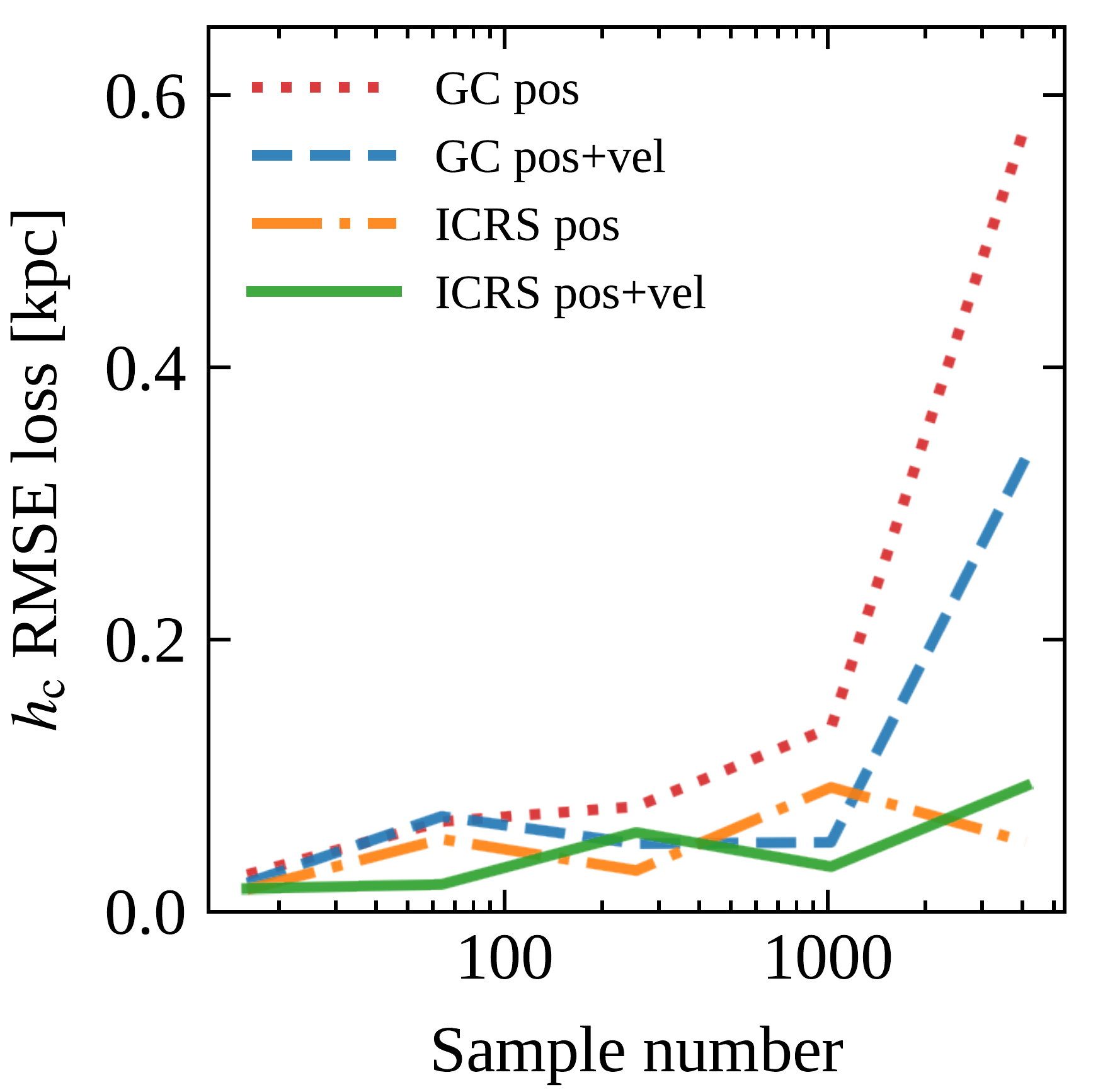}
\caption{\label{fig:2par_cnn_tests}{\acs{CNN} best training \acs{RMSE} values for the training process on the two parameters $\sigma_{\rm k}$ of the Maxwell kick-velocity distribution (\textit{left panel}) and characteristic scale height $h_{\rm c}$ (\textit{right panel}) as a function of the training data-set size for the four different input configurations T1 (GC position), T2 (GC position + velocity), T3 (ICRS position) and T4 (ICRS position+ velocity).}}
\end{figure}  

\begin{figure*}
\centering
\includegraphics[width = 0.32\textwidth]{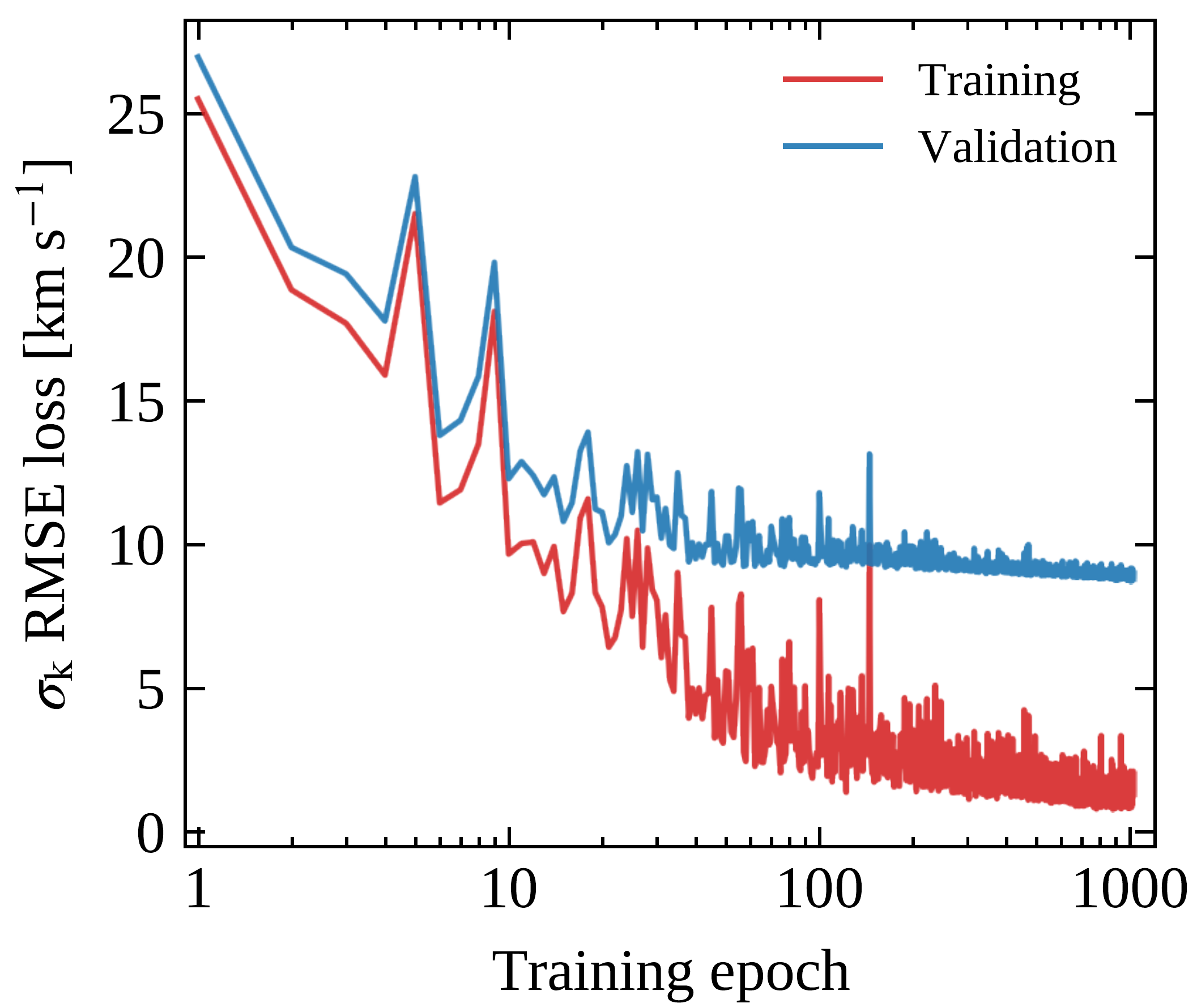}
\includegraphics[width = 0.32\textwidth]{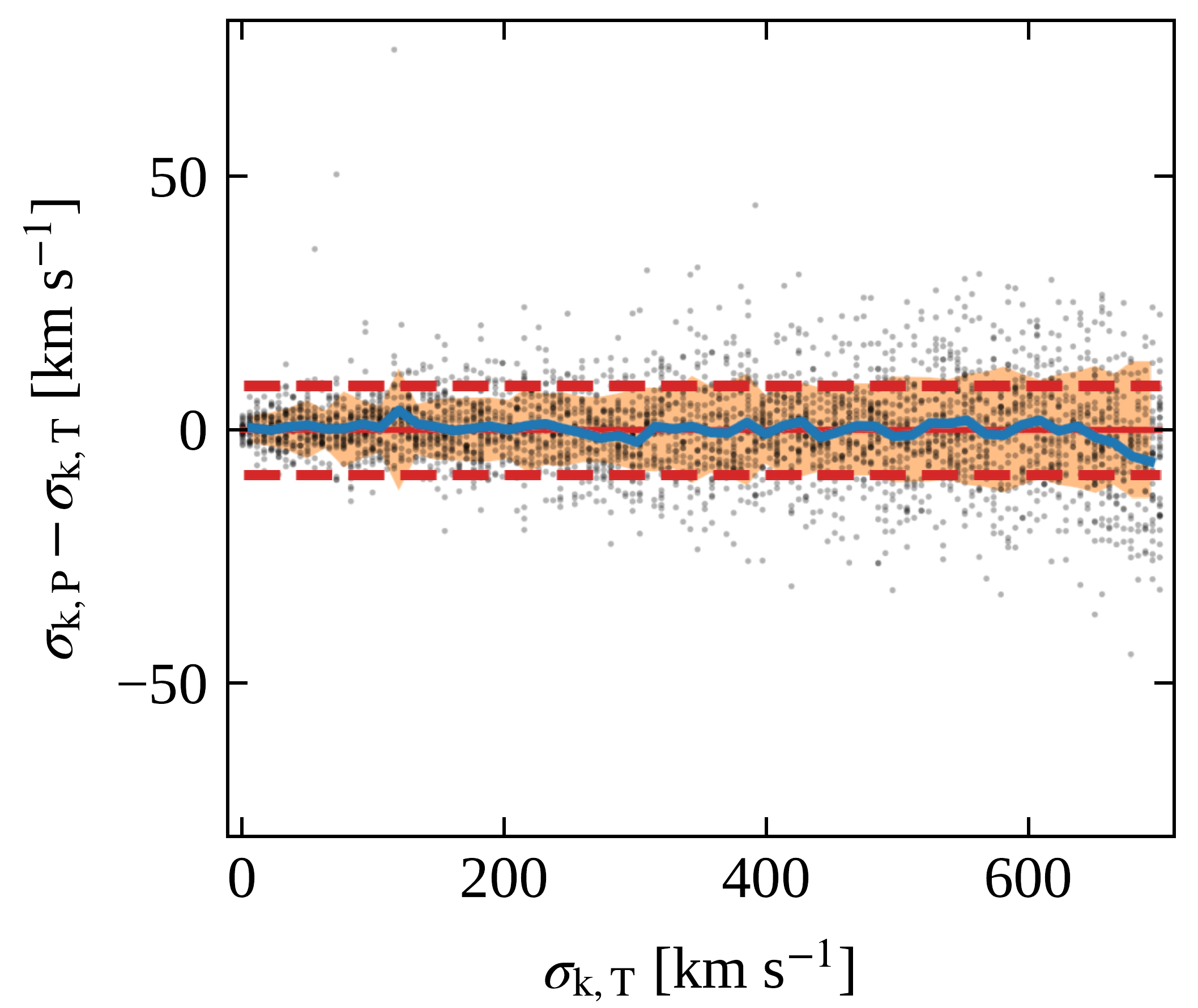}
\includegraphics[width = 0.32\textwidth]{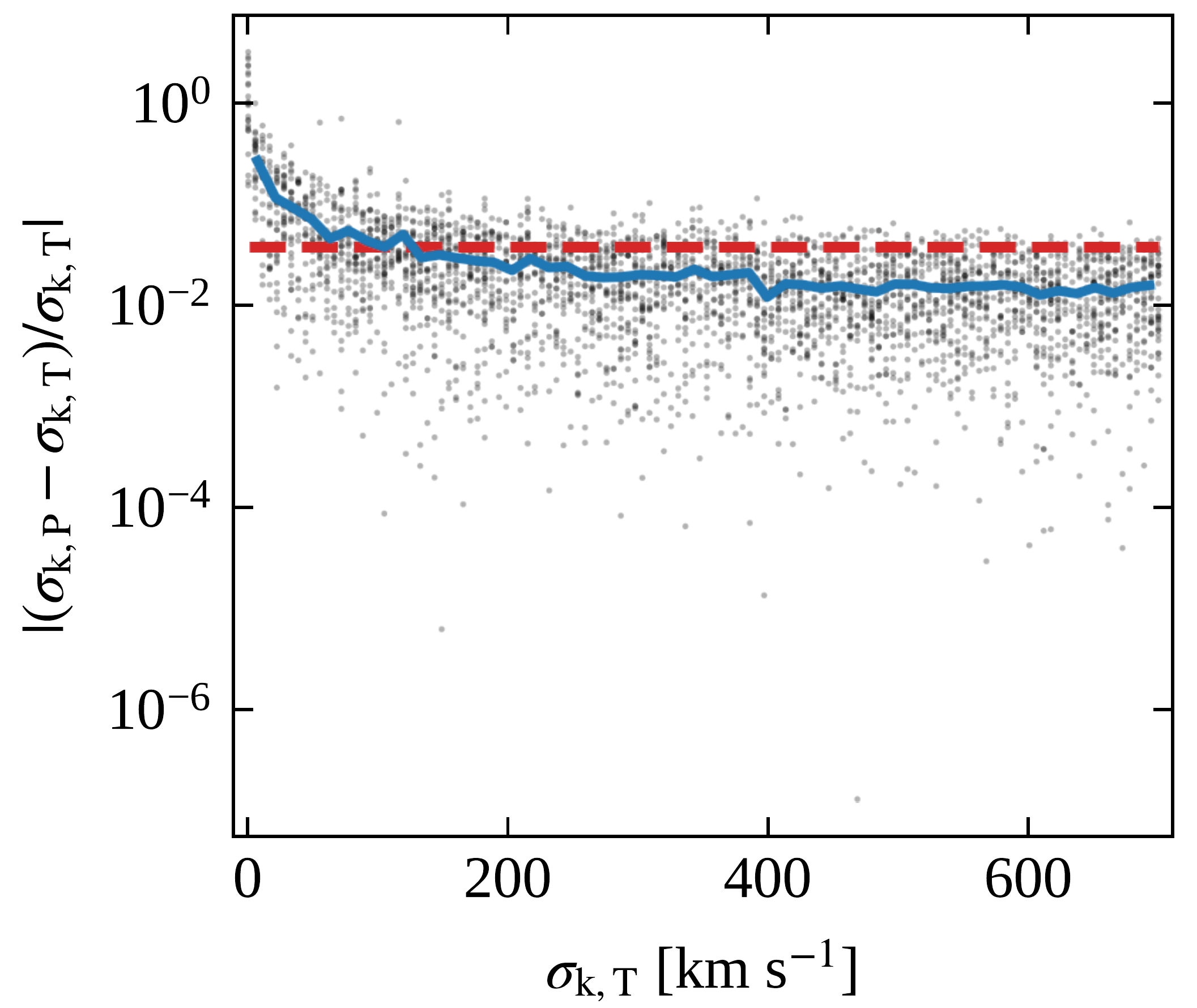}\\
\vspace{0.25cm}
\includegraphics[width = 0.32\textwidth]{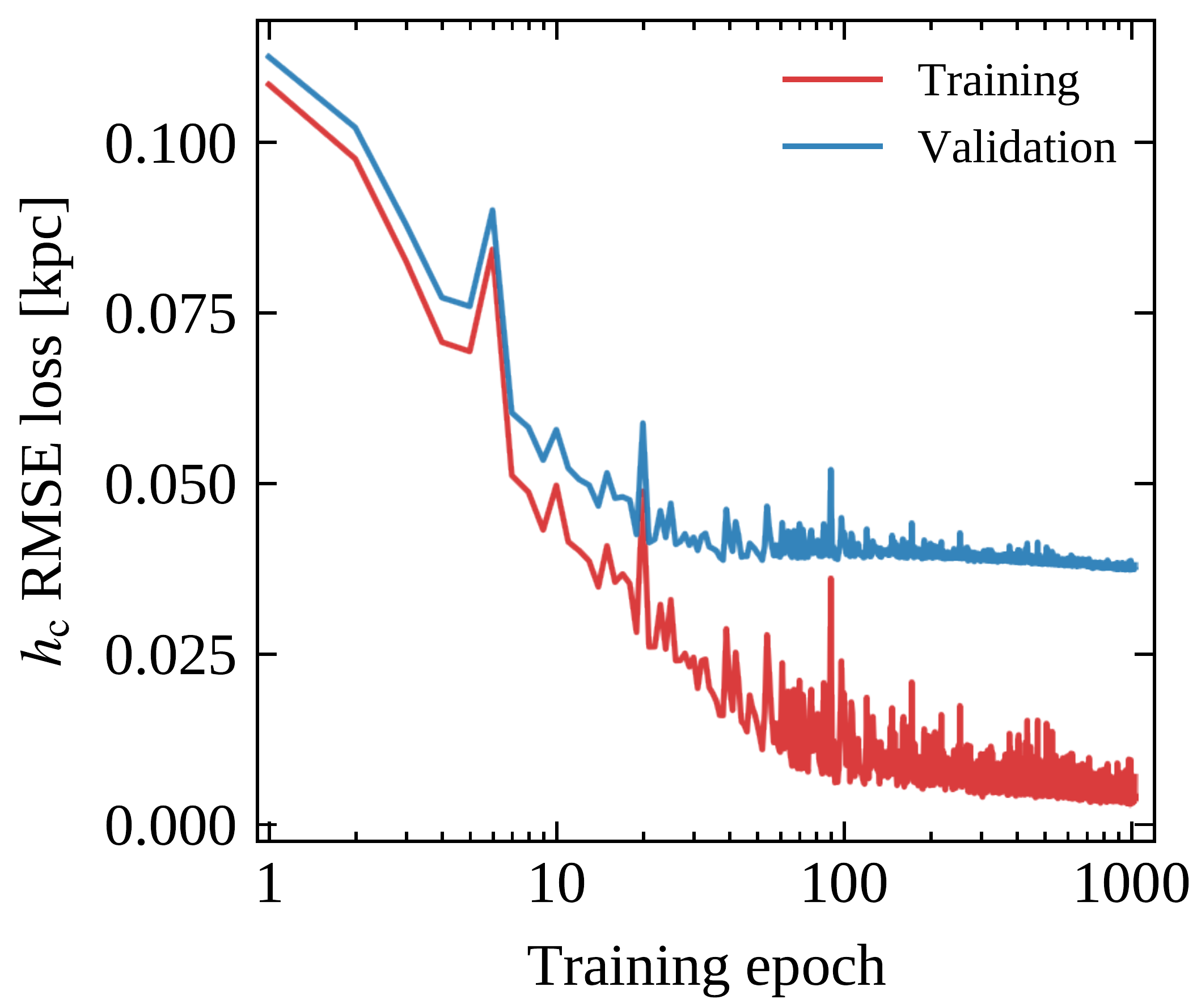}
\includegraphics[width = 0.32\textwidth]{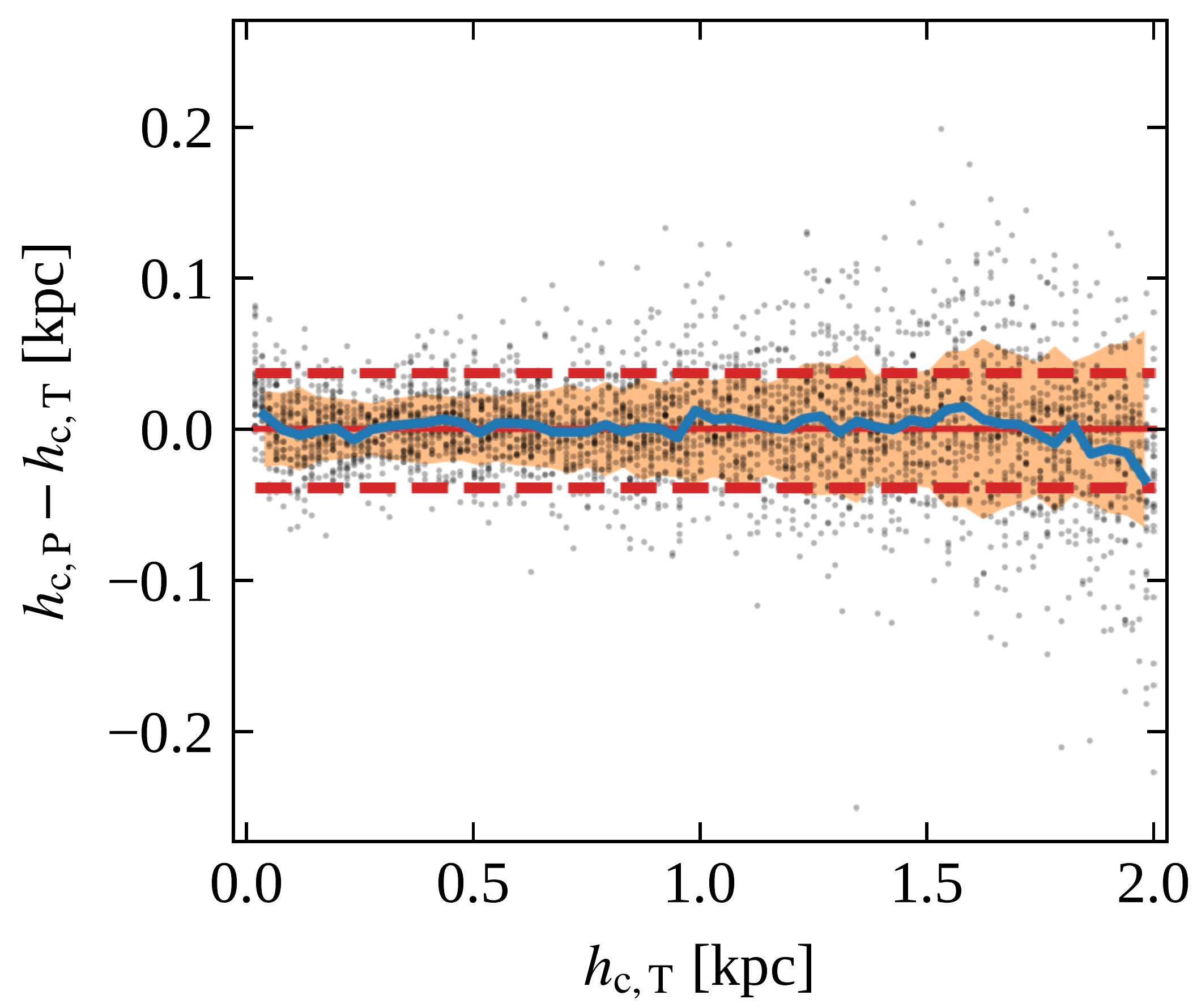}
\includegraphics[width = 0.32\textwidth]{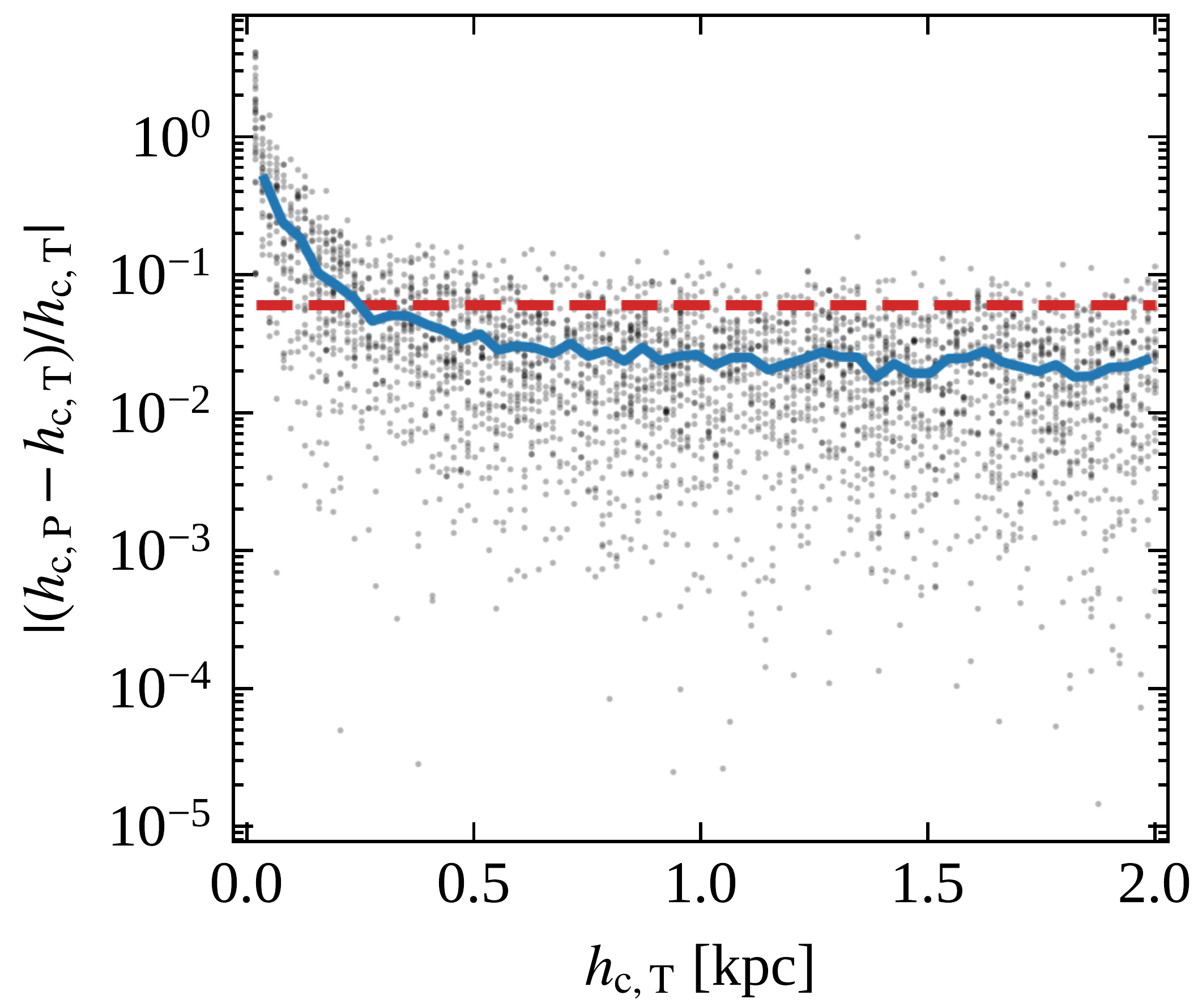}
\caption{\label{fig:2par_cnn_inference_sigmak_hc}Results of the \acs{CNN} two-parameter prediction for the kick-velocity parameter $\sigma_{\rm k}$ and scale height $h_{\rm c}$ for the corresponding validation sets. {\it Top (bottom) left panel}: evolution of the \acs{RMSE} training ({\it red}) and validation  ({\it blue}) losses as a function of the training/validation epoch for the $\sigma_{\rm k}$ ($h_{\rm c}$) parameters. {\it Top (bottom) central panel}: residuals of the prediction as a function of the target $\sigma_{\rm k}$ ($h_{\rm c}$) value; the subscripts P and T refer to the predicted and target values, respectively. The red dashed lines delimit the $68$\% uncertainty region corresponding to a \acs{RMSE} $=\unit[8.8]{km \, s^{-1}}$ ($\unit[0.038]{kpc}$) computed over the whole range $[1, 700]$ $\unit[]{km \, s^{-1}}$ ($[0.02, 2]$ $\unit[]{kpc}$). The {\it orange} region delimits the $68$\% uncertainty region computed as a running \acs{RMSE} for which we have divided the full $\sigma_{\rm k}$ ($h_{\rm c}$) range into $50$ bins of equal size. The {\it blue} line shows the trend of the average residuals which are well centered around the value 0. {\it Top (bottom) right panel}: relative error of the prediction as a function of the target $\sigma_{\rm k}$ ($h_{\rm c}$) value. The red dashed line corresponds to a \acs{MRE} $=0.039$ ($0.061$) computed over the whole range $[1, 700]$ $\unit[]{km \, s^{-1}}$ ($[0.02, 2]$ $\unit[]{kpc}$). The {\it blue} line shows the trend of the running \acs{MRE} computed over $50$ bins of equal size into which we have divided the full $\sigma_{\rm k}$ ($h_{\rm c}$) range.}
\end{figure*}  


\subsubsection{Generalization Results}

As for the single-parameter analysis, we assess the actual performance of the network when simultaneously predicting $\sigma_{\rm k}$ and $h_{\rm c}$ by training the \acs{CNN} on the biggest data-set with $16384 = 128 \times 128$ simulations (see run S3). As discussed above, we use the 3-channel ICRS representation with one density map and two proper motion maps with 128 resolution as input features. We split the entire data-set into training and validation subsets with a relative percentage of $80/20$\%, respectively, leading to 13107 samples in the training and 3277 samples in the validation data-set and use the same configuration as in \S \ref{sec:1par_generalization_result}. The evolution of the individual training and validation losses is shown in the left panels of Fig.~\ref{fig:2par_cnn_inference_sigmak_hc}. 

The results for the trained network's prediction on the validation set for both parameters are shown in Fig.~\ref{fig:2par_cnn_inference_sigmak_hc} and summarized in Table~\ref{tab:genralization_results}. The network is able to predict $\sigma_{\rm k}$ and $h_{\rm c}$ with an average \acs{RMSE} uncertainty of $\unit[8.8]{km \, s^{-1}}$ and $\unit[0.038]{kpc}$, respectively, which is approximately doubled compared to the single-parameter experiment. These \acs{RMSE} values are computed over the full target ranges and represented by the red dashed lines in the residuals plots (see central panels of Fig.~\ref{fig:2par_cnn_inference_sigmak_hc}). As in the single-parameter case, we observe that the \acs{RMSE} uncertainties increase with increasing target parameters as indicated by the orange regions in the residuals plots. The relative errors represented in the right panels of Fig.~\ref{fig:2par_cnn_inference_sigmak_hc} show the same decreasing trend with the target value as for the single-parameter predictions, albeit with larger relative errors. When computing the \acs{MRE} over the whole range of the two target parameters, we obtain $0.039$ and $0.061$ for $\sigma_{\rm k}$ and $h_{\rm c}$, respectively. These values are highlighted by the red dashed lines in the right panels of Fig.~\ref{fig:2par_cnn_inference_sigmak_hc}.

We then evaluate the generalization capability of the trained network on the test set with 1000 samples. We find RMSEs of $\unit[9.1]{km \, s^{-1}}$ and $\unit[0.041]{kpc}$ and MREs of $0.041$ and $0.057$ for $\sigma_{\rm k}$ and $h_{\rm c}$, respectively. As before, we evaluate the confidence intervals of the two estimators over 1000 bootstrapped sets of the test set and find that the RMSE and MRE variations are around $3$\% and  $8$\%, respectively, for both predicted parameters. This indicates that also in the two-parameter prediction the trained network is stable and guarantees a good level of generalization power. 

\begin{figure}
\centering
\includegraphics[height = 0.34\textwidth]{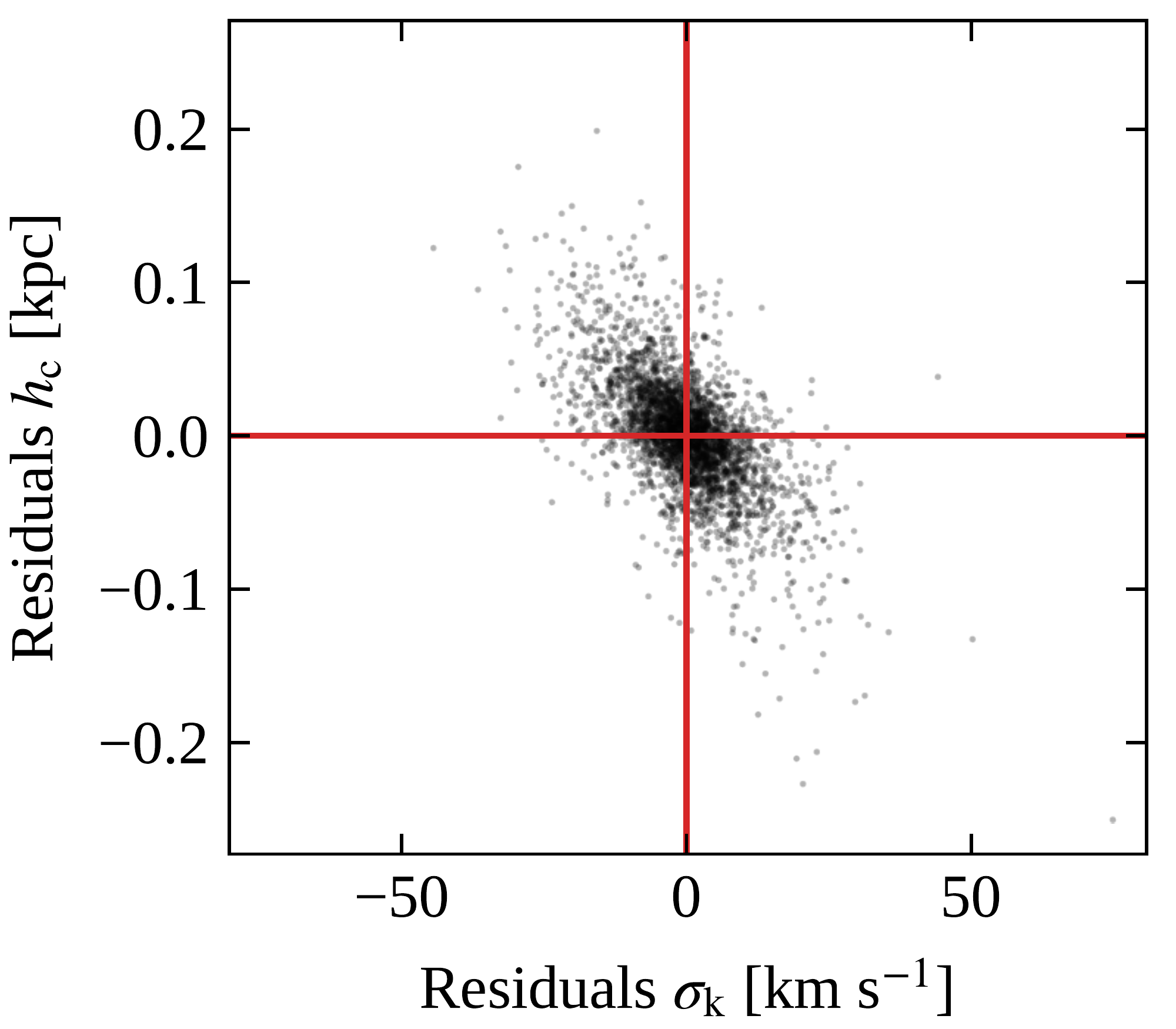}
\caption{\label{fig:2par_cnn_inference_res_corr}{Scatter plot of residuals of the predicted scale-height parameter $h_{\rm c}$ versus residuals of the predicted kick-velocity parameter $\sigma_{\rm k}$ for the validation data-set} of our two-parameter generalization experiment. An anticorrelation can be observed.}
\end{figure}

However, we find that the \acs{CNN} trained to simultaneously predict $\sigma_{\rm k}$ and $h_{\rm c}$ is not able to reach the same level of accuracy as in the experiments where a single parameter was predicted at a time. This could be due to three distinct causes: (i) either our neural network is not sophisticated enough to discern between the effects of both parameters on the simulation outcomes represented in the ICRS maps, (ii) our choice of ICRS maps as input does not provide sufficient information for the network to distinguish between both parameters, or (iii) this is a physical (real) degeneracy, and there is a limit to what we can measure. To investigate this issue we train the \acs{CNN} to predict \textit{only} the parameter $h_{\rm c}$ using the same two-parameter data-set used above where both $\sigma_{\rm k}$ and $h_{\rm c}$ are varied. After predicting on the validation data-set we obtain a \acs{RMSE} accuracy of $\unit[0.038]{kpc}$ which is equal to the result obtained above for the two-parameter prediction. This suggests that the network complexity is suitable to predict either one or two parameters simultaneously. Limitations in performance are therefore either due to an inadequate input representation or a physical degeneracy that imposes a natural accuracy threshold. While we cannot distinguish these two with our current simulation and \acs{ML} pipeline, we can illustrate the underlying problem in the following way: Fig.~\ref{fig:2par_cnn_inference_res_corr} shows the residuals of the scale-height parameter $h_{\rm c}$ versus the residuals of the kick-velocity parameter $\sigma_{\rm k}$ for the predictions over the validation set. The overall negative slope indicates that the network tends to overpredict $h_{\rm c}$ in those simulations where $\sigma_{\rm k}$ is underestimated and vice versa; i.e., large (small) $h_{\rm c}$ values have the same overall effect on the phenomenology of the pulsar population as large (small) $\sigma_{\rm k}$ values and the network struggles to distinguish these cases. This highlights the degeneracy between the two parameters already discussed above, which might be broken if the data itself were represented in a different way or additional input information about each neutron star (beyond position and velocity) were provided.

\begin{deluxetable*}{ccccc}
\tablecaption{Summary of the \acs{CNN} generalization results on the validation and test (in parenthesis) data-sets for the single-parameter and two-parameter cases.
\label{tab:genralization_results}}
\tabletypesize{\small}
\tablecolumns{5}
\tablenum{4}
\tablewidth{0pt}
\tablehead{
&
\multicolumn{2}{c}{1-par. generalization} &
\multicolumn{2}{c}{2-par. generalization} \\
\colhead{Parameter} &
\colhead{\acs{RMSE}} &
\colhead{\acs{MRE}} &
\colhead{\acs{RMSE}} &
\colhead{\acs{MRE}} }
\startdata
$\sigma_{\rm k}$ & $\unit[4.4 \,(4.8)]{km~s^{-1}}$ & 0.014 (0.018) & $\unit[8.8 \,(9.1)]{km~s^{-1}}$ & 0.039 (0.033) \\
$h_{\rm c}$ & $\unit[0.017 \,(0.019)]{kpc}$ & 0.024 (0.029) & $\unit[0.038 \,(0.041)]{kpc}$ & 0.061 (0.057) 
\enddata

\end{deluxetable*}


\section{Discussion}
\label{sec:discussion}

In this paper we have studied the potential of an artificial neural network to estimate with high accuracy the dynamical characteristics of a mock population of isolated pulsars. Implementing a simplified population-synthesis framework we focused on the pulsar natal kick-velocity distribution and the distribution of birth distances from the Galactic plane. Taking into account the Galaxy gravitational potential and evolving the pulsar motions in time, we generate a series of simulations that are used to train and validate a suitably structured convolutional neural network.

The generalized results presented in the previous sections are obtained in a very idealized and simplified scenario, implying that caution is required when the uncertainties for the prediction of the kick-velocity dispersion $\sigma_{\rm k}$ and birth scale height $h_{\rm c}$ are taken at face value and conclusions for the real pulsar population are drawn. 
\begin{figure*}
\centering
\includegraphics[width = \textwidth]{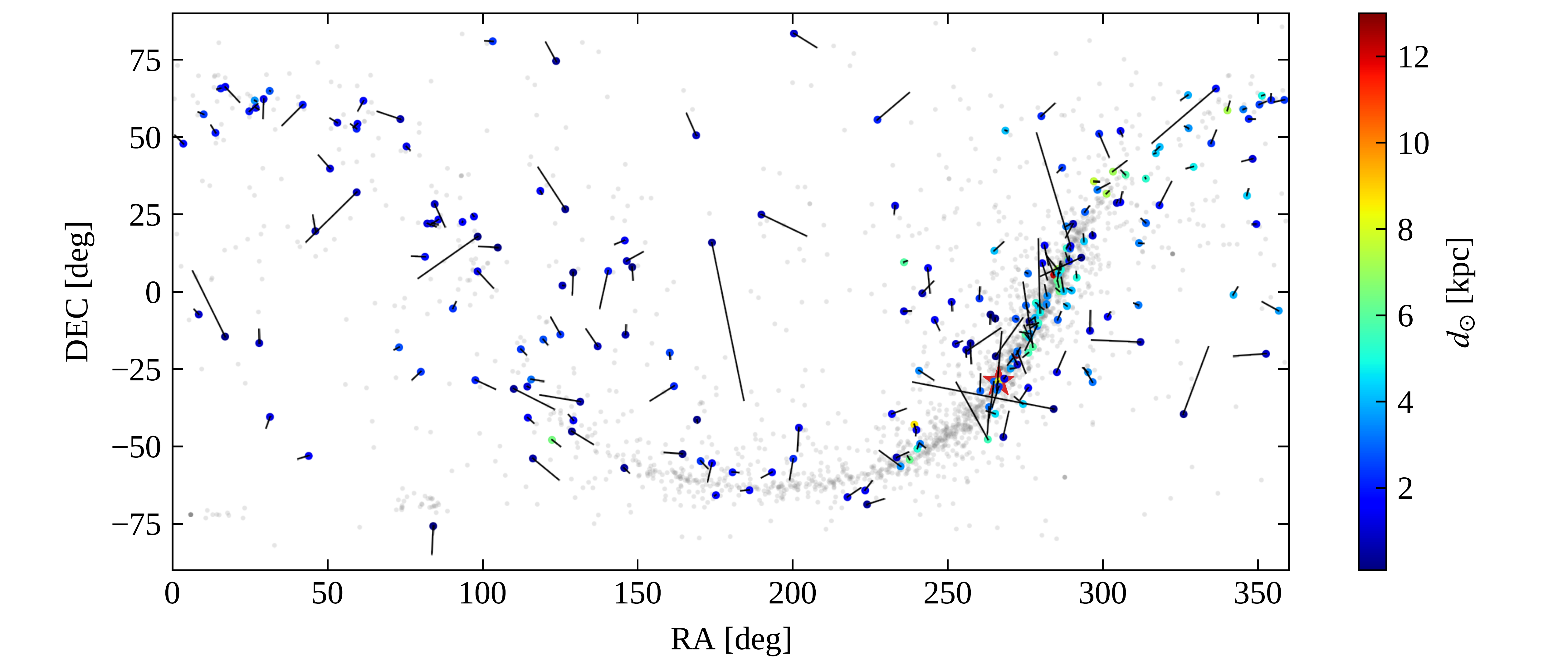}
\caption{\label{fig:icrs_proper_motion}{Proper motions of the selected 216 neutron stars in the sky represented in the ICRS reference frame. The current locations of the neutron stars are indicated by the colored circles, whereas the tracks indicate their motion for the past $\unit[0.5]{Myr}$, assuming no radial velocity and neglecting the effects of the Galactic potential. The color encodes the heliocentric distance $d_\odot$ of the neutron stars. The corresponding data is provided in Table~\ref{tab:proper_motion}. In the background, we show in gray all non-binary pulsars in the ATNF catalogue (those in the Small and Large Magellanic Clouds as well as those in globular clusters are included). The red star highlights the position of the Galactic Center.}}
\end{figure*}  
In particular, our simulations assume that the distribution of neutron star progenitors in Galactic height is represented by the exponential thin-disk model, and that the kick-velocity magnitudes follow a Maxwellian distribution. While the choice of an exponentially thin disk is commonly adopted \citep{Wainscoat1992, Polido2013, Li2019} and can be justified theoretically as the outcome of a self-gravitating isothermal disk \citep{Spitzer1942}, the choice of a Maxwellian model for the kick-velocity distribution is difficult to motivate from a theoretical standpoint. The Maxwellian model has found empirical support as it has been shown to well describe the proper motions of observed pulsars \citep{Hobbs2005}. It is for this reason, and its rather simple mathematical form, that the Maxwell kick-velocity distribution is often adopted in population synthesis studies of compact stars \citep{Sartore2010, Cieslar2020}. However, the real functional form of the kick-velocity distribution is still unknown and debated. Several authors have studied the kick-velocity problem and concluded that other models explain observed proper motions of neutron stars equally well. For example, by using maximum-likelihood methods \citet{Arzoumanian2002} found that a bimodal Gaussian distribution with a low-velocity and a high-velocity component is the preferred model to describe the observed proper motion of a sample of 79 radio pulsars. \citet{Faucher2006} studied the velocity component along the Galactic longitude for a sample of 34 pulsars observed with interferometric techniques \citep{Brisken2002, Brisken2003}. After testing a two-component Gaussian model as well as a variety of single-component models, they opted for a single-component description with an exponential shape, although a two-component model could not be ruled out due to the poor statistics of their sample. More recently, \citet{Verbunt2017} and \citet{Igoshev2020} analyzed a sample of isolated young pulsars and found that a two-component Maxwellian model explained the observed sample best. 

In general, the presence of a low-velocity and a high-velocity component could indicate different progenitor properties as well as birth scenarios for the pulsar population. Numerical simulations of supernova explosions have for example suggested that neutron stars with lower kick velocities could be generated in the core-collapse supernovae of progenitors with small iron cores or in electron-capture supernovae \citep{Podsiadlowski2004, Mandel2020}. While also possible scenarios for isolated systems \citep{Janka2017}, such conditions might generally affect those neutron stars born in binaries \citep{Giacobbo2020}, where mass-loss episodes could strip their progenitors off their hydrogen envelopes. This might favor the formation of small iron cores or accretion-induced electron-capture supernovae, resulting in weaker natal kicks \citep{Schwab2010, Tauris2013b}. Only for the strongest kicks can the binaries be disrupted by the supernova and both companions expelled; otherwise the two stars remain gravitationally bound. Such effects are neglected in our model but could in principle generate an imprint on the observed neutron star population. The ability of \acs{ML} algorithms to recognize features and patterns in the processed data could help to better constrain the bi-modality of the kick-velocity distribution. In future work, we will investigate these aspects in more detail and explore ways to implement a ML framework to reconstruct the shape of the kick-velocity distribution underlying an evolved population of pulsars. 

Up to this point, we have not considered any kind of selection effects or observational biases and effectively assumed that all the neutron stars in our simulation are detectable. While this provides direct insight into how various initial conditions affect the evolved population of neutron stars, a direct comparison with observational data in principle requires a careful treatment of biases. For example, due to beaming effects not all Galactic radio pulsars are visible from Earth \citep{Tauris1998, Melrose2017}, while survey sensitivity thresholds and instrumental limitations might hamper the detection of faint or far
away sources \citep{Manchester2001, Johnston2008, Stovall2014, Coenen2014, Good2021}. Additionally, timing noise can significantly limit the sensitivity and precision in the detection of pulsar proper motions via timing analysis techniques \citep{Hobbs2004, Lentati2016, Parthasarathy2019}. 
With the aim of obtaining a rough idea on how selection effects and biases could potentially influence a future comparison with observational we perform the following experiment. We first collect those neutron stars that have measured proper motions. As the main resource we use the ATNF pulsar catalogue\footnote{\url{https://www.atnf.csiro.au/research/pulsar/psrcat/}} \citep{Manchester2005}, but in some cases we provide proper motion results from more recent analyses (see Appendix~\ref{app:proper_motion} for details). We find a total of 417 neutron stars whose angular positions, proper motions in ICRS coordinates, spin periods, spin-period derivatives, \acs{DM} values and distance estimates are reported in Table~\ref{tab:proper_motion}. Out of these objects, we remove those stars that belong to the Magellanic Clouds, are associated with globular clusters or have a binary companion. We further select only those neutron stars with a spin-period derivative $\dot{P} > 10^{-17}$ to exclude those that have potentially been recycled. Finally, we consider only those for which an estimate of the heliocentric distance $d_{\odot}$ is available; in the case of radio pulsars we quote values that are derived from their respective \acs{DM}s using the free-electron density model of \citet{Yao2017} (YMW16 model hereafter). As some neutron stars have a $DM$ that exceeds the maximum Galactic $DM$ allowed by the YMW16 model, these cases are assigned a default distance of $\unit[25]{kpc}$. We exclude those cases unless an alternative distance measurement is available. Applying these filters we obtain a sample of 216 Galactic, likely isolated neutron stars, whose positions and proper motions are illustrated in Fig.~\ref{fig:icrs_proper_motion}. 

\begin{figure*}
\centering
\includegraphics[height = 0.34\textwidth]{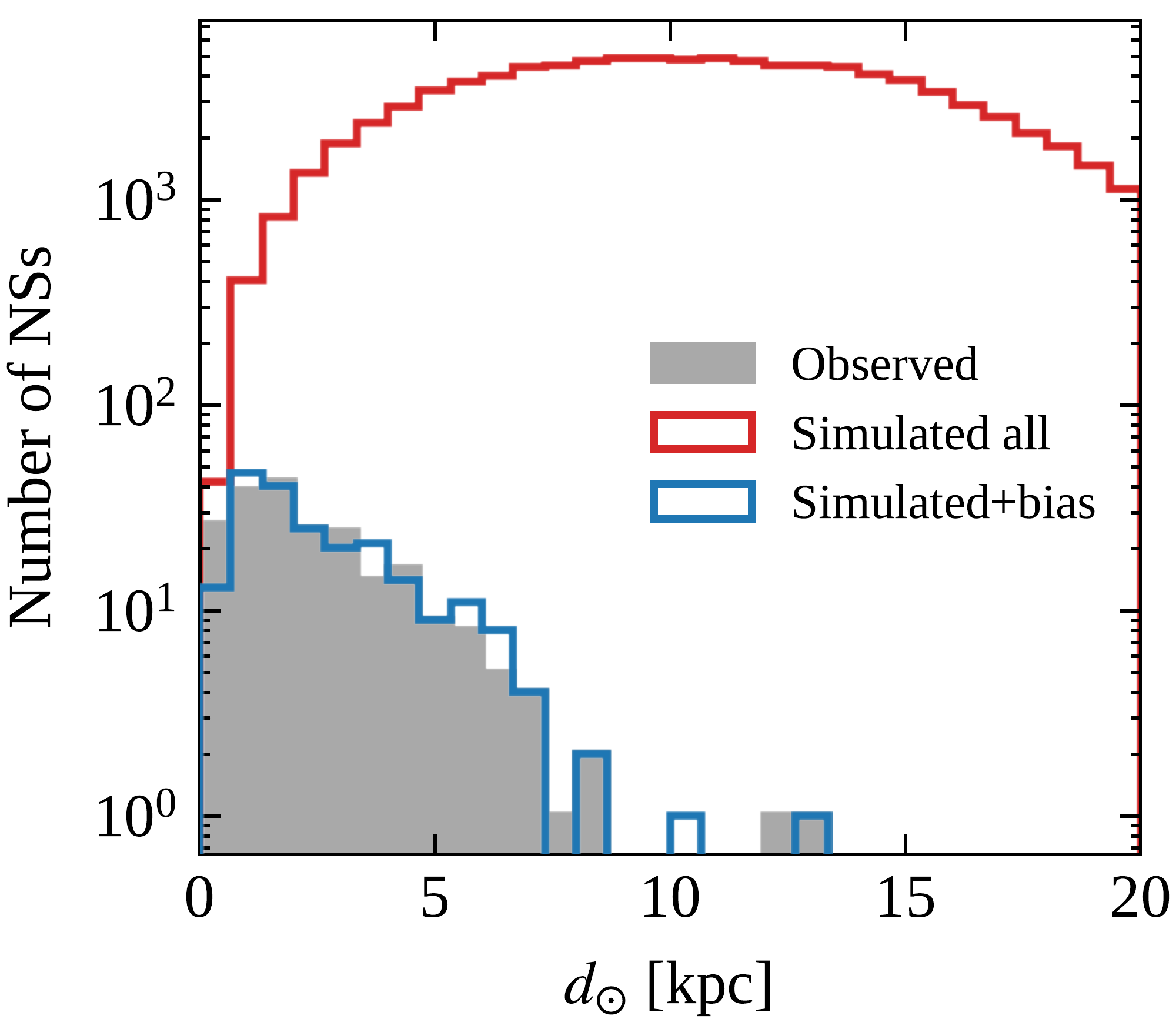}
\hspace{0.5cm}
\includegraphics[height = 0.34\textwidth]{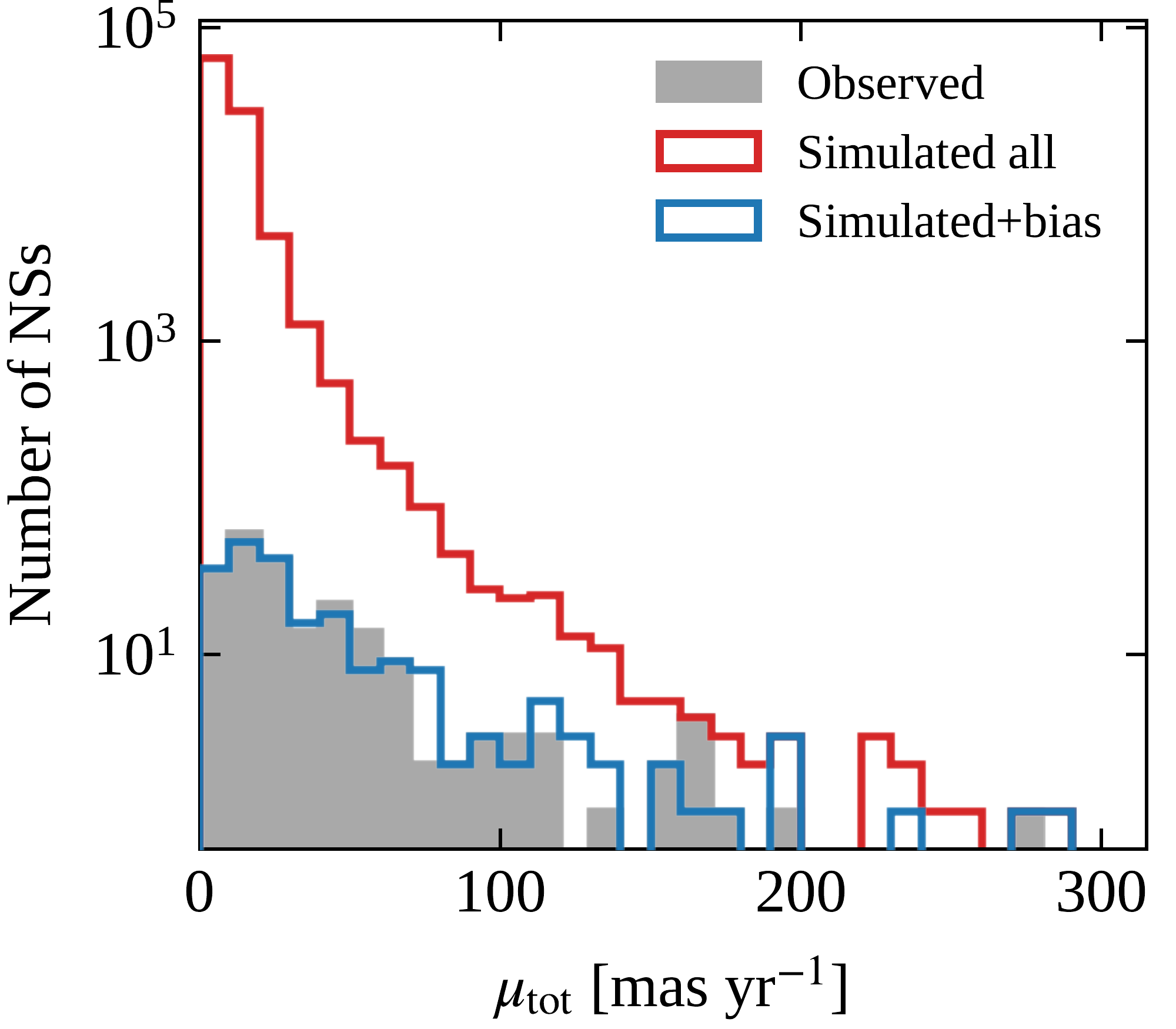}
\caption{\label{fig:dist_pm_histo}{Distribution of heliocentric distances $d_{\odot}$ (\textit{left panel}) and proper-motion magnitudes $\mu_{\rm tot}$ (\textit{right panel}) for the 216 neutron stars with measured proper motion ({\it gray histograms}). For comparison, we also show the distances and proper-motion magnitudes for 216 simulated neutron stars ({\it blue histograms}) randomly sampled from a mock population generated with the fiducial parameters $\sigma_{\rm k} = \unit[265]{km \, s^{-1}}$ and $h_{\rm c} = \unit[0.18]{kpc}$ ({\it red histograms}). For the weighted sampling, we use the weight function $f(d_{\odot}) = d_{\odot}^{-1} \exp (-0.5 d_{\odot})$.}}
\end{figure*}  

In the left panel of Fig.~\ref{fig:dist_pm_histo}, the gray histogram shows the distribution of their heliocentric distances. Even if subject to some uncertainties due to imprecisions in the YMW16 model, the distance distribution peaks around $\unit[1]{kpc}$, followed by a sharp exponential decrease. For a realistic pulsar distribution and in the absence of selection effects, we would expect the number of neutron stars to increase with distance due to an increase in the explored volume, until reaching a maximum at a distance of about $\unit[10]{kpc}$, which comprises the region around the Galactic Center (see the red histogram in the left panel of Fig.~\ref{fig:dist_pm_histo}). Thus, the shape of the gray distribution in Fig.~\ref{fig:dist_pm_histo}, as expected, indicates that our observed sample of neutron stars with measured proper motions is incomplete in distance and subject to selection biases. By looking at the right panel of Fig.~\ref{fig:dist_pm_histo}, we also note that a selection bias on distance is also reflected in the distribution of the total proper motion magnitudes (gray histogram), computed as $\mu_{\rm tot} \equiv \sqrt{\mu_{\rm RA}^2 + \mu_{\rm DEC}^2}$. Indeed, since the nearest stars are also characterized by larger angular proper motions, there is a tendency to detect high proper-motion stars with higher probability.

In this first empirical approach, instead of attempting to identify these underlying biases (which will be the subject of a subsequent paper in preparation), we follow a more agnostic approach to introduce a comparable selection effect in our simulated populations. Specifically, we use a weighted random-sampling routine to select $n$ pulsars from our mock populations, where each simulated star is assigned a weight according to a function $f(d_{\odot})$ of its heliocentric distance. This weight function has to assign larger weights to closer neutron stars in order to ensure their higher detection probabilities and has to be chosen such that we recover the observed distance distribution with sufficient accuracy. To find $f(d_{\odot})$ we simulate a mock population with the fiducial values of the kick velocity and scale height, that is  $\sigma_{\rm k} = \unit[265]{km \, s^{-1}}$ and $h_{\rm c} = \unit[0.18]{kpc}$, respectively. After using a given $f(d_{\odot})$ to weight the simulated neutron stars we sample 216 mock stars and compare their distance and proper-motion distributions (shown as blue histograms in Fig.~\ref{fig:dist_pm_histo}) with those of the observed sample by performing two-sample Kolmogorov-Smirnov (KS) tests. After testing various functional forms we find that $f(d_{\odot}) = d_{\odot}^{-1} \exp (-0.5 d_{\odot})$ is able to reproduce the observed distributions with a good level of accuracy. More precisely, for this choice of $f(d_{\odot})$, the KS tests performed over $1000$ distinct comparisons give average p-values of $\sim 0.3$ and $\sim 0.6$ for the distance and proper-motion comparison, respectively. This means that at $95$\% confidence level, we cannot reject the null hypothesis that the simulated and observed samples are drawn from the same underline distribution.
We have also verified that for this weight function, the KS tests always provide p-values $>0.05$ when comparing the observed sample with the simulated populations for reasonable values of $\sigma_{\rm k}$ and $h_{\rm c}$. Only in the cases where $\sigma_{\rm k}$ and $h_{\rm c}$ assume extreme values (near the edges of their respective ranges) the p-values might drop below $0.05$. However, these cases are associated with simulations with extreme initial conditions that are unlikely to reproduce the observations. For our basic experiment, we further make the simplified assumptions that $f(d_{\odot})$ emulates all selection biases and that it is the same for every number $n$ of sampled neutron stars. We stress that for the purpose of this initial analysis we do not aim to accurately constrain the selection function that affects the observed population of neutron stars. Instead we study how the introduction of realistic selection biases will alter the predictive power of our machine learning framework. Although one might intuitively attribute the exponential factor in $f(d_{\odot})$ to scattering in the interstellar medium at large distances, the underlying nature and the precise form of the true selection function is certainly more complicated. We expect it to encompass a series of effects due to the physics of the interstellar medium and the pulsar emission itself, as well as selection effects of pulsar surveys and pulsar searches. We reserve a more accurate study disentangling the different effects that contribute to $f(d_{\odot})$ to future work.  

We then analyze how the predictive power of the \acs{CNN} evolves as a function of the number $n$ of neutron stars, sampled with the above weight function $f(d_{\odot})$. To do so, we vary $n$ from 200 to 2000 in steps of 200, and in each case re-sample the 16384 simulated populations from run S3, where both $\sigma_{\rm k}$ and $h_{\rm c}$ are varied. After applying a $80/20$\% training-validation split, we retrain the \acs{CNN} on each of the down-sampled simulations. As before, we use the 3-channel ICRS input maps (i.e., density plus proper motion information) but instead opt for a resolution of $32\times16$ bins to accommodate the smaller number of stars represented in our maps. We have verified that a higher resolution of $128 \times 64$ bins does not affect the training results significantly, but slows down the training process; we therefore choose the lower resolution. We use the same training hyperparameters as in \S \ref{sec:1par_generalization_result}, that is an initial learning rate of $10^{-4}$, a batch size of 64 and an early stop at 128 epochs. Once trained for each $n$ value, we apply the \acs{CNN} to the validation sets and compute the \acf{RMSE} and \acf{MRE} for the $\sigma_{\rm k}$ and $h_{\rm c}$ predictions as a function of $n$.

\begin{figure*}
\centering
\includegraphics[height = 0.34\textwidth]{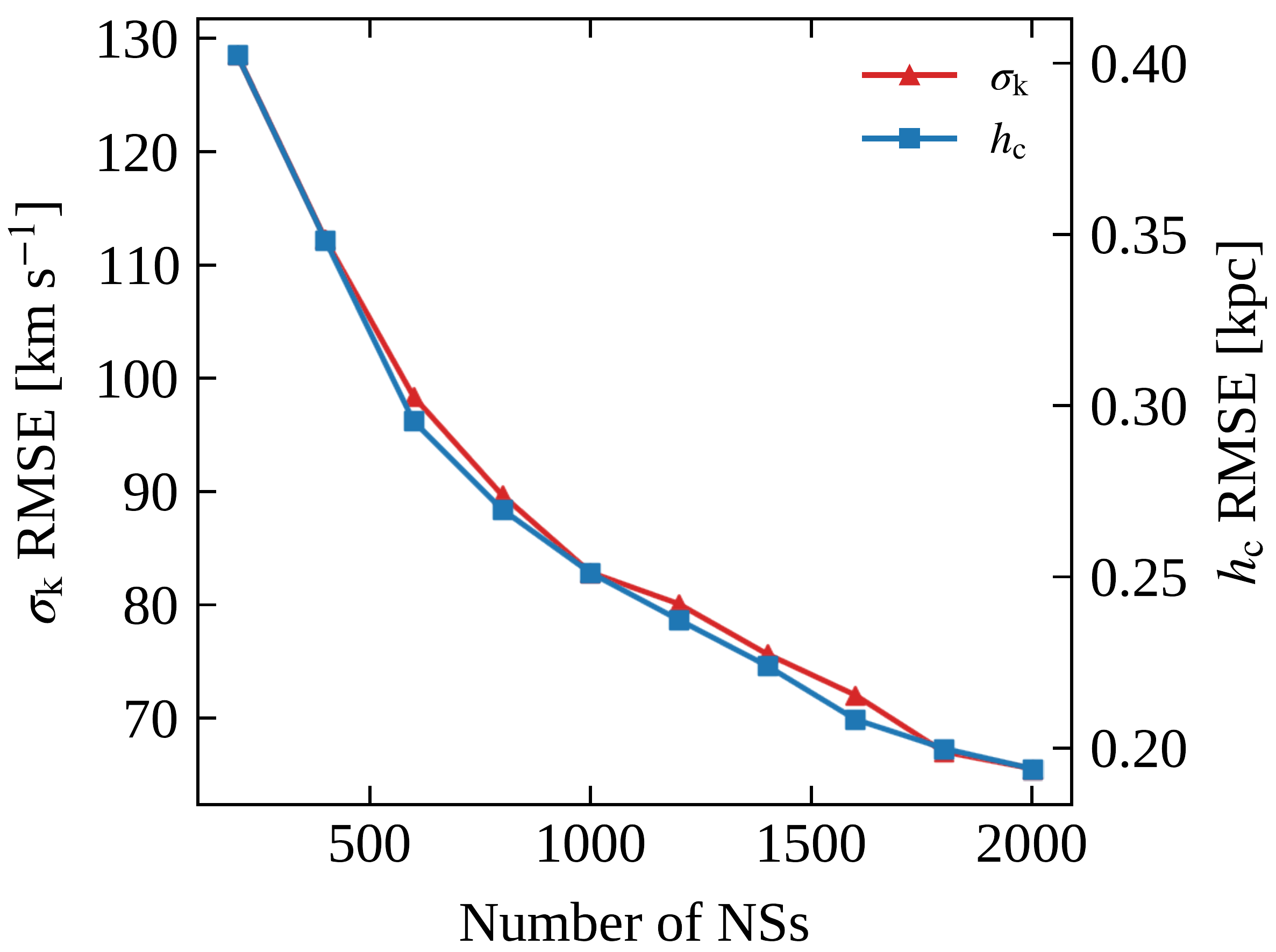}
\hspace{0.5cm}
\includegraphics[height = 0.34\textwidth]{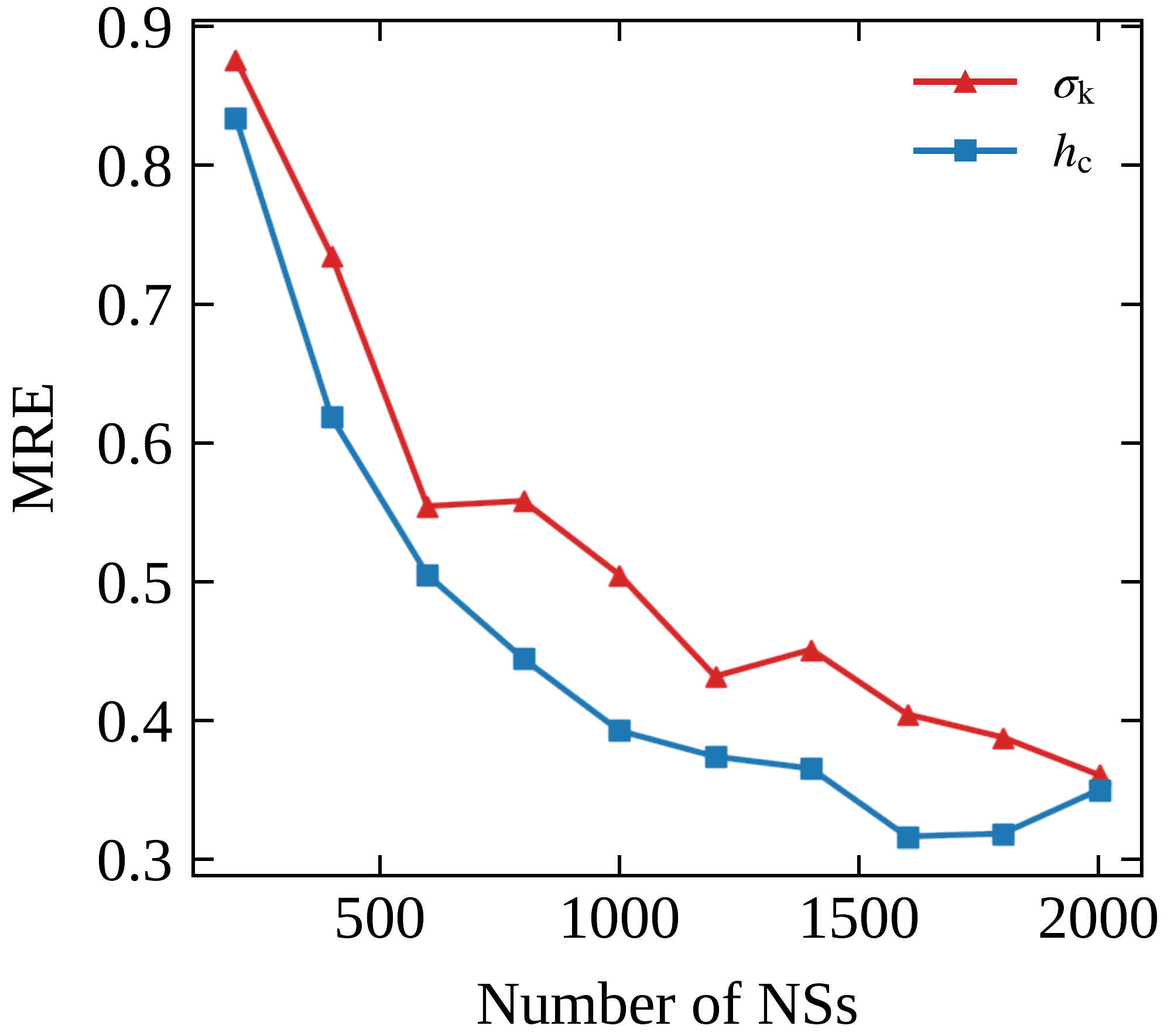}
\caption{\label{fig:rmse_mre_nsnumber}{Trend of the \acs{RMSE} ({\it left panel}) and \acs{MRE} ({\it right panel}) uncertainties on the prediction of the two parameters $\sigma_{\rm k}$ ({\it blue}) and $h_{\rm c}$ ({\it red}) as a function of the number of neutron stars $n$ in the resampled simulations in the validation sets. To train and validate the \acs{CNN} we use the 16384 simulated populations from simulation run S3 (with a $80/20$\% training/validation split), where both $\sigma_{\rm k}$ and $h_{\rm c}$ are varied, and which are resampled with increasing number of stars $n$ according to the weight function $f(d_{\odot}) = d_{\odot}^{-1} \exp (-0.5 d_{\odot})$.}}
\end{figure*}

In the left panel of Fig.~\ref{fig:rmse_mre_nsnumber}, we show how the \acs{RMSE} uncertainties for the predictions of the two parameters $\sigma_{\rm k}$ ({\it blue}) and $h_{\rm c}$ ({\it red}) diminish with increasing number of neutron stars $n$ sampled from the simulations. We observe that both curves (with the appropriate rescaling) follow very similar trends. On the right, we show instead how the \acs{MRE}s evolve with $n$, indicating how the precision of the two-parameter prediction improves with the number of detected neutron stars. This plot shows that, under the assumptions that selection effects are unaltered and the underlying kick-velocity and birth-height distributions have a Maxwellian and exponential shape, respectively, our trained \acs{CNN} is able to predict $\sigma_{\rm k}$ and $h_{\rm c}$ with a relative error of $\sim 0.35$ for a sample of 2000 stars. 

These results highlight that the number of neutron stars plays a crucial role for the level of accuracy that the \acs{CNN} can reach. As expected, a larger number of stars provides more information, which allows the \acs{CNN} to pinpoint differences between populations evolved from different initial conditions, also when selection effects are introduced. Future observational efforts aimed at detecting and characterizing new pulsars will thus play an important role in constraining the birth properties of neutron stars. Specifically, the advent of the \acf{SKA} will represent an important step forward into this direction. Due to its large sensitivity (a factor of 10 better than other radio telescopes) and its long baseline (up to $\unit[3000]{km}$), \acs{SKA} has the potential to increase the number of discovered pulsars by a factor $10$ \citep{Smits2009, Smits2011}. This will allow more precise timing and astrometric measurements of pulsar positions as well as distances and proper motions. A larger and more precise data-set could also help to better constrain the shape of the kick-velocity distribution and differentiate between models that try to explain the origin of the natal kicks \citep{Tauris2015}.


\section{Summary}
\label{sec:summary}

In this work we have analyzed the possibility of using machine-learning (ML) techniques to reconstruct the dynamical birth properties of an evolved population of isolated neutron stars. For this purpose, we developed a simplistic population-synthesis pipeline to simulate the dynamical evolution of Galactic neutron stars and subsequently used these simulations to train and validate two different neural networks. We specifically focused on their ability to predict two parameters that strongly impact on the phenomenology of the evolved population: the dispersion $\sigma_{\rm k}$ of a Maxwell kick-velocity distribution and the scale height $h_{\rm c}$ of an exponential distribution for the Galactic birth heights. This was achieved by providing the networks with two-dimensional stellar density and velocity maps in galactocentric and equatorial (ICRS) coordinate frames. We found that a \acf{CNN} is able to estimate the physical parameters with high accuracy when multiple input channels, i.e., position and velocity information, are provided. In particular, when simultaneously predicting $\sigma_{\rm k}$ and $h_{\rm c}$ from ICRS maps, the network is able to reach absolute uncertainties lower than $\unit[10]{km \, s^{-1}}$ and $\unit[0.05]{kpc}$, respectively, corresponding to a relative error of around $10^{-2}$ for both parameters. Albeit obtained under simplified assumptions, our feasibility study, the main focus of this paper, has thus demonstrated that \acs{ML} techniques are indeed suitable to infer information about the pulsar population. Our phenomenological analysis incorporating proper-motion measurements (an attempt at including observational biases in an agnostic way) has further highlighted that increasing the sample of known pulsars and accurately measuring their current characteristics with future telescopes is crucial to tightly constrain the birth properties of the neutron stars in the Milky Way. In particular, our trained \acs{CNN} is able to predict $\sigma_{\rm k}$ and $h_{\rm c}$ with a relative error of $\sim 0.35$ for a sample of 2000 pulsars with measured proper motions.

We also demonstrated that one of the main factors in limiting the accuracy of the \acs{CNN}'s predictions in our set-up is the degeneracy between $\sigma_{\rm k}$ and $h_{\rm c}$; as they both affect the evolved populations in a similar way, the network struggles to disentangle their effects. This limitation is a direct consequence of our simplified population-synthesis framework. In future works, we will go beyond modeling the dynamical evolution and focus on incorporating additional physics such as magneto-thermal and spin-period evolution. We will further model their emission in different electromagnetic bands and study corresponding detectability limits by addressing selection effects as well as observational survey biases. Such additional input information could potentially break the degeneracy between the kick-velocity and the Galactic height distributions and provide more accurate model constraints on $\sigma_{\rm k}$ and $h_{\rm c}$ as well as other input parameters. 

The ultimate goal will be to use multi-wavelength observations of the Galactic neutron star population and take advantage of \acs{ML}, combined with population synthesis, to recover their birth properties, such as the natal kick-velocity, spin-period or magnetic-field distribution. The potential to learn complex patterns from multi-dimensional data and the power of generalizing to unseen data-sets, makes \acs{ML} algorithms a powerful tool to tackle such multi-parameter optimization studies.


\section*{Acknowledgements}

We would like to thank Alice Borghese and Francesco Coti Zelati for their help with the collection of the observed pulsar proper motions, Daniele Vigan\`o for providing comments on the manuscript, and Alessandro Patruno for helpful conversations related to this paper. MR, VG, AG and NR acknowledge support from the H2020 ERC Consolidator Grant ``MAGNESIA'' under grant agreement Nr. 817661 (PI: Rea), and grants SGR2017-1383 and PGC2018-095512-B-I00. JAP acknowledges support from the Generalitat Valenciana grant PROMETEO/2019/071, AEI grant PGC2018-095984-B-I00 and the Alexander von Humboldt Foundation via a Humboldt Research Award. This work has also been partially supported by the PHAROS COST Action (CA16214). 
The data production, processing and analysis tools for this paper have been developed, implemented and operated in collaboration with the Port d'Informació Científica (PIC) data center. PIC is maintained through a consortium of the Institut de Física d'Altes Energies (IFAE) and the Centro de Investigaciones Energéticas, Medioambientales y Tecnológicas (Ciemat). We particularly thank Christian Neissner, Ricard Cruz, Carles Acosta, Gonzalo Merino, and Pau Tallada for their support at PIC. 
Finally, we acknowledge the use of the following software: \texttt{Astropy} \citep{astropy13, astropy18}, \texttt{Hydra} \citep{Yadan2019}, \texttt{IPython} \citep{PerezGranger07}, \texttt{Matplotlib} \citep{Hunter07}, \texttt{Numba} \citep{Lam2015}, \texttt{NumPy} \citep{Oliphant06, vanderWalt-etal11, Harris-etal20}, \texttt{Pandas} \citep{McKinney10}, \texttt{PyTorch} \citep{Paszke2019}, and \texttt{SciPy} \citep{Jones-etal01, Virtanen2020}. 


\appendix

\section{Hardware and Software}

Our test machine features an Intel(R) Xeon(R) Gold 6230R CPU at 2.10GHz with a single NVIDIA GTX 2080 Ti GPU, 16 GiB RAM, and SSD drives. The system is running CentOS Linux release 7.8.2003 (Core) with PyTorch 1.2.0, CUDA toolkit 10.0.130 and GPU driver 455.32.00.


\begin{figure*}
\centering
\includegraphics[width = 0.245\textwidth]{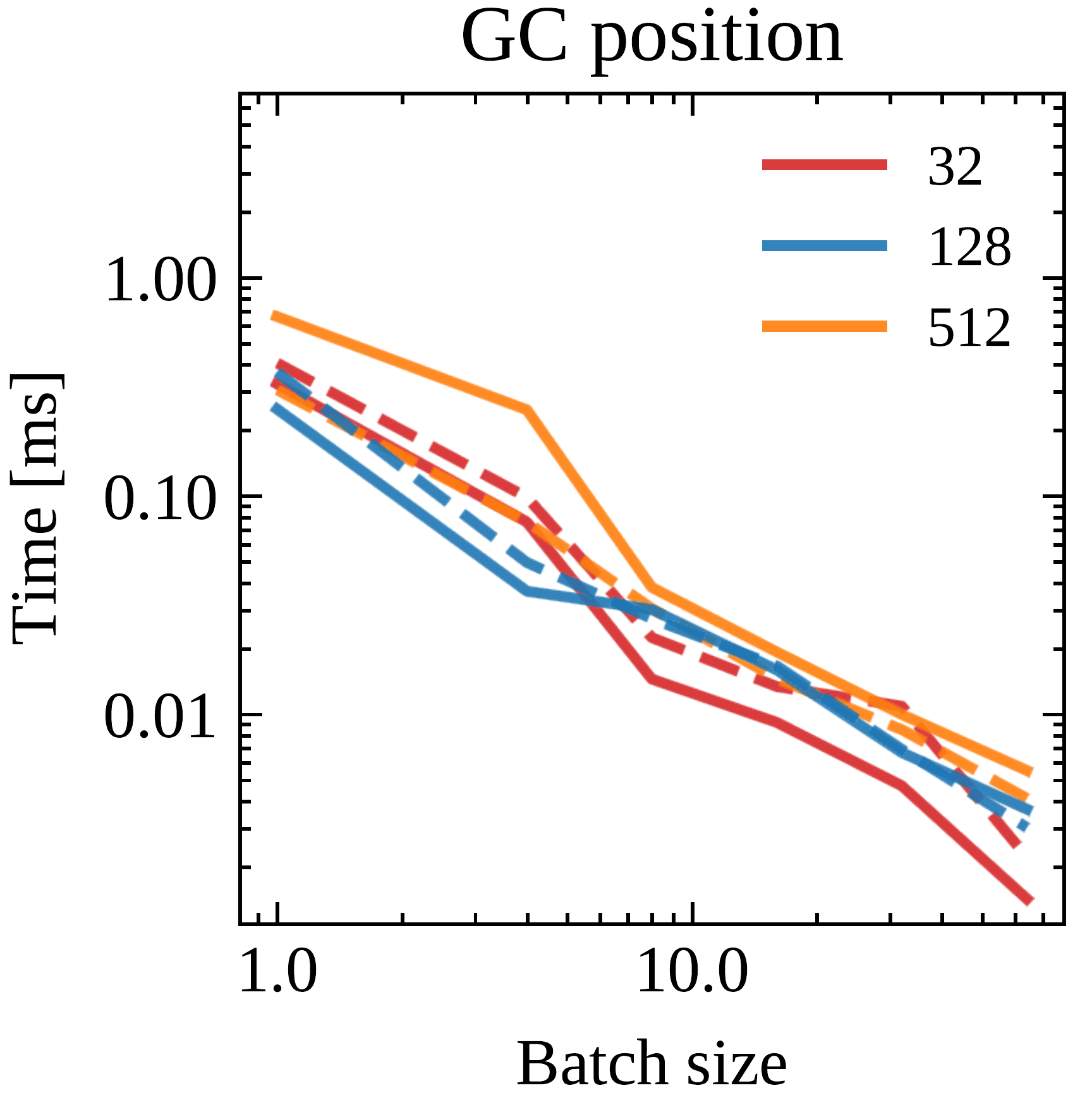}
\includegraphics[width = 0.245\textwidth]{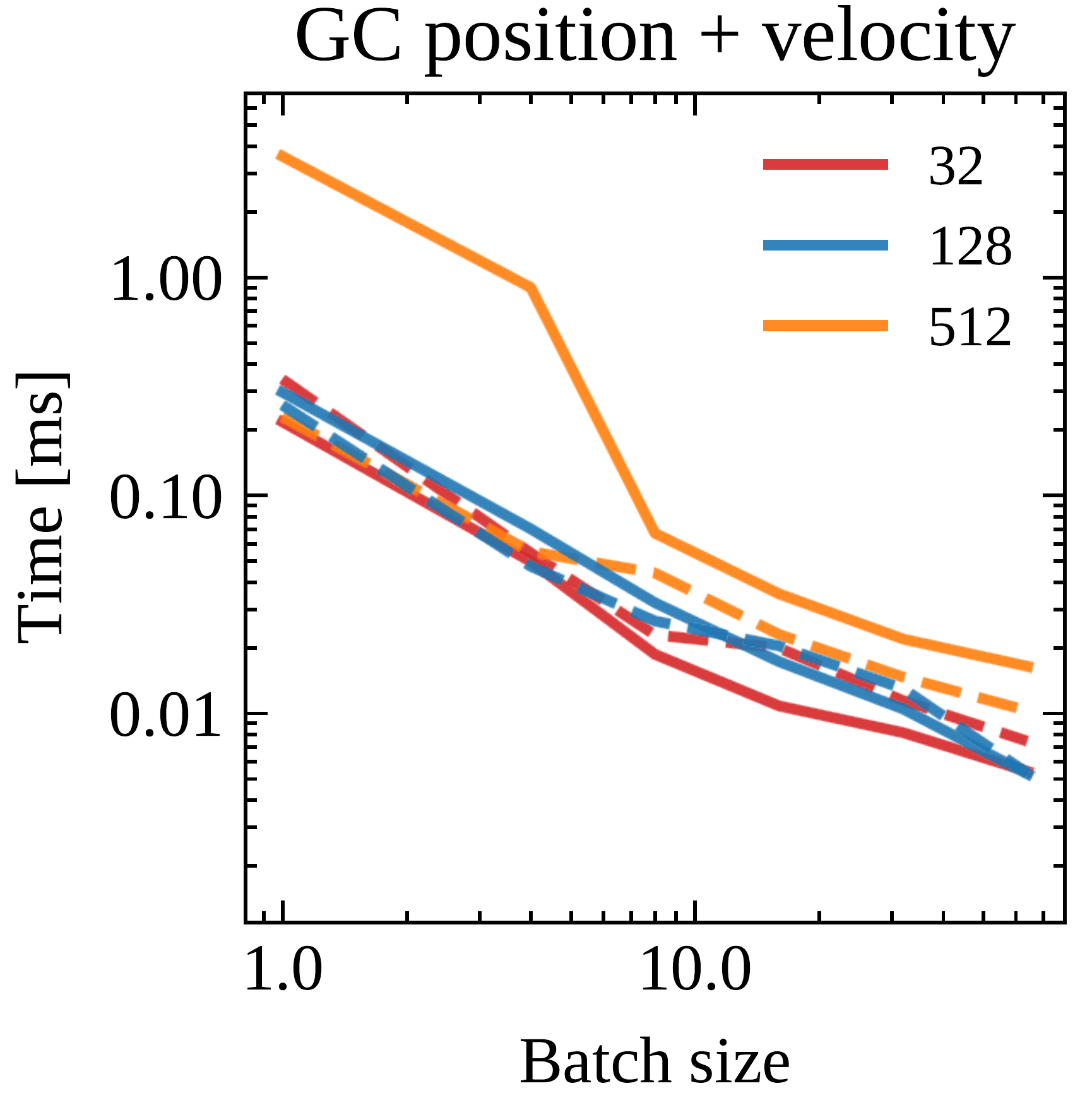}
\includegraphics[width = 0.245\textwidth]{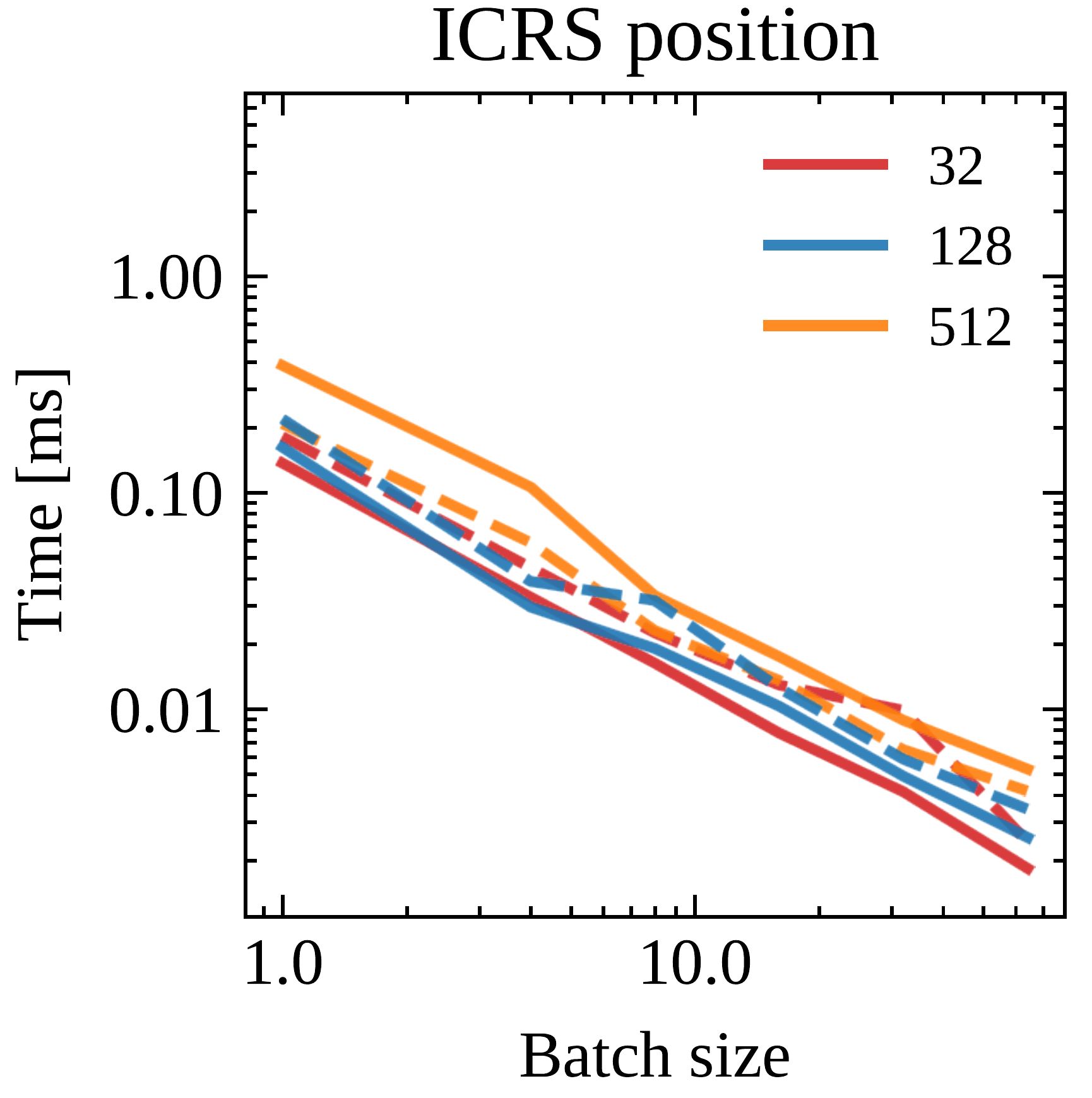}
\includegraphics[width = 0.245\textwidth]{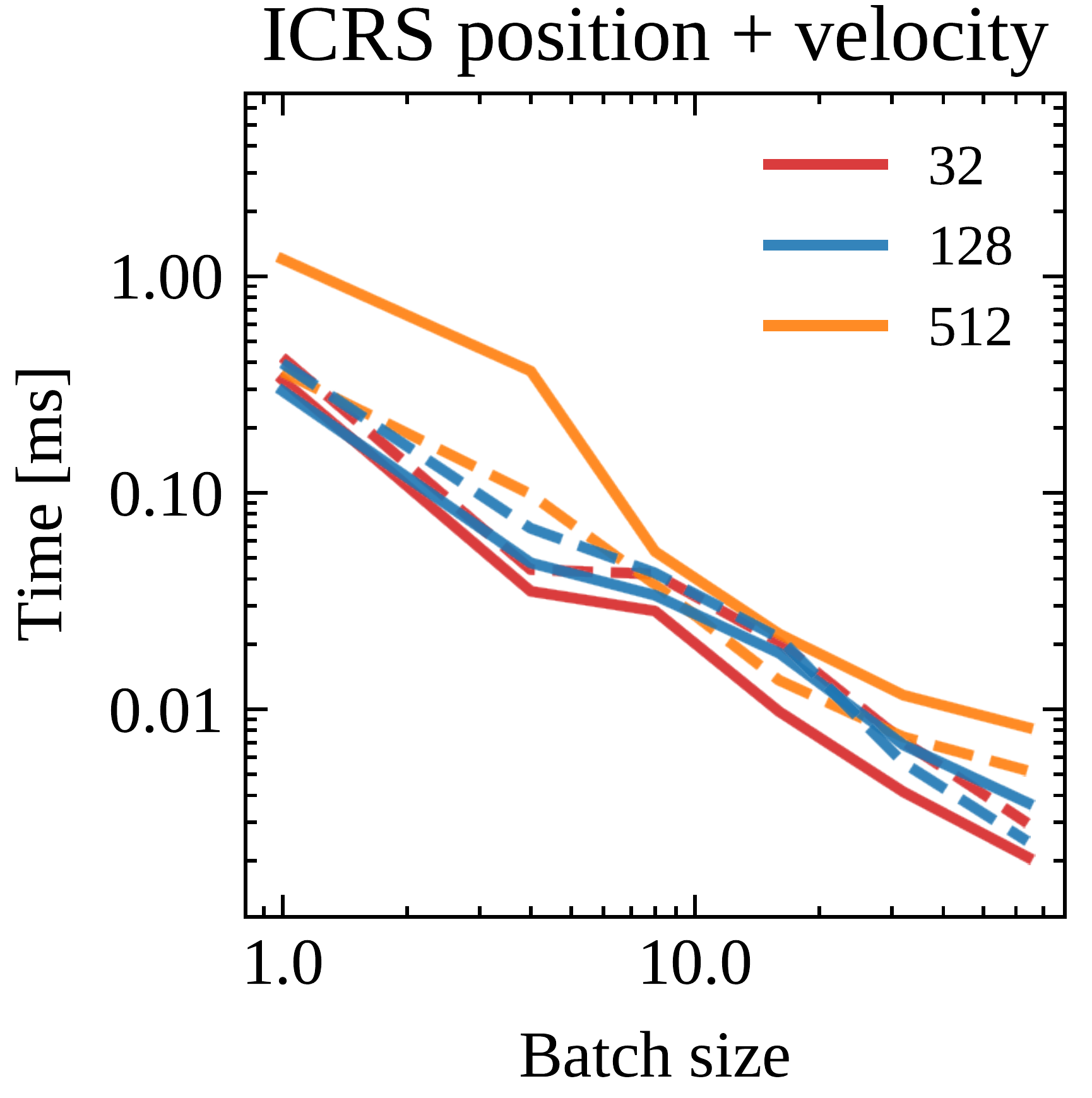}
\caption{{\acs{MLP} forward ({\it solid lines}) and backward ({\it dashed lines}) pass times per sample in ms for the training process on the single parameter $\sigma_{\rm k}$ of the Maxwell kick-velocity distribution, as a function of the batch size and the resolution ({\it red}, {\it blue}, and {\it orange} curves for 32, 128 and 512 respectively) using the four different input configurations T1 (GC position), T2 (GC position + velocity), T3 (ICRS position) and T4 (ICRS position + velocity).}}
\label{fig:lnn_time_experiment}
\end{figure*}  
\begin{figure*}
\centering
\includegraphics[width = 0.245\textwidth]{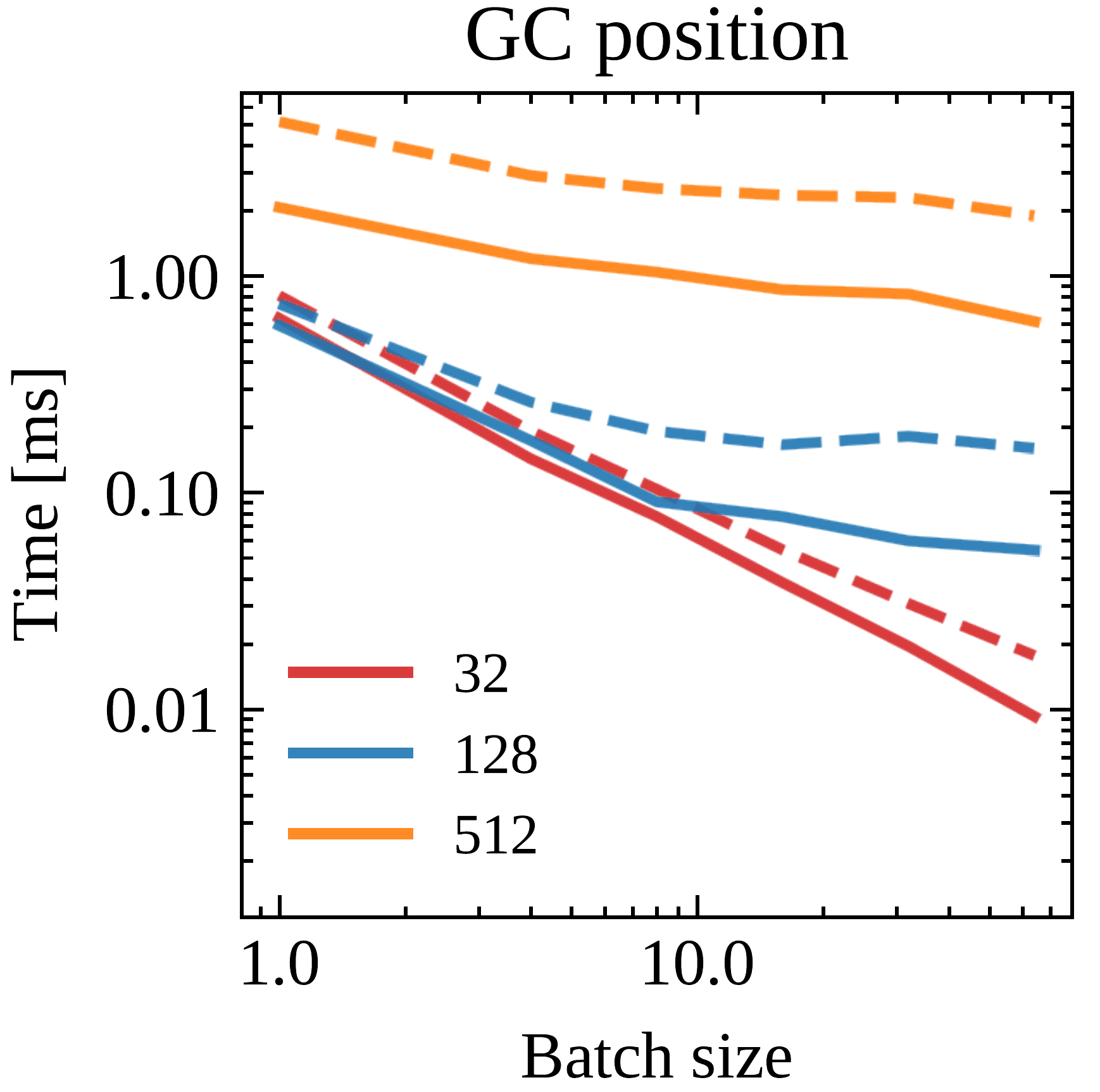}
\includegraphics[width = 0.245\textwidth]{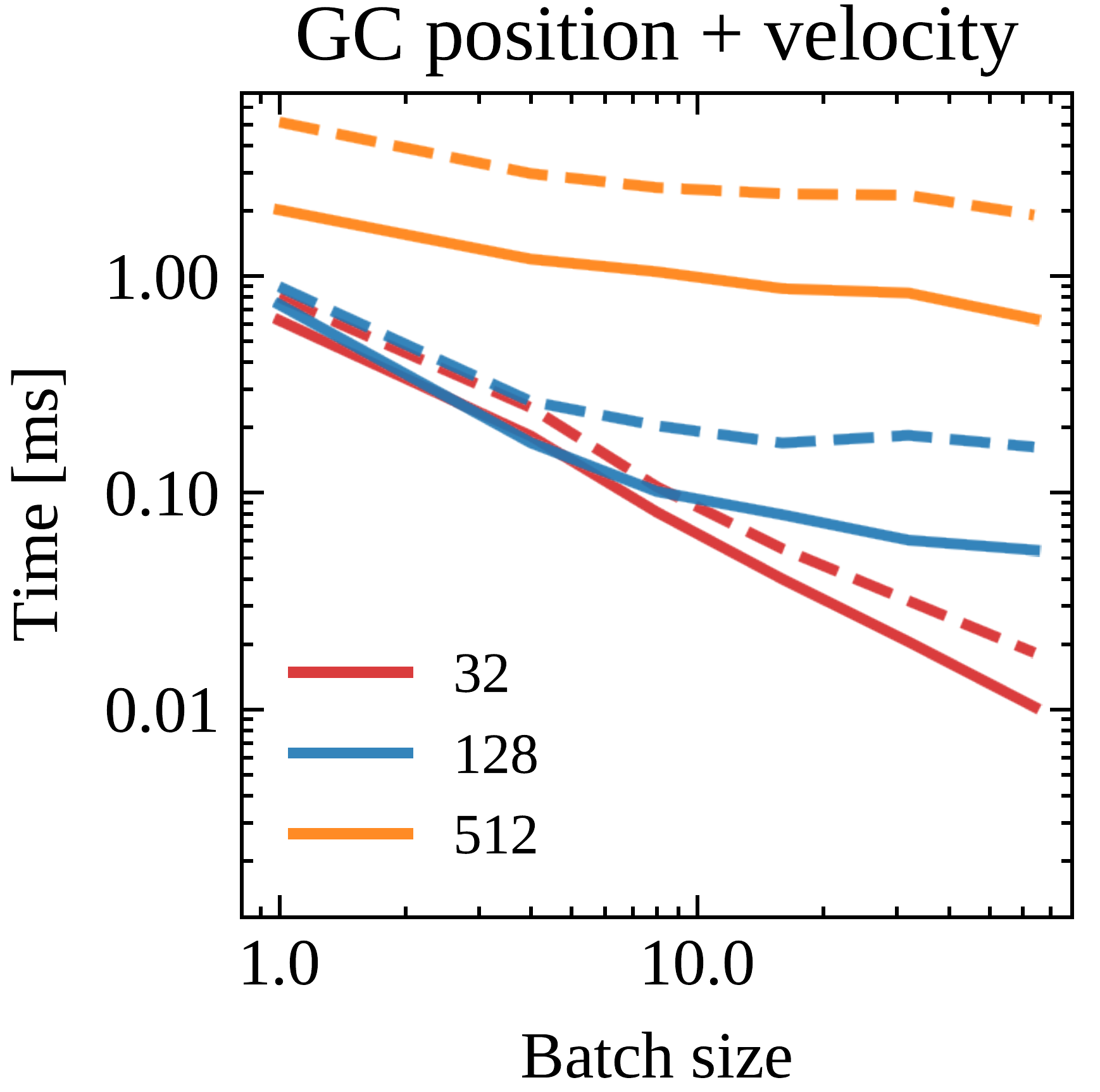}
\includegraphics[width = 0.245\textwidth]{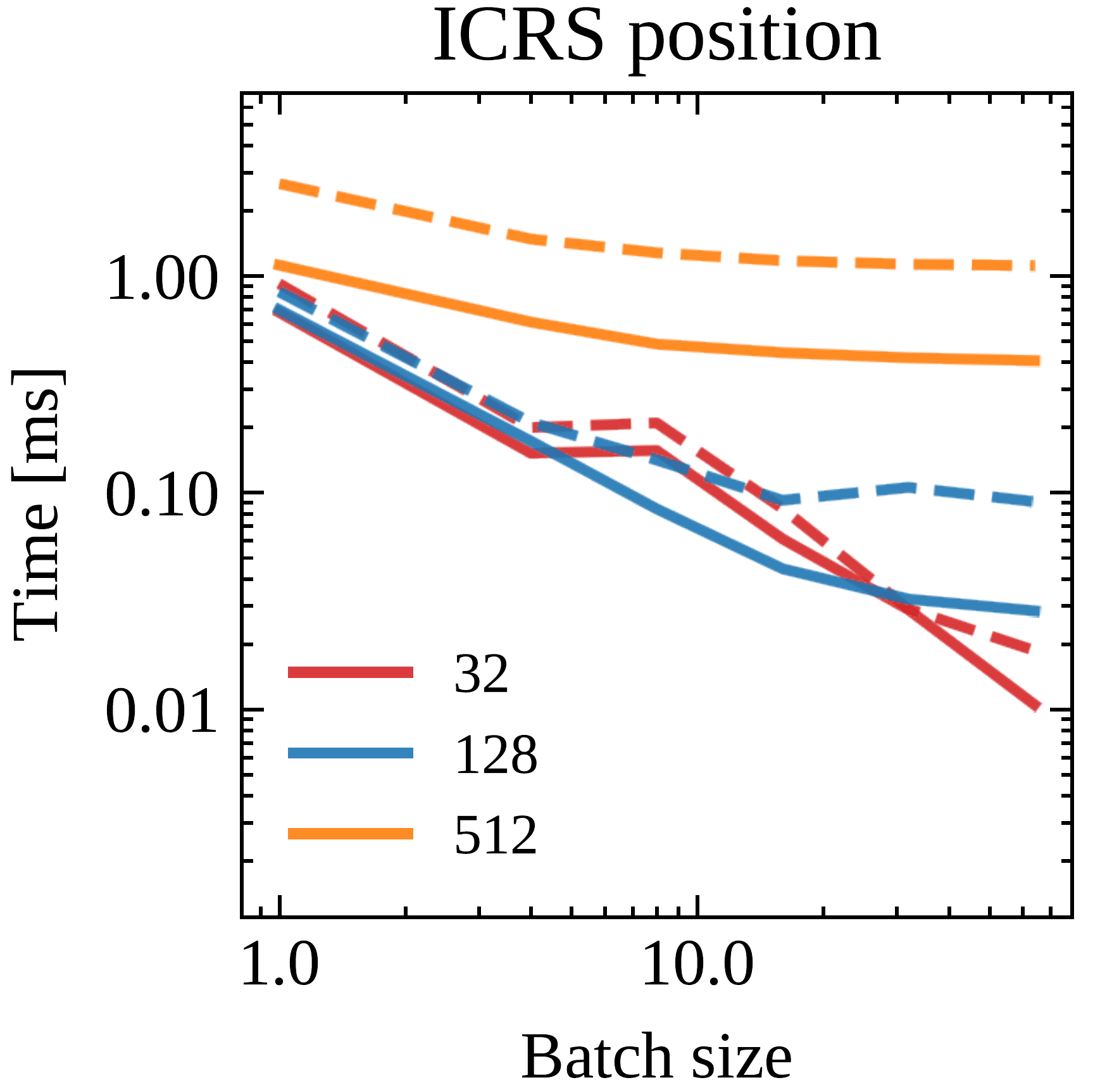}
\includegraphics[width = 0.245\textwidth]{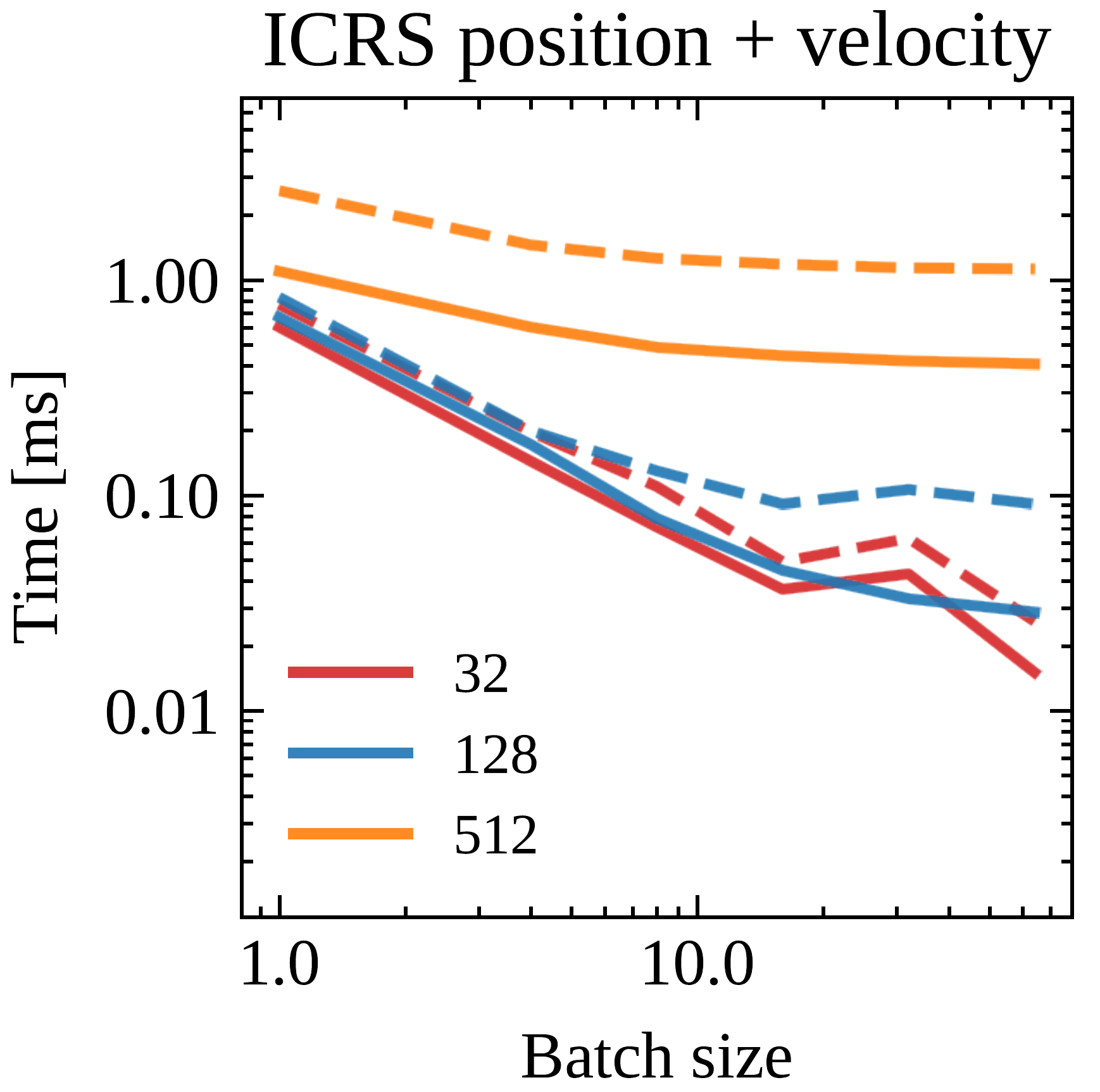}
\caption{{\acs{CNN} forward ({\it solid lines}) and backward ({\it dashed lines}) pass times per sample in ms for the training process on the single parameter $\sigma_{\rm k}$ of the Maxwell kick-velocity distribution, as a function of the batch size and the resolution ({\it red}, {\it blue}, and {\it orange} curves for 32, 128 and 512 respectively) using the four different input configurations T1 (GC position), T2 (GC position + velocity), T3 (ICRS position) and T4 (ICRS position + velocity).}}
\label{fig:cnn_time_experiment}
\end{figure*}  

\section{Timing Tests}
\label{app:timing_tests}

\subsection{Timing for Single-parameter Predictions}

We report here the timing benchmarks for the \acs{MLP} and \acs{CNN} during the single-parameter training experiments discussed in \S \ref{sec:1par_tests}. We run our experiments on the test machine and individually record the forward pass time (the time needed to go through the samples in a batch and compute a prediction) and the backward pass time (the time to compute all the gradients and perform a single optimization step) as a function of the batch size and resolution. Our benchmarks for the training data-sets from simulation run S1 using the four training configurations T1, T2, T3 and T4 are shown in Figs.~\ref{fig:lnn_time_experiment} and \ref{fig:cnn_time_experiment}, respectively.

The timing benchmark shows that the \acs{MLP} is slightly faster in performing an optimization step. This is expected due to fewer trainable parameters when compared to the more complex \acs{CNN}. We can also see that the forward and backward pass times per sample decrease with increasing batch size for both the \acs{MLP} and \acs{CNN}. For a larger batch size, several input samples are transferred from the CPU to the GPU in one step, reducing the overall number of calls between the two. Thus, on average, the processing time for an individual sample reduces when the batch size is increased. Moreover, a higher resolution generally implies an increase in computational cost, albeit being more pronounced in the case of the \acs{CNN} than the \acs{MLP}. The number of input channels itself has very little effect on our timings. Finally, we note that ICRS maps are slightly faster to process (in particular for the higher resolutions), due to the fact that their size is smaller compared to the galactocentric maps (they have half the height in bins).

\subsection{Timing for Two-parameter Predictions}

\begin{figure}
\centering
\includegraphics[width = 0.245\textwidth]{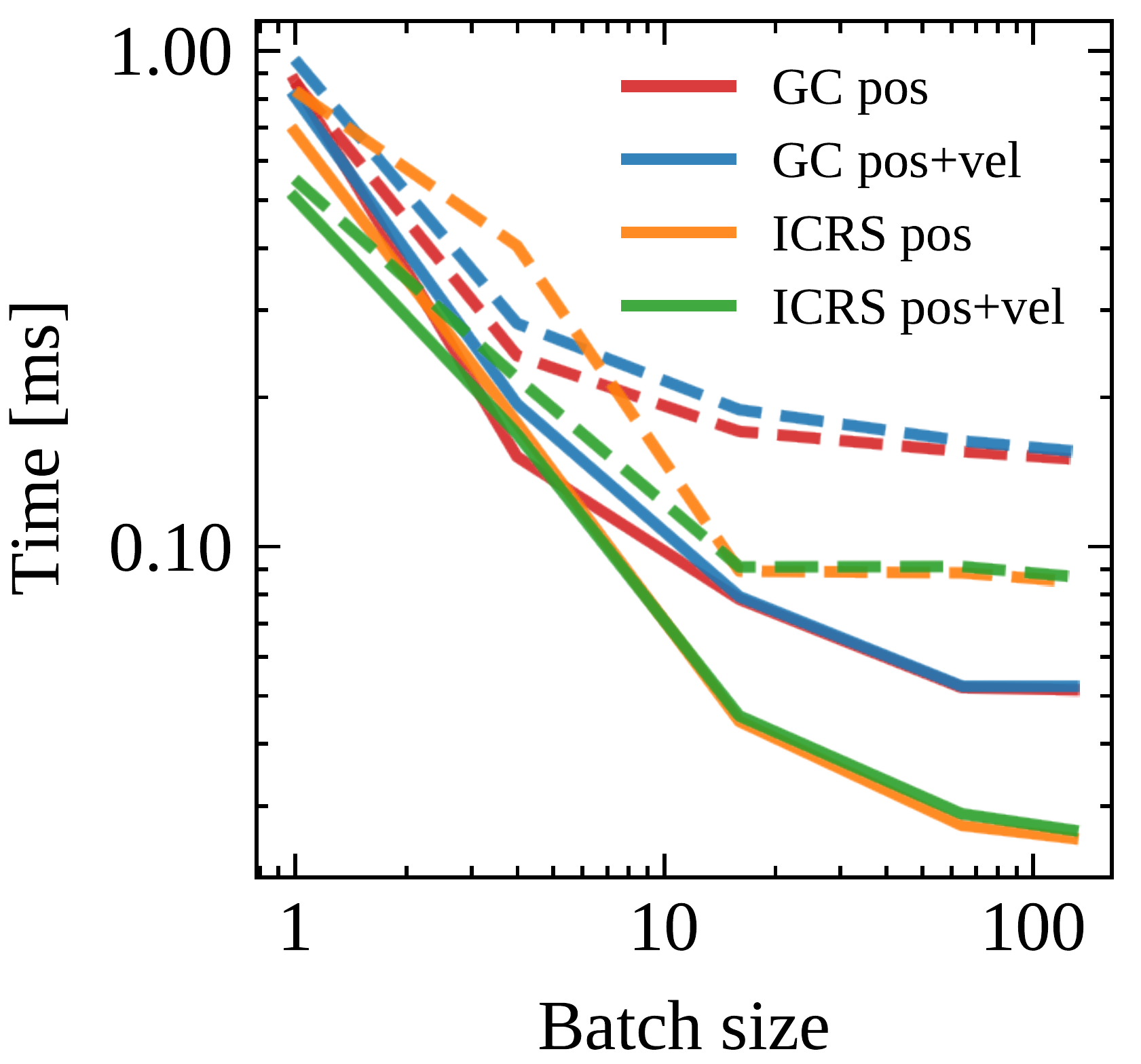}
\caption{\label{fig:2par_cnn_time}{\acs{CNN} forward pass ({\it solid line}) and backward pass time ({\it dashed line}) per sample for the two-parameter experiments as a function of the batch size for the four different input configurations T1 (GC position), T2 (GC position + velocity), T3 (ICRS position) and T4 (ICRS position + velocity).}}
\end{figure}

Following the results for the single-parameter experiments, we restrict our two-parameter predictions to the \acs{CNN} model only and fix the resolution of the input maps to 128. The results of our timing benchmarks using the training data-sets from simulation run S3 with the four configurations T1, T2, T3 and T4 are shown in Fig.~\ref{fig:2par_cnn_time}. We again report the timings for the forward and backward passes per sample as a function of the batch size and the type of input channels provided. As for the single-parameter case, we conclude that using ICRS maps ensures the lowest forward and backward pass times.


\section{Neutron stars with measured proper motion}
\label{app:proper_motion}

In Table~\ref{tab:proper_motion}, we report the properties of 417 neutron stars with measured proper motions in RA and DEC. Data for these neutron stars are primarily collected from the ATNF catalogue\footnote{\url{https://www.atnf.csiro.au/research/pulsar/psrcat/}} \citep{Manchester2005}. In some cases, updated estimates are available and those values quoted and the corresponding references specified. Note that in those cases, where multiple proper-motion estimates are available, we choose the ones with the lowest absolute error. The columns report in order: (i) the object name based on J2000 coordinates; (ii) the right ascension (RA) in hour angle and (iii) declination (DEC) in degrees with the last digit uncertainty given in parentheses; the proper motion in (iv) RA and (v) DEC in milliarcseconds per year with corresponding uncertainties; (vi) the parallax measured in milliarcseconds with uncertainty where available; (vii) the position epoch in modified Julian days; (viii) the spin period in seconds; (ix) the spin-period derivative in seconds over seconds; (x) the dispersion measure in $\unit[]{[pc \, cm^{-3}}]$ with the last digit uncertainty given in parentheses; (xi) the heliocentric distance derived from the \acs{DM} using the YMW16 free-electron density model (for some objects the \acs{DM} exceeds the maximum Galactic \acs{DM} allowed by the YMW16 model, which assigns a default value of $\unit[25]{kpc}$; when available, we quote other distance estimates; * indicates a distance derived from other techniques especially for X-rays and gamma-ray sources, which have no measured \acs{DM}); (xii) the classification of the object, i.e., radio pulsar (PSR), binary pulsar (binary PSR), gamma-/X-ray pulsar (Gamma-/X-ray PSR), magnetar (MAG), X-ray dim isolated neutron star (XDINS); if the object is associated with a globular cluster (GC) or the Small Magellanic Cloud (SMC) this is reported in brackets; and (xiii) the reference for the proper motion measurements, indicated only if different from the ATNF catalogue, i.e., [1] \citet{Motch2009}, [2] \citet{Eisenbeiss2010}, [3] \citet{Walter2010}, [4] \citet{Stovall2014}, [5] \citet{Jennings2018}, [6] \citet{Perera2019}, [7] \citet{Dang2020}, [8] \citet{Danilenko2020}. 


\bibliographystyle{aasjournal}
\bibliography{biblio}


\begin{longrotatetable}
\movetabledown=9mm
\begin{deluxetable*}{ccccccccccccc}
\tablecaption{Up-to-date list of 417 neutron stars with measured proper motions in RA and DEC.
\label{tab:proper_motion}}
\tabletypesize{\tiny}
\tablecolumns{13}
\tablenum{5}
\tablewidth{0pt}
\tablehead{
\colhead{\multirow{2}{*}{JName}} &
\colhead{RA} &
\colhead{DEC} &
\colhead{$\mu_{\rm RA}$} & 
\colhead{$\mu_{\rm DEC}$} & 
\colhead{parallax} &
\colhead{Pos. Epoch} &
\colhead{$P$} &
\colhead{$\dot{P}$} & 
\colhead{\acs{DM}} & 
\colhead{$d_{\odot}$} & 
\colhead{\multirow{2}{*}{Class(Assoc.)}} & 
\colhead{\multirow{2}{*}{Ref.}} \\
& 
\colhead{[hms]} & 
\colhead{[$^{\circ}$ ' '']} & 
\colhead{[mas yr$^{-1}$]} & 
\colhead{[mas yr$^{-1}$]} & 
\colhead{[mas]} & 
\colhead{[MJD]} & 
\colhead{[s]} & 
\colhead{[s s$^{-1}$]} & 
\colhead{[pc cm$^{-3}$]} & 
\colhead{[kpc]} & 
& 
}
\startdata
J0014+4746   & 00:14:17.75(4)	     &  +47:46:33.4(3)      &	  $   19.3     \pm  1.8   $   &   $   -19.7    \pm   1.5   $   &   $   \_		     $   &  49664      &   $	1.24070     $	&    $    5.6446 \times 10^{-16	   } 	    $	&     $     30.405     (13)    $   &	 1.776       &  PSR		  &	    \\  		 					   
J0023+0923   & 00:23:16.87910(2)     &  +09:23:23.8689(8)   &	  $   -12.4    \pm  0.5   $   &   $   -6.1     \pm   0.1   $   &   $   0.4	\pm   0.3    $   &  56567      &   $	0.00305     $	&    $    1.14234 \times 10^{-20   } 	    $	&     $     14.32810  (4)      $   &	 1.248       &  PSR		  &   [6]   \\  		 					   
J0024-7204ab & 00:24:08.1615(5)      &  -72:04:47.602(2)    &	  $   4.2      \pm  0.6   $   &   $   -2.9     \pm   0.5   $   &   $   \_		     $   &  51600      &   $	0.00370     $	&    $    9.820 \times 10^{-21	   } 	    $	&     $     24.37     (2)      $   &	 2.54	     &  PSR(GC)	  	  &	    \\       		 					   
J0024-7204C  & 00:23:50.3546(1)      &  -72:04:31.5048(4)   &	  $   5.2      \pm  0.1   $   &   $   -3.1     \pm   0.1   $   &   $   \_		     $   &  51600      &   $	0.00576     $	&    $    -4.98503 \times 10^{-20  } 	    $	&     $     24.5955   (8)      $   &	 2.594       &  PSR(GC)	  	  &	    \\  		 					   
J0024-7204D  & 00:24:13.88092(6)     &  -72:04:43.8524(2)   &	  $   4.24     \pm  0.07  $   &   $   -2.24    \pm   0.05  $   &   $   \_		     $   &  51600      &   $	0.00536     $	&    $    -3.4220 \times 10^{-21   } 	    $	&     $     24.7432   (9)      $   &	 2.631       &  PSR(GC)	  	  &	    \\  		 					   
J0024-7204E  & 00:24:11.10528(5)     &  -72:05:20.1492(2)   &	  $   6.15     \pm  0.03  $   &   $   -2.35    \pm   0.06  $   &   $   \_		     $   &  51600      &   $	0.00354     $	&    $    9.85103 \times 10^{-20   } 	    $	&     $     24.236    (2)      $   &	 2.509       &  PSR(GC)	  	  &	    \\
\enddata

\end{deluxetable*}

\end{longrotatetable}

\end{document}